\documentclass[aps,prb,twocolumn,superscriptaddress,preprintnumbers,amsmath,amssymb,axodraw]{revtex4-2}

\usepackage{latexsym}
\usepackage{comment}
\usepackage{color}
\usepackage{amssymb}
\usepackage{mathtools}
\usepackage{amsmath}
\usepackage{bm}
\usepackage{amsfonts}
\usepackage{bbold}
\usepackage{cancel}
\usepackage{slashed}
\usepackage{appendix}
\usepackage{float}
\usepackage{cancel}
\usepackage{graphicx} 
\usepackage{dcolumn} 
\usepackage{braket}
\usepackage{tikz}
\usepackage{tikz-feynman}
\usepackage{axodraw2}
\usepackage{subcaption}
\usepackage{graphicx}
\usepackage[font={small}]{caption}
\usepackage{amsbsy}
\usepackage{orcidlink}
\usepackage{soul}
\DeclareMathOperator*{\SumInt}{%
\mathchoice%
{\ooalign{$\displaystyle\sum$\cr\hidewidth$\displaystyle\int$\hidewidth\cr}}
  {\ooalign{\raisebox{.14\height}{\scalebox{.7}{$\textstyle\sum$}}\cr\hidewidth$\textstyle\int$\hidewidth\cr}}
  {\ooalign{\raisebox{.2\height}{\scalebox{.6}{$\scriptstyle\sum$}}\cr$\scriptstyle\int$\cr}}
  {\ooalign{\raisebox{.2\height}{\scalebox{.6}{$\scriptstyle\sum$}}\cr$\scriptstyle\int$\cr}}
}

\newcommand{\pd}{\partial}
\newcommand{\bea}{\begin{eqnarray}}
\newcommand{\eea}{\end{eqnarray}}
\newcommand{\be}{\begin{equation}}
\newcommand{\ee}{\end{equation}}

\newcommand{\mbf}{\mathbf}

\newcommand{\bs}{\boldsymbol}
\newcommand{\nn}{\nonumber}
\newcommand{\ii}{\mathrm{i}}

\newcommand{\p}{\pi}

\newcommand{\lb}{\lambda}

\newcommand{\sign}{\text{sign}}

\begin{document}

\title{Electron-phonon interactions and instabilities in Weyl semimetals under magnetic fields and torsional strain}
\author{Fabi\'an Jofr\'e Parra}
\affiliation{Facultad de F\'isica, Pontificia Universidad Cat\'olica de Chile, Vicu\~{n}a Mackenna 4860, Santiago, Chile}

\author{Daniel A. Bonilla}
\email{daniel.bonillam@correo.nucleares.unam.mx}
\affiliation{Instituto de Ciencias Nucleares, Universidad Nacional Aut\'{o}noma de M\'{e}xico, 04510 Ciudad de M\'{e}xico, M\'{e}xico}

\author{Enrique Mu\~noz~\orcidlink{0000-0003-4457-0817}}
\email{ejmunozt@uc.cl}
\affiliation{Facultad de F\'isica, Pontificia Universidad Cat\'olica de Chile, Vicu\~{n}a Mackenna 4860, Santiago, Chile}
\affiliation{Center for Nanotechnology and Advanced Materials CIEN-UC, Avenida Vicuña Mackenna 4860, Santiago, Chile}

\date{\today}

\begin{abstract}
We study the presence of an external magnetic field, in combination with torsional strain, over the electron-phonon interactions in a type I Weyl semimetal. This particular superposition of field and strain, modeled in the continuum approximation by an effective gauge field, leads to an asymmetric pseudo-magnetic field at each Weyl node of opposite chirality. Therefore, we also studied the role of nodal asymmetry in the properties of the system by means of the Kadanoff-Wilson renormalization group and the corresponding flow equations. By solving those, we discuss the evolution of the coupling parameters of the theory, and analyze possible fixed points and lattice (Peierls) instabilities emerging from interactions between phonons with the chiral Landau level in the very strong pseudo-magnetic field regime.
\end{abstract}

\maketitle 

\section{Introduction}
Among the so-called "quantum materials", Weyl semimetals (WSMs) were discovered nearly a decade ago in TaAs crystals\cite{Xu613}, after being proposed as a theoretical concept~\cite{Xiangang2011,Fang2012,Ruan2016,Dirac,Felser,3dWeyl,Burkov}. WSMs are gapless three-dimensional materials with non-trivial topological properties, as their band structure displays an even number of Weyl nodes. Near each node, charge carriers are massless quasi-particles with linear dispersion and pseudo-relativistic properties\cite{Dirac,Felser,3dWeyl,Burkov}. In particular, each node represents a monopole source of Berry curvature and, therefore, is protected from being distorted because their topological charge (chirality) is an invariant \cite{Burkov}. Therefore in Weyl Fermions chirality, i.e. the projection of spin over their momentum direction, is preserved, a condition usually named as "spin-momentum locking". Some additional remarkable properties of WSMs are the presence of Fermi arcs\cite{Xu613}, the chiral anomaly, and the chiral magnetic effect\cite{vanderbilt}. Beyond this single-particle picture, the effects of electronic interactions and strong correlations~\cite{Yang_2019} in WSMs have been explored using quantum field theory methods~\cite{Roy_2017,Laubach_2016,Xue_2017,Carlstrom_2018,Wei_2012,Witczak-Krempa_2014,Maciejko_2014,Wei_2014}, and there is theoretical evidence suggesting a rich structure that may lead to novel topological and magnetic phases~\cite{Munoz_2020,Munoz_2024,Zhai_2016,Wei_Chao_2014,Wang_2017}. 

The effects of intense magnetic fields on the electronic properties of materials with Dirac or Weyl spectra is a subject of great interest. In this regime, the electronic spectrum is semi-discrete in the form of pseudo-relativistic Landau levels, thus acquiring an effective dimensionality reduction from 3D to 1D~\cite{Zhu_NatPhys_2010}. The nesting property of 1D electronic systems thus precludes the emergence of instabilities that may give rise to strongly correlated phases such as the charge density wave (CDW)~\cite{Yakovenko_PhysRevB_1993} or the spin density wave (SDW)~\cite{TAKAHASHI1994384}. This phenomenon has been observed experimentally in graphite, where the existence of a field-induced many-body state beyond a critical magnetic field intensity was attributed to a CDW~\cite{Fauque_PhysRevLett_2013,Yaguchi_2009}. Moreover, the theoretical interpretation involved the interplay of only the lowest Landau levels (LLL) with $(n=0,\uparrow)$ and $(n=-1,\downarrow)$\cite{Arnold_PhysRevLett_2017}. The presence of very intense magnetic fields also affects the magnetoelectric response in WSMs~\cite{Zyunin_PhysRevB_2012}, partly due to the mixing of the LLLs, thus potentially breaking down the chiral anomaly~\cite{Kim_PhysRevLett_2017}. Many-body effects induced by the presence of intense magnetic fields are also reflected in the renormalization of the phonon spectrum~\cite{Rinkel_PhysRevB_2019, Kundu_2022}. Remarkably, an axial electron-phonon coupling can induce a dynamical chemical potential difference between Weyl nodes of opposite chirality, thus generating a dynamical realization of the chiral anomaly\cite{Hui_PhysRevB_2019}.
The effect of strong magnetic fields in interacting models for Weyl semimetals has been studied from a field theory perspective and renormalization group methods~\cite{Alicea_PhysRevB_2009,Kundu_2022}, indicating the presence of instabilities towards CDW phases. It has been reported~\cite{Kundu_2022} that the participation of the Cooper channel can prevent electronic instabilities but, in contrast, enhance lattice instabilities that are not of BCS type~\cite{Kundu_2022}. Moreover, breaking the mirror symmetry that links the Weyl nodes suppresses the Cooper channel, thus leading to an increase of the critical temperature for the instability~\cite{Kundu_2022}. Therefore, electron-phonon interactions, as suggested in~\cite{Kundu_2022}, seem to constitute an essential ingredient in the emergence of interacting phases induced by the presence of intense magnetic fields in topological materials.

The effects of mechanical strain as dislocations and disclinations in WSMs have been explored to a much lesser extent in the context of electronic properties. 
Those defects can be modeled in a continuum approximation by means of gauge fields~\cite{Cortijo_2015,Cortijo_2016,Arjona_Vozmediano_PRB2018,nature_1}, thus leading to the emergence of pseudo-magnetic fields in WSMs. The origin of those strain-induced pseudo-magnetic fields is understood from a basic lattice model, of the form~\cite{Pikulin_2016}
\bea
H^{latt} = \left( \begin{array}{cc} \epsilon_{\mathbf{k}}+h^{latt}(\mathbf{k}) & 0\\0 & \epsilon_{\mathbf{k}}-h^{latt}(\mathbf{k}) \end{array}\right),
\eea
with $\epsilon_{\mathbf{k}} = c_0 + c_1\cos(a k_z) + c_2\left( \cos(a k_x) + \cos(a k_y) \right)$, and
\bea
h^{latt}(\mathbf{k}) = \tau^{z}m_{\mathbf{k}} + \Lambda \left( \tau^{x}\sin(a k_x) + \tau^{y}\sin(a k_y) \right),
\eea
where $m_{\mathbf{k}} = t_0 + t_1\cos(a k_z) + t_2\left(\cos(a k_x) + \cos(a k_y)  \right)$, and $t_j$ (for $j=0,1,2$) and $\Lambda$ parameters chosen in order to match, via a $\mathbf{k}\cdot\mathbf{p}$ approximation near the $\Gamma$ point, a first-principles band structure calculation~\cite{Pikulin_2016,Wang_PhysRevB_2012,Wang_PhysRevB_2013}. This minimal model exhibits a pair of Weyl nodes~\cite{Pikulin_2016} $\mathbf{K}_{\xi} = (0,0,\xi Q)$ (for $\xi = \pm$), with $Q = a^{-1}\arccos\left( -\left( t_0 + 2 t_2 \right)/t_1\right)$. A low-energy expansion in the limit $a Q \ll 1$ near each node $h^{latt}\left(\mathbf{K}_{\xi}\pm\mathbf{k}\right)$, leads to the effective Hamiltonian representation
\bea
h_{\xi}(\mathbf{k}) = \hbar v_{\xi}^{j}\tau^{j}k_j,
\eea
with the velocity components 
\bea
\mathbf{v}_{\xi} = \hbar a^{-1}\left( \Lambda,\Lambda,-\xi t_1 \sin(a Q) \right).
\eea
The corresponding chiral charge at each node is then given by $\sign(v_{\xi}^x v_{\xi}^y v_{\xi}^{z})= -\xi$.
In the context of this lattice model, elastic deformations are characterized by the displacement vector field $\mathbf{u} = (u_1,u_2,u_3)$, and the corresponding symmetric component of the strain tensor $u_{ij} \equiv (1/2)\left( \partial_i u_j + \partial_j u_i \right)$. Those in turn induce a change in the hopping term, as follows~\cite{Pikulin_2016}
\bea
t_1\tau^{x} \rightarrow t_1(1 - u_{33})\tau^z + \ii\Lambda \sum_{j\ne3}u_{3j}\tau^j.
\eea
This introduces a modification of the lattice Hamiltonian~\cite{Pikulin_2016}
\bea
\delta h^{latt}(\mathbf{k}) &=& -t_1 u_{33}\tau^z\cos(a k_z) + \Lambda\left( u_{13}\tau^x\right.\nn\\
&&\left.- u_{23}\tau^y \right)\sin(a k_z).
\eea
After expanding this modified Hamiltonian, as before, in the vicinity of each Weyl node $\mathbf{K}_{\xi}$, one obtains the effective Hamiltonian~\cite{Pikulin_2016}
\bea
H_{\xi}(\mathbf{k}) = v_{\xi}^j\tau^j \left( \hbar k_j - \xi \frac{e}{c} A_j^S \right),
\eea
where we can identify the presence of the strain-induced "gauge field"~\cite{Pikulin_2016}
\bea
\mathbf{A}^S &=& -\frac{\hbar c}{e a}\left( u_{13}\sin(a Q),u_{23}\sin(a Q),u_{33}\cot(a Q) \right).
\label{eq_Astrain1}
\eea
A particular case, applicable to a crystal in the shape of a cylindrical rod of length $L$ along the z-axis, and subjected to a uniform mechanical torsion by a total angle $\theta$ around the axis, is given by the vector displacement field~\cite{Pikulin_2016,Arjona_Vozmediano_PRB2018},
\bea
\mathbf{u} = \theta \frac{z}{L} \left(\mathbf{r}\times\hat{z}\right).
\eea
Here $\mathbf{r}$ represents the radial vector from the z-axis of the cylindrical rod of material. The only non-zero components of the strain tensor for this case are $u_{13} = \left( \theta/2L \right)y$ and $u_{23} = -\left(  \theta/2L\right)x$, respectively. For this particular torsional strain pattern, the corresponding gauge field (in the limit $Q a \ll 1$) is given by~\cite{Pikulin_2016,Arjona_Vozmediano_PRB2018}
\bea
\mathbf{A}^S = \frac{B_S}{2}\left( -y,x,0 \right),
\label{eq_Astrain2}
\eea
with $B_S = \left(\hbar c/e\right)\left(\theta Q/L\right)$ the magnitude of the associated "pseudo-magnetic field" defined via the curl~\cite{Pikulin_2016,Arjona_Vozmediano_PRB2018}
\bea
\mathbf{B}_S \equiv \nabla\times\mathbf{A}^S = \hat{z} B_S.
\label{eq_Bstrain}
\eea
Clearly, this result indicates that even though the torsional strain-induced gauge field grows linearly with distance after Eq.~\eqref{eq_Astrain1} or Eq.~\eqref{eq_Astrain2}, the associated pseudo-magnetic field that carries over the physical consequences of this effect turns out to be constant and uniform across the cylindrical sample. There is of course a gauge freedom involved in the particular representation for the gauge field that induces this constant pseudo-magnetic field. For instance, a $U(1)$-transformation to the Landau gauge $\mathbf{A}^S \rightarrow B_S (0,x,0)$ will represent exactly the same pseudo-magnetic field~\cite{Pikulin_2016,Arjona_Vozmediano_PRB2018}.

In previous works~\cite{Munoz2019,Soto_Garrido_2020,Bonilla_Munoz_2021}, we studied quasi-ballistic transport through a nano-junction in a WSM with a single torsional dislocation, in combination with an external physical magnetic field. For this system, we obtained the electronic~\cite{Munoz2019,Soto_Garrido_2020} and thermoelectric~\cite{Munoz2019,Bonilla_Munoz_2021} transport coefficients, using the Landauer ballistic formalism in combination with a mathematical analysis for the quantum mechanical scattering cross-section~\cite{Munoz_2017}. More recently, we also considered the case of a diluted, uniform concentration of torsional dislocations and their effects on the electrical conductivity of type I WSMs~\cite{Bonilla_Munoz_2022}, in the Kubo linear-response formalism combined with vertex corrections~\cite{Bonilla_Munoz_2022} in a Bethe-Salpeter equation. 

In the present work, our purpose is to study, from a field theoretical renormalization group perspective, the effects of a pseudo-magnetic field arising from a combination of an external magnetic field and the presence of torsional strain in a generic type I Weyl semimetal with a minimal number of two nodes. As discussed in the text, such a combined magneto-strain pseudo-magnetic field breaks the symmetry between the Weyl nodes and in principle can achieve extremely high intensities tanks to the strain component. Therefore, our calculations are mainly focused on the role of LLLs. We shall explore the emergence of Peierls instability leading to a CDW phase, and how the presence of strain, particularly in its role at breaking the nodal symmetry, may affect the emergence of such strongly-correlated phase.

The paper is organized as follows: In section~\ref{sec_model}, we present the effective Hamiltonian model at the single-particle level, while in section~\ref{sec_interactions} we define the different interaction terms. In section~\ref{field_theory_ren} we present the Kadanoff-Wilson scheme for the renormalization group (RG) analysis of the model, and in section~\ref{sec_rnLLL} we specialize it to the restriction of the LLL subspace, in order to obtain the corresponding RG flow equations for the couplings. In section~\ref{sec_Peierls_instab} we solve the RG equations, in order to analyze the emergence of the Peierls instability and its corresponding critical temperature. Finally, we present our discussion and conclusions. All mathematical derivations and details are presented in a set of Appendixes.

\section{The model}
\label{sec_model}
\begin{figure}[t]
    \centering
    \includegraphics[scale=0.5]{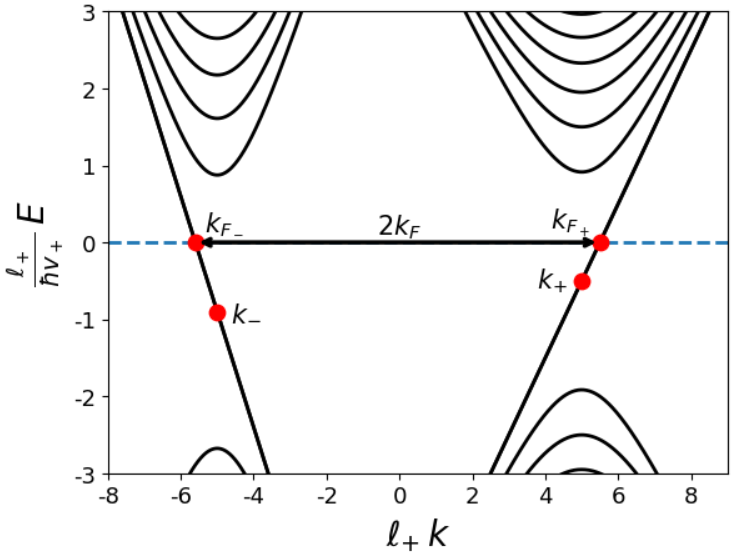}
    \caption{Non-interacting energy dispersion in Eq.~\eqref{eq:eigenvalues_2}, illustrated for the choice of parameters $b_0 = 0.2 \hbar v_+/\ell_+$, $v_- = 1.5 v_+$, $\ell_- = \ell_+/\sqrt{0.7}$, and $k_{\pm} = \pm\, 5/\ell_+$. The corresponding Fermi energy is $0.5 \hbar v_+ / \ell_+$ above $k_+$.}
    \label{fig:dispersion-relation}
\end{figure}

We start by considering a minimal low-energy continuum model for a Weyl semimetal involving two nodes of opposite chirality, represented by the Hamiltonian
\begin{align}
    \hat{H}_{\xi}=\xi b_0 \sigma_0 + \xi v_{\xi} \bs{\sigma}\cdot\left(\mbf{p}-\hbar\mathbf{K}_{\xi} \right). \label{eq:hamiltonian_1}
\end{align}
Here $\xi=\pm$ is the chirality index of each Weyl node, $v_{\xi}$ is the Fermi velocity at each node, $b_0$ is the local shift from the zero energy, $\sigma_{i}$ with $i=1,2,3$ are the Pauli matrices, and $\sigma_0$ is the identity matrix. The location of each Weyl node in momentum space is $\mathbf{K}_{\xi} = k_{\xi}\mbf{\hat{z}}$.
We shall consider the presence of an external magnetic field along the $\hat{z}$-direction $\mathbf{B}_0 = \hat{z} B_0$, that corresponds to the curl of a gauge field (in Landau gauge) $\mathbf{A}_0 = \hat{y}x B_0$. In addition, we shall also consider the presence of torsional strain, modeled by a pseudo-gauge field~\cite{Cortijo_2015,Cortijo_2016,Arjona_Vozmediano_PRB2018} $\mathbf{A}^S_{\xi} = \xi \hat{y} x B_S $ that exhibits opposite directions depending on the chirality of each node $\xi = \pm$. The corresponding superposition of the magnetic and elastic gauge fields is thus given by
\bea
\mathbf{A}_{\xi} = \mathbf{A}_0 + \mathbf{A}^S_{\xi} = \hat{y}x\left(  B_0 + \xi B_S\right) \equiv \hat{y}x B_{\xi},
\eea
thus clearly leading to an effective pseudo-magnetic field $B_{\xi} = B_0 + \xi B_S$ with different magnitudes depending on the chirality of each node. This effect is incorporated by minimal coupling to the Hamiltonian Eq.~\eqref{eq:hamiltonian_1}
\begin{align}
     \hat{H}_{\xi}=\xi b_0 \sigma_0 + \xi v_{\xi} \bs{\sigma}\cdot\left[\mbf{p}+e\mbf{A}_{\xi}(x)\right], \label{eq:hamiltonian_2}
\end{align}
where, for convenience, we define the momentum operators at each node
$\mathbf{p}-\hbar\mathbf{K}_{\xi}\rightarrow\mathbf{p}$. Since this effective single-particle Hamiltonian depends on the spatial coordinate $x$, we need to define an appropriate eigenbasis to construct the corresponding many-body theory. As shown in detail in Appendix~\ref{App_Landau_basis}, this turns out to be the spinor eigenfunctions
\bea
    \Psi(\mbf{x}) = \frac{e^{- i\left(X_{\xi}/\ell_{\xi}^2\right)y + i(k+k_{\xi})z}}{\sqrt{L_y L_z}}\begin{pmatrix}
        u_{\xi \lb nk}\,\varphi_{n-1}(x-X_{\xi})  \\  v_{\xi \lb nk}\,\varphi_{n}(x-X_{\xi})
    \end{pmatrix} \label{eq:spinor_2}
\eea
where we defined the coefficients
\bea
    u_{\xi\lb nk}= \frac{i\lb}{\sqrt{2}}\sqrt{1+\frac{\hbar v_{\xi} k}{\lb \epsilon_{\xi nk}}},\,\,\,v_{\xi\lb nk}=\frac{1}{\sqrt{2}}\sqrt{1-\frac{\hbar v_{\xi} k}{\lb \epsilon_{\xi nk}}}, \label{eq:def_uv}
\eea
with $\lambda = \pm$ the band index and
\begin{equation}
    \epsilon_{\xi nk} = \hbar v_{\xi} \sqrt{k^2 + \frac{2n}{\ell_{\xi}^2}}
    \label{eq:eigenvalues_2}
\end{equation}
the energy eigenvalues (for $n\ge 0$) for discrete Landau levels. Here, $k\in\mathbb{R}$ is the continuum component of the momentum eigenvalue in the $z$-direction. We also define here the "magnetic length" (or Landau radius) associated with the magnitude of the pseudo-magnetic field at each Weyl node
\be
\ell_{\xi} = \sqrt{\hbar/|e B_{\xi}|}.
\label{eq:lchi}
\ee
The corresponding eigenfunctions representing each spinor component, and centered at the "nodal guiding center" $X_{\xi} = \hbar k_2/(e B_{\xi})$ are given by (see Appendix~\ref{App_Landau_basis})
\begin{align}
    \varphi_{n}(x)=\frac{(-1)^n}{\sqrt{2^{n}n!}} \left( \frac{1}{\pi \ell_{\xi}^2} \right)^{1/4} e^{-x^2/2\ell_{\xi}^2}\, H_{n}\left( \frac{x}{\ell_{\xi}} \right),\label{eq:def_phi_n}
\end{align}
with $H_{n}(z)$ a Hermite polynomial of order $n\ge 0$. The single-particle spectrum at the non-interacting level of the model is depicted in Fig.~\ref{fig:dispersion-relation}, where $\lambda = \pm$ represents the band index in the non-interacting energy spectrum. 
It is important to distinguish the particular role of the lowest Landau level $n = 0$, $\lambda = -1$, whose spinor eigenfunction is given by
\bea
    \Psi(\mbf{x})_{n=0,\lb=-1} = \frac{e^{- i\left(X_{\xi}/\ell_{\xi}^2\right)y + i(k+k_{\xi})z}}{\sqrt{2 L_y L_z}}\begin{pmatrix}
        0  \\  \varphi_{0}(x-X_{\xi})
    \end{pmatrix}. \label{eq:spinor_3}
\eea
In terms of this eigenbasis, we can now construct the corresponding many-particle Hamiltonian by defining annihilation $c_{\alpha}$ and creation $c_{\alpha}^{\dagger}$ fermion operators, satisfying anti-commutation relations $\left\{ c_{\alpha},c^{\dagger}_{\alpha'} \right\} = \delta_{\alpha,\alpha'}$, for a multi-indexed label 
$\alpha \equiv \left(\lambda,\xi,n,X_{\xi},k\right)$, such that
\bea
H_e &=& \sum_{\xi=\pm}\sum_{\lambda=\pm}\sum_{n=0}^{\infty}\int_{-\infty}^{\infty}\frac{dk}{2\pi}\int_{-\infty}^{\infty} dX_{\xi}\, c_{\alpha}^{\dagger}\left[\lambda\,\epsilon_{\xi n k}\right]c_{\alpha}\nn\\
&\equiv& \SumInt_{\alpha} c_{\alpha}^{\dagger}\epsilon_{\alpha}c_{\alpha}.
\eea
\section{Interactions}
\label{sec_interactions}
Let us now consider the fully interacting many-body Hamiltonian for our system, 
\bea
H = H_e + H_{ph} + H_{e-ph} + H_{ee}. 
\eea
Here, we introduced the phonon Hamiltonian 
\bea
H_{ph} = \sum_{\mathbf{q},j}\hbar\nu_{\mathbf{q},j}^{0}\left(  
a_{\mathbf{q},j}^{\dagger}a_{\mathbf{q},j} + \frac{1}{2}
\right),
\eea
where $a_{\textbf{q},j}$ and $a_{\textbf{q},j}^{\dagger}$ are the annihilation and creation operators for the phonon modes with momentum $\textbf{q}$ and polarization $j$, which satisfy the bosonic commutation relations $\left[ a_{\textbf{q},j}, a_{\textbf{q}',j'}^{\dagger}\right] = \delta_{\mathbf{q},\mathbf{q}'}\delta_{j,j'}$. Likewise, $\nu_{\mathbf{q},j}^{0}$ is the corresponding unrenormalized (or bare) frequency for such vibrational modes. 

In particular, we shall focus on acoustic phonon branches such that 
the electron-phonon interaction is described by the Hamiltonian
\be
    H_{\text{e-ph}} = \SumInt_{\alpha} \SumInt_{\alpha'} \sum_{\textbf{q},j} \Tilde{g}_{\alpha \alpha' , j}^{\text{ep}}(\textbf{q}) c^{\dag}_{\alpha} c_{\alpha'} \left( a^{\dag}_{\textbf{q},j} +  a_{-\textbf{q},j} \right),
    \label{eq:He-ph}
\ee
where the couplings 
\bea
    \Tilde{g}^{ep}_{\alpha \alpha' , j}(\textbf{q}) &=& \Tilde{g}_{\xi \xi',j}(\textbf{q})\mathcal{M}_{\alpha}^{\alpha'}(\textbf{q})
    \label{eq:electron-phonon-matrix-elements_1}
\eea
are determined by the matrix elements in the magnetic Landau eigenbasis as follows
\bea    \mathcal{M}_{\alpha}^{\alpha'}(\textbf{q}) =\int d^3x \; \Psi^{\dag}_{\alpha}(\textbf{x}) e^{i\textbf{q} \cdot \textbf{x}} \Psi_{\alpha'}(\textbf{x}).    \label{eq:electron-phonon-matrix-elements}
\eea
Here $\Tilde{g}_{\xi \xi',j}(\textbf{q})$ is the deformation potential coefficient~\cite{Kundu_2022}
\begin{equation}
    \Tilde{g}_{\xi \xi',j}(\textbf{q}) \sim \frac{\hbar |\textbf{q}|^2 d^2_{\text{ac},j}}{2\rho \mathcal{V} \nu_{\mathbf{q},j}^0} = \frac{\hbar |\textbf{q}| d^2_{\text{ac},j}}{2\rho \mathcal{V} c_j}  \,, 
\end{equation}
and $d_{\text{ac},j}$ represents the acoustic deformation potential, for $\nu_{\mathbf{q},j}^0 = c_j |\textbf{q}|$ the unperturbed phonon frequency in the acoustic branch $j$. We also defined $\mathcal{V} = L_x L_y L_z$ as the volume of the system and $\rho$ as the ion mass density.

Finally, the electron-electron interaction is given by the Hamiltonian
\begin{equation}
    H_{ee} = \frac{1}{2\mathcal{V}} \sum_{\textbf{q}} \SumInt_{\alpha} \Tilde{g}^{\alpha_3 \alpha_4}_{\alpha_1 \alpha_2} c^{\dag}_{\alpha_1} c_{\alpha_3} c^{\dag}_{\alpha_2} c_{\alpha_4} \, ,
\end{equation}
with a coupling constant given in terms of the interaction matrix elements in the magnetic Landau eigenbasis, as follows:
\begin{eqnarray}
    \Tilde{g}^{\alpha_3 \alpha_4}_{\alpha_1 \alpha_2} (\textbf{q}) &=& V(\textbf{q}) 
     \mathcal{M}_{\alpha_1}^{\alpha_3}(\textbf{q})\mathcal{M}_{\alpha_2}^{\alpha_4}(-\textbf{q}).
    \label{eq:electron-electron-matrix-element}
\end{eqnarray}
Here, $V(\textbf{q}) = e^2/(\epsilon_0 \epsilon_{\infty} \textbf{q}^2)$ is the Fourier transform of the unscreened Coulomb potential and $\epsilon_{\infty}$ is the high-frequency dielectric constant.
\section{Field theoretical representation and renormalization analysis}
\label{field_theory_ren}
Since our main interest is to study the renormalization flow of the coupling parameters involved in the different terms of the interacting Hamiltonian, it is technically necessary to switch to a field-theoretical representation. In this language, and by means of the standard techniques involving coherent states for both fermions and bosons, the functional integral representation for the partition function of the model at finite temperature $T > 0$ is given by
\begin{equation}
    \mathcal{Z} = \int \mathcal{D} \Psi^{\dag} \mathcal{D} \Psi \mathcal{D} \Phi \, e^{-\frac{1}{\hbar} S\left[ \Psi^{\dag}, \Psi, \Phi \right]} , \label{eq:partition_function}
\end{equation}
where the action $S$ corresponds to the sum of the following terms:
\bea
    S^0_e\left[ \Psi^{\dag}, \Psi \right] &=& \SumInt_{\alpha} \sum_{\omega_n} \left[ \mathcal{G}^0_{\alpha}(\omega_n) \right]^{-1} \Psi_{\alpha}^{\dag}(\omega_n) \Psi_{\alpha}(\omega_n),\label{eq:S0_e}
\eea
\bea
    S^0_{ph}\left[ \Phi \right] &=& \sum_{\mathbf{q}, j} \sum_{\nu_m}\mathcal{D}_{j,0}^{-1}(\mathbf{q},\nu_m) \left| \Phi_j(\textbf{q},\nu_m) \right|^2,\label{eq:S0_ph}
\eea
\bea
&&S_{ee}\left[ \Psi^{\dag}, \Psi \right] = \frac{\pi v_F}{\beta \mathcal{V}} \SumInt_{\alpha}\sum_{\omega_n}\sum_{\textbf{q}} g_{\alpha_1 \alpha_2}^{\alpha_3 \alpha_4}(\textbf{q}) \Psi_{\alpha_1}^{\dag}(\omega_{n_1}) \nn\\
&&\times \Psi_{\alpha_2}^{\dag}(\omega_{n_2})\Psi_{\alpha_3}(\omega_{n_3}) \Psi_{\alpha_4}(\omega_{n_1 + n_2 - n_3})  ,\label{eq:S_ee}
\eea
\bea
    &&S_{ep}\left[ \Psi^{\dag}, \Psi, \Phi \right] = \sqrt{\frac{\pi v_F}{\beta \mathcal{V}}} \sum_{\omega_n} \sum_{\nu_m} \sum_{\textbf{q} j} \SumInt_{\alpha}\SumInt_{\alpha'} g^{ep}_{\alpha \alpha'}(\textbf{q})\nn\\ 
    &&\times\Psi_{\alpha}^{\dag}(\omega_n + \nu_m) \Psi_{\alpha'}(\omega_n) \Phi_j(\textbf{q},\nu_m),\label{eq:S_ep}
\eea
with $\beta=1/(k_B T)$ the inverse temperature, $\omega_{n} = \pi (2n+1)/\beta$ and $\nu_{m} = 2\pi m/\beta$ fermionic and bosonic Matsubara frequencies for $n,\,m\in\mathbb{Z}$ respectively. 
For technical reasons, we redefine the couplings as follows:
\begin{eqnarray}
    g^{ep}_{\alpha \alpha', j}(\textbf{q}) &=& \frac{2}{\hbar} \sqrt{\frac{\nu_{\mathbf{q},j}^0 \mathcal{V}}{\pi v_F}} \Tilde{g}^{ep}_{\alpha \alpha', j}(\textbf{q}) \\
    g_{\alpha_1 \alpha_2}^{\alpha_3 \alpha_4}(\textbf{q}) &=& \frac{\Tilde{g}_{\alpha_1 \alpha_2}^{\alpha_3 \alpha_4}(\textbf{q})}{2\pi \hbar v_F} \, ,
\end{eqnarray}
where we have introduced the Fermi velocity at the $\xi = +$ node as $v_F \equiv v_+$ so that at the opposite node we have
\begin{equation}
    v_- \equiv v_F + \delta v \, ,   \label{eq:v_F-delta-v}
\end{equation}
with $\delta v = v_{-} - v_{+}$ representing a possible nodal asymmetry.
We also defined the electron and phonon bare Green's functions in Matsubara space by
\bea
    \mathcal{G}^0_{\alpha}(\omega_n) &=& (-i\hbar \omega_n + \epsilon_{\alpha} - \mu)^{-1} \label{eq:G0} \\
    \mathcal{D}_{\mathbf{q},j}^0 (\nu_m) &=& \left[ \nu_m^2 + \left[\nu_{\mathbf{q},j}^{0}\right]^2 \right]^{-1}. \label{eq:D0}
\eea
Now, we shall establish a renormalization flow analysis of the model. Following Wilson's procedure, we introduce a cutoff for the $z$-component of the momentum $k$, defined by $\Lambda_n(l) = \Lambda_n e^{-l}$ for each Landau level $n$, with $l>0$. We separate the momentum components of the fermion fields accordingly, such that
\bea
    \Psi_{\alpha}(\omega_m) &\longrightarrow& \begin{cases}
          \Psi_{\alpha}(\omega_m) \quad &, \, k < \Lambda_n e^{-l} \\
          \Bar{\Psi}_{\alpha}(\omega_m) \quad &, \, \Lambda_n e^{-l} \leq k < \Lambda_n \\
     \end{cases} 
\eea
Hence, the partition function is now given by
\begin{equation}
    \mathcal{Z} = \int \mathcal{D} \Psi^{\dag} \mathcal{D} \Psi \mathcal{D} \phi \int_{>} \mathcal{D} \Bar{\Psi}^{\dag} \mathcal{D} \Bar{\Psi} \, e^{-\frac{1}{\hbar} S\left[ \Psi^{\dag} + \Bar{\Psi}^{\dag}, \Psi + \Bar{\Psi}, \phi \right]} \, ,
\end{equation}
where the subindex $>$ stands for integrating over the large momentum $\Lambda_n e^{-l} \le k < \Lambda_n$ field components.
By taking into account the orthogonality between the low-$\Psi_{\alpha}(\omega_n)$ and high momentum $\Bar{\Psi}_{\alpha}(\omega_n)$ components of the field, the terms in the action can be separated to obtain
\begin{eqnarray}
    \mathcal{Z} &=& \int \mathcal{D} \Psi^{\dag} \mathcal{D} \Psi \mathcal{D} \phi \, e^{-S\left[ \Psi^{\dag}, \Psi, \phi \right]} \int_{>} \mathcal{D} \Bar{\Psi}^{\dag} \mathcal{D} \Bar{\Psi} \, e^{-S^0_e\left[ \Bar{\Psi}^{\dag}, \Bar{\Psi} \right]} e^{-S_I}  \nonumber \\
    &=& \mathcal{N} \int \mathcal{D} \Psi^{\dag} \mathcal{D} \Psi \mathcal{D} \phi \, e^{-S\left[ \Psi^{\dag}, \Psi, \phi \right]} \braket{e^{-S_I}} \nonumber \\
    &=& \mathcal{N} \int \mathcal{D} \Psi^{\dag} \mathcal{D} \Psi \mathcal{D} \phi \, e^{-S\left[ \Psi^{\dag}, \Psi, \phi \right]} e^{-\braket{S_I} \, + \, \frac{1}{2}\left( \braket{S_I^2} - \braket{S_I}^2 \right) \, + \, ...} \,\nn\\ \label{eq:cumulant-expansion}
\end{eqnarray}
where we defined the effective interaction
\bea
S_{I} = S_{ee}\left[ \Psi^{\dagger} + \Bar{\Psi}^{\dagger}, \Psi + \Bar{\Psi}\right] + S_{e-ph}\left[\Psi^{\dagger} + \Bar{\Psi}^{\dagger}, \Psi + \Bar{\Psi},\Phi\right]\nn\\
\eea
Here we used the cumulant expansion, where $\braket{\cdot\cdot}$ denotes the ensemble average for the given operator, integrated over the field components $\Bar{\Psi}_{\alpha}(\omega_n)$ in the high-momentum sector, \textit{i.e.,} $\Lambda_ne^{-l} \leq k < \Lambda_n$, and $\mathcal{N}$ is a global constant. Each of these ensemble averages, leading to correlation functions of different orders, is computed by means of a Feynman diagram expansion, as explained in detail in the Appendix~\ref{app_rengen}.

 \section{Explicit renormalization in the high pseudo-magnetic field regime}
\label{sec_rnLLL}
Along this section, we shall apply the general one-loop renormalization scheme discussed before, to obtain explicit results in the regime of very strong pseudo-magnetic fields at both nodes. This physical condition is in principle possible to achieve in the presence of strain, since the pseudo-magnetic fields related to strain are predicted to be much stronger~\cite{Cortijo_2015,Cortijo_2016,Arjona_Vozmediano_PRB2018,nature_1} than the actual magnetic fields that can be imposed under normal laboratory conditions $B_0/B_S \ll 1$, and hence we have $\ell_{\xi}^2 \sim B_S^{-1} \ll k_F^{-1}$. In this regime, the expansion over Landau levels is highly dominated (due to the presence of a large gap) by the lowest Landau level. 

We first need to evaluate the matrix elements according to Eq.~\eqref{eq:electron-phonon-matrix-elements} for the lowest Landau level, with $\alpha = \left(\lambda = -1,\xi,n = 0,X_{\xi},k\right)$. For backward tunneling, 
where $\Bar{\xi} \equiv -\xi$, $\alpha' = (\lambda' = -1,\Bar{\xi},n' = 0,X'_{\Bar{\xi}},k')$, we obtain (see Appendix~\ref{App_Landau_Matrix}) 
\begin{eqnarray}
&&\mathcal{M}_{\alpha}^{\alpha'}(\textbf{q}) 
    = \frac{\ell_G}{\ell_A} \delta_{k'+k_{\Bar{\xi}}, k+k_{\xi}-q_z} \delta_{X'_{\Bar{\xi}}/\ell^2_{\Bar{\xi}}, X_{\xi}/\ell^2_{\xi}+q_y}  \nonumber \\
    && \times e^{-\frac{1}{4} \frac{\ell_G^4}{\ell_A^2} q_x^2} e^{-\frac{1}{4} \frac{\ell_{\bar{\xi}}^4}{\ell_A^2} q_y^2} e^{\frac{i}{2} \frac{\ell_G^4}{\ell_A^2} q_x q_y}  e^{i\frac{\ell_{\bar{\xi}}^2}{\ell_A^2}q_x X_{\xi}},  \label{eq:M_backward}
\end{eqnarray}
where for notational convenience we defined the magnetic length scales
\begin{equation}
    \ell_A \equiv  \sqrt{\frac{\ell_{+}^2 + \ell_{-}^2}{2}} \hspace{4mm} \text{and} \hspace{4mm} \ell_G \equiv  \sqrt{\ell_{+} \ell_{-}} \, .   \label{eq:length-scales-2}
\end{equation}

On the other hand, for forward tunneling (where $\xi' = \xi$) the corresponding matrix element reduces to the expression (see Appendix~\ref{App_Landau_Matrix}) 
\begin{eqnarray}
    &&\mathcal{M}_{\alpha}^{\alpha'}(\textbf{q}) = \delta_{k', k-q_z} \delta_{X'_{\xi}, X_{\xi} + \ell^2_{\xi} q_y}\nn\\ 
    &&\times e^{-\frac{1}{4} \ell_{\xi}^2 (q_x^2 + q_y^2)} e^{\frac{i}{2} \ell_{\xi}^2 q_x q_y} e^{i q_x X_{\xi} }   \, .  \label{eq:M_forward}
\end{eqnarray}
We shall consider only the LLL ($n=0$, $\lambda=-1$). Using the deltas to sum over $X_{2,\Bar{\xi}}$ and $k_2$, and defining
\begin{equation}
    \xi b \equiv k_{\xi} - k_{\bar{\xi}} \, ,
\end{equation}
with $b \equiv k_+ - k_- >0$, we obtain the simplified form of the electron-phonon component of the action: 
\begin{eqnarray}
    S_{ep} &=& \sqrt{\frac{\pi v_F}{\beta \mathcal{V}}} \frac{\ell_G}{\ell_A} \sum_{\xi} \sum_{\omega_n \nu_m} \sum_{\textbf{q}, j} \sum_{X_{1,\xi} } \sum_{k_1} \nonumber   \\
    && \times z_{\xi\bar{\xi}}\,g_{\xi \Bar{\xi},j} e^{i \frac{\ell_{\bar{\xi}}^2}{\ell_A^2} q_x X_{1,\xi}}  \Psi^{\dag}_{\xi ,X_{1,\xi}}(k_1, \omega_n + \nu_m) \nn\\ &&\times\Psi_{\Bar{\xi}, \varepsilon_{\Bar{\xi}}^2 X_{1,\xi} + \ell_{\Bar{\xi}}^2 q_y}(k_1 + \xi b - q_z, \omega_n)\Phi_j(\textbf{q},\nu_m), \,  \label{eq:S_ep_final}
\end{eqnarray}
where we included the vertex renormalization factor $z_{\xi\bar{\xi}} = \sqrt{z_{\xi}z_{\bar{\xi}}}$ arising from the field strength renormalization $\Psi_{\xi X_{\xi}}\rightarrow \sqrt{z_{\xi}}\Psi_{\xi X_{\xi}}$ and $\Psi_{\bar{\xi} X_{\bar{\xi}}}\rightarrow \sqrt{z_{\bar{\xi}}}\Psi_{\bar{\xi} X_{\bar{\xi}}}$, respectively.
Let us now consider the electron-electron interaction. For this purpose, we will distinguish the short-wavelength part of the Coulomb interaction, involved in backward scattering, from the long-wavelength part, which is mainly involved in forward scattering.

Let us start from the backward scattering contribution by defining the distance in momentum space between both nodal momenta,
\begin{equation}
    2k_F \equiv k_+ - k_- \, .
\end{equation}
We further define the backward Coulomb coupling $g_1$ (at the bare level) as an average over the sphere $S$ of radius $\ell_A/\ell_G^2$, as this length scale defines the support in $q_x$ of the gaussian exponentials in Eq.~\eqref{eq:M_backward},
\bea
g_1 &=& \frac{1}{\pi \hbar v_F} \frac{\int_{S} d^2q_{\perp} \, V(2k_F \mbf{\hat{z}} + \mbf{q}_{\perp})}{\int_{S} d^2q_{\perp}}\nn\\
&=& \frac{e^2 \ell_G^4}{\pi \hbar v_F \ell_A^2 \epsilon_0 \epsilon_{\infty}} \log \left( 1 + \frac{\ell_A^2}{4k_F^2 \ell_G^4} \right).\label{eq:g1-definition}
\eea

\begin{figure}[t!]
    \centering
    \includegraphics[scale=0.5]{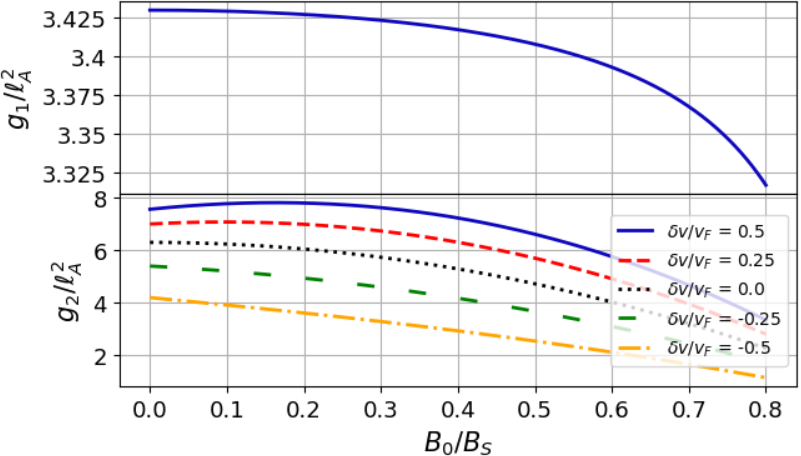}
    \caption{Bare values of $g_1$ (top panel) and $g_2$ (bottom panel) as functions of $B_0/B_S$. For $g_2$, we display the curves for different values of $\delta v / v_F$.}
    \label{fig:initial-conditions}
\end{figure}

In Fig.~\ref{fig:initial-conditions} (top panel), we show a plot of the bare $g_1$, showing that it decreases monotonically with the ratio $B_0/B_S$. Therefore, the bare value of $g_1$ decreases as the physical magnetic field $B_0$ becomes comparable in magnitude to the pseudo-magnetic field $B_S$ due to mechanical strain.

The "backward" electron-electron interaction contribution to the action is thus given by the expression 
\begin{widetext}
\begin{eqnarray}
    S_{ee}^{B} 
    &=& g_1 \frac{\pi v_F}{2 \beta \mathcal{V}} \frac{\ell_G^2}{\ell_A^2} \sum_{\xi \textbf{q}} \sum_{\{ \omega_{n} \}} \sum_{X_{1,\xi} X_{2,\Bar{\xi}}} \sum_{k_1 k_2} e^{i \frac{\ell_{\bar{\xi}}^2}{\ell_A^2} q_x \left( X_{1,\xi} - \varepsilon_{\xi}^2 X_{2,\Bar{\xi}} \right)}  \Psi^{\dag}_{\xi X_{1,\xi}}(k_1, \omega_{n_1}) \Psi^{\dag}_{\Bar{\xi} X_{2,\Bar{\xi}}}(k_2, \omega_{n_2})\nonumber \\
    && \times \, \Psi_{\xi, \varepsilon_{\xi}^2 X_{2,\Bar{\xi}} - \ell_{\xi}^2 q_y}(k_2-\xi b+q_z, \omega_{n_3})   \Psi_{\Bar{\xi}, \varepsilon_{\Bar{\xi}}^2 X_{1,\xi} + \ell_{\Bar{\xi}}^2 q_y}(k_1+\xi b-q_z, \omega_{n_1 + n_2 - n_3} ) \, .
\end{eqnarray}
\end{widetext}

Performing a similar analysis for the long-wavelength component of the Coulomb interaction, named forward scattering, the corresponding contribution to the action is
\begin{eqnarray}
&&S_{ee}^{F} 
= g_2 \frac{\pi v_F}{2 \beta \mathcal{V}} \sum_{\xi \textbf{q}} \sum_{\{ \omega_{n} \}} \sum_{X_{1,\xi} X_{2,\Bar{\xi}}} \sum_{k_1 k_2}  e^{i q_x \left( X_{1,\xi} - X_{2,\Bar{\xi}} \right)}  \nonumber \\
    && \times \Psi^{\dag}_{\xi X_{1,\xi}}(k_1, \omega_{n_1})  \Psi^{\dag}_{\Bar{\xi} X_{2,\Bar{\xi}}}(k_2, \omega_{n_2}) \Psi_{\Bar{\xi}, X_{2,\Bar{\xi}} - \ell_{\Bar{\xi}}^2 q_y}(k_2 +q_z, \omega_{n_3}) \nonumber \\
&& \times  \Psi_{\xi, X_{1,\xi} + \ell_{\xi}^2 q_y}(k_1-q_z, \omega_{n_1 + n_2 - n_3} ) \, , 
\end{eqnarray}
where the forward (long-range) Coulomb coupling $g_2$ is defined by
\begin{equation}
    \lim_{q\to 0} \frac{V(\mbf{q})}{\epsilon(\mbf{q},0)} \equiv g_2 \pi \hbar v_F \, , \label{eq:g2-definition}
\end{equation}
with 
\begin{equation}
    \epsilon(\mbf{q}, \nu_m) = 1 - V(\mbf{q})\Pi_{\text{RPA}}(\mbf{q},\nu_m)
\end{equation}
being the dielectric function in the random phase approximation (RPA), and 
\bea
    \Pi_{\text{RPA}}(\mbf{q},\nu_m) &=& \frac{1}{\mathcal{V}} \SumInt_{\alpha \alpha'} \left| \int d^3x \; \Psi^{\dag}_{\alpha}(\textbf{x}) e^{i\textbf{q} \cdot \textbf{x}} \Psi_{\alpha'}(\textbf{x}) \right|^2\nn\\ 
    &&\times\frac{f_0(E_{\alpha}) - f_0(E_{\alpha'})}{E_{\alpha} - E_{\alpha'} + \hbar \nu_m + i\eta} \label{eq:polarization-definition}
\eea
is the electronic polarization function in RPA, $f_0(E) = \left( \exp(\beta (E - \mu)) + 1 \right)^{-1}$ is the Fermi distribution function, and $\eta \to 0^+$ is an infinitesimal regulator. In the high pseudo-magnetic field regime, we can approximate our calculations by restricting ourselves to the chiral Landau level, by considering $\xi = \xi'$ and $n=n'=0$ in the sets of indices $\alpha$ and $\alpha'$.

As explained in detail in Appendix~\ref{app_RPA}, we obtain
\begin{equation}
    g_2 = 4\pi \ell_G^2 \, \frac{1 + \frac{\delta v}{v_F}}{2\frac{\ell_A^2}{\ell_G^2} + \frac{\delta v}{v_F}\varepsilon_-}  \, .
\end{equation}

In Fig.~\ref{fig:initial-conditions} (bottom panel), we show a plot of this bare value for $g_2$ as a function of the ratio $B_0/B_S$ between the magnetic field and the pseudo-magnetic field due to strain, for different values of the Fermi velocity asymmetry ratio $\delta v/v_F$.

Even though bare $g_2$ is not strictly monotonic and depends on $\delta v/v_F$, it is clear from Fig.~\ref{fig:initial-conditions} that for $B_0/B_S \gtrapprox 0.3$, it decreases quite fast as a function of this ratio.

\subsection{One-loop renormalization of the electron-electron vertex}

The electron-electron inter-nodal vertex will introduce corrections for the backward $g_1$ and forward $g_2$ couplings arising from 2 different channels. The Peierls' channel ($\sim\Bar{\Psi}_+^{\dag} \Psi_-^{\dag} \Bar{\Psi}_- \Psi_+ + \Psi_+^{\dag} \Bar{\Psi}_-^{\dag} \Psi_- \Bar{\Psi}_+$) represents the pairing of one electron and one hole with opposite chiralities, while the Cooper channel ($\sim\Bar{\Psi}_+^{\dag} \Bar{\Psi}_-^{\dag} \Psi_- \Psi_+ + \Psi_+^{\dag} \Psi_-^{\dag} \Bar{\Psi}_- \Bar{\Psi}_+$) represents the pairing of two electrons (or two holes) with opposite chiralities. In addition, the Landau channel ($\sim\Bar{\Psi}_+^{\dag} \Psi_-^{\dag} \Psi_- \Bar{\Psi}_+ + \Psi_+^{\dag} \Bar{\Psi}_-^{\dag} \Bar{\Psi}_- \Psi_+$) only contributes to intranodal scattering and therefore does not contribute to the renormalization of the $g_1$ and $g_2$ couplings.

\subsubsection{Peierls channel}

Let us first analyze the vertex corrections arising from the Peierls channel. Applying the general procedure described in Section~\ref{field_theory_ren}, and restricting ourselves to the LLL subspace, the contribution from forward scattering is given by 
\begin{widetext}
\bea
    &&\frac{1}{2}\braket{\left(S^F_{ee}\right)_P^2} = 2 g_2^2 \left( \frac{\pi v_F}{2 \beta \mathcal{V}} \right)^2 \sum_{\xi \textbf{q}} \sum_{X'_{1,\xi} X_{2,\Bar{\xi}}} \sum_{k_1 k'_2} \sum_{q_x q_y} \sum_{\omega_{n'_1} \omega_{n_2} \omega_{n'_3}} e^{iq_x(X'_{1,\xi} - X_{2,\bar{\xi}})}  \Psi^{\dag}_{\xi X'_{1,\xi}}(k'_1, \omega_{n'_1}) \Psi^{\dag}_{\Bar{\xi} X_{2,\Bar{\xi}}}(k_2, \omega_{n_2})   \\
    && \times \Psi_{\Bar{\xi}, X_{2,\Bar{\xi}} - \ell_{\Bar{\xi}}^2 q_y}(k_2+q_z, \omega_{n'_3}) \Psi_{\xi, X'_{1,\xi} + \ell_{\xi}^2 q_y}(k'_1-q_z, \omega_{n'_1 + n_2- n'_3})    \sum_{q'_x q'_y} \sum_{k \omega_{n_1}} \mathcal{G}^0_{\xi}(k + k'_1 - k_2 - q_z, \omega_{n_1}) \mathcal{G}^0_{\bar{\xi}}(k, \omega_{n_1 - n'_1 + n'_3}) \nn.  \label{eq:correction-F_P^2-aux1}
\eea
\end{widetext}
(see Appendix~\ref{app_rengen} for details). Since we are computing the action within the LLL subspace, after Eq.~\eqref{eq:eigenvalues_2} the corresponding energy eigenvalue (for $n=0$) is $\epsilon_{n=0,\xi}=\xi\hbar v_F (k - k_{F_{\xi}})$, and hence the corresponding explicit expression for the free Fermion Green's functions is
\begin{equation}
    \mathcal{G}^0_{\xi}\left(k, \omega_{n}\right) = \frac{1}{i\hbar \omega_n - \xi \hbar v_{\xi} (k - k_{F_{\xi}})} \, .
\end{equation}
We can approximate the relative difference between external momenta in Eq.~\eqref{eq:correction-F_P^2-aux1} as $k'_1 - k_2 \approx k_{\xi} - k_{\bar{\xi}} \equiv \xi b$, since each momenta in this expression is very close to its respective Weyl node.  Also, introducing $\nu_m = \omega_{n'_1} - \omega_{n'_3}$, the Matsubara sums and the $k$-integrals can be expressed as (for $\xi b - q_z  \approx 2 k_{F}$)
\bea 
&&\sum_{k} \sum_{\omega_{n}} \mathcal{G}^0_{\xi}\left(k+\xi b-q_{z}, \omega_{n}\right) \mathcal{G}^0_{\bar{\xi}}\left(k, \omega_{n}-\nu_{m}\right) \nonumber \\
&&=\sum_{k} \sum_{\omega_{n}} \mathcal{G}^0_{\xi}\left(k+2 k_{F}, \omega_{n}+\nu_{m}\right) \mathcal{G}^0_{\bar{\xi}}\left(k, \omega_{n}\right), \label{eq:peierls-symmetrization-1}
\eea
where in the second line we performed the frequency shift $\omega_n \to \omega_n + \nu_m$.
This operation can be symmetrized, in order to define the polarization insertion $\Pi_P(\nu_m)$ for the Peierls channel (details in Appendix~\ref{App_Peierls})

\bea
    \Pi_P(\nu_m) &\equiv& \frac{1}{2} \sum_{k} \sum_{\omega_{n}}\left[\mathcal{G}^0_{+}\left(k, \omega_{n}\right) \mathcal{G}^0_{-}\left(k - 2 k_{F}, \omega_{n} - \nu_m\right)\right.\nn\\
    &&\left.+ \mathcal{G}^0_{+}\left(k+2 k_{F}, \omega_{n} + \nu_m \right) \mathcal{G}^0_{-}\left(k, \omega_{n}\right) \right]\nn\\
    &=&-\frac{\beta L_z}{2 \pi \hbar v_F} \cdot \frac{dl}{1+ \delta v/2v_F} \lambda_P(T,\delta v, \nu_m, l) , \label{eq:I_P}
\eea
Here, we defined
\bea
    &&\lambda_P(T,\delta v, \nu_m, l) \equiv \frac{1}{1-i\frac{\hbar \nu_m}{(1+\gamma)\Lambda(l)}} \frac{1}{4} \left[ 2\tanh\left( \frac{\beta \Lambda(l)}{2} \right)\right.\nn\\
    &&\left.+ \tanh\left( \frac{\beta \Lambda(l)}{2} \gamma \right) + \tanh\left( \frac{\beta \Lambda(l)}{2\gamma}  \right) \right] \, . \label{eq:lambda_P}
\eea
In Fig.~\ref{fig:Landau-Cooper-loops}, we represent $\lambda_P(\nu_m=0)$ as a function of the cutoff $\beta \Lambda(l)$, for different values of $\gamma - 1 = \delta v/v_F$. Note that $0 \leq \lambda_P(T,\delta v, 0,l) < 1$.

Defining $\nu_m = \omega_{n'_1} - \omega_{n'_3}$, we conclude that the contribution to the beta function from this channel is  
\begin{eqnarray}
    \left( \frac{dg_2}{dl} \right)_P = g_2^2 \frac{1}{4\pi\ell_A^2} \frac{\lambda_P(l)}{1 + \delta v/2v_F}.  \label{eq:beta-ee-P-F2}
\end{eqnarray}
Let us now consider the contribution to the Peierls channel arising from backward scattering.  Following the general procedure described in Section~\ref{field_theory_ren} and restricting ourselves to the LLL subspace, we obtain 
\begin{eqnarray}
    &&\frac{1}{2}\braket{\left(S^B_{ee}\right)_P^2} = -2 g_1^2 \left( \frac{\pi v_F}{2 \beta \mathcal{V}} \right)^2 \frac{\ell_G^4}{\ell_A^4} \sum_{\xi \textbf{q}} \sum_{k'_1 k_2} \sum_{\{\omega_{n}\}}    \nonumber \\
    && \times  \sum_{X'_{1,\xi} X_{2,\Bar{\xi}}}\sum_{X'_{2,\bar{\xi}}} e^{i \frac{\ell_{\xi}^2}{\ell_A^2} q_x(X'_{2,\bar{\xi}} - X_{2,\bar{\xi}})} \sum_{q'_x} e^{i \frac{\ell_{\bar{\xi}}^2}{\ell_A^2} q'_x(X'_{1,\xi} - \varepsilon_{\xi}^2 X'_{2,\bar{\xi}})}    \nonumber \\
    && \times  \sum_{k'_2 \omega_{n'_1}} \mathcal{G}^0_{\xi}(k'_2 - \xi b + q_z, \omega_{n'_1}) \mathcal{G}^0_{\bar{\xi}}(k'_2, \omega_{n'_1+ n_2 - n_3})  \nonumber \\
    && \times
    \Psi^{\dag}_{\xi X'_{1,\xi}}(k'_1, \omega_{n_1})
    \Psi^{\dag}_{\Bar{\xi} X_{2,\Bar{\xi}}}(k_2, \omega_{n_2})\nonumber\\
    &&\times\Psi_{\xi, \varepsilon_{\xi}^2 X_{2,\Bar{\xi}} - \ell_{\xi}^2 q_y}(k_2-\xi b+q_z, \omega_{n_3})  \nonumber \\
    && \times  \Psi_{\bar{\xi}, \varepsilon_{\bar{\xi}}^2 X'_{1,\xi} + \ell_{\bar{\xi}}^2 q_y}(k'_1+ \xi b -q_z, \omega_{n_1 + n_2 - n_3}) .
    \label{eq:See2Peierls}
\end{eqnarray}
By a similar analysis as in the previous case, we conclude that  
the contribution to the beta function for the $g_1$ coupling arising from this channel is  
\begin{eqnarray}
    \left( \frac{dg_1}{dl} \right)_{P,B^2} &=& -g_1^2 \frac{1}{4\pi \ell_G^2} \frac{\lambda_P(l)}{1 + \delta v/2v_F} \, , \label{eq:beta-ee-P-B2}
\end{eqnarray}
where we applied the definitions of the magnetic lengths defined in Eq.~\eqref{eq:length-scales-2}.

The last contribution arising from the Peierls channel corresponds to the product between backward and forward couplings $g_1 g_2$, given by 
\begin{eqnarray}
    &&\braket{\left( S^F_{ee} S^B_{ee} \right)_P} = 4\left( \frac{\pi v_F}{2 \beta \mathcal{V}} \right)^2 \frac{\ell_G^2}{\ell_A^2} \sum_{\xi \textbf{q}'} \sum_{X'_{1,\xi} X_{2,\Bar{\xi}}} \sum_{k'_1 k_2}\sum_{\{ \omega_n \}}   \nonumber \\
    && \times  e^{i \frac{\ell_{\bar{\xi}}^2}{\ell_A^2} q'_x(X'_{1,\xi} - \varepsilon_{\xi}^2 X_{2,\bar{\xi}})} \Psi^{\dag}_{\xi X'_{1,\xi}}(k'_1, \omega_{n'_1}) \Psi^{\dag}_{\Bar{\xi} X_{2,\Bar{\xi}}}(k_2, \omega_{n_2})  \nonumber \\
    &&\times \Psi_{\xi, X_{1,\xi} + \ell_{\xi}^2 q_y}(k-\xi b+q'_z, \omega_{n_3}) \nn\\
    && \times \, \Psi_{\bar{\xi}, \varepsilon_{\bar{\xi}}^2 X'_{1,\xi} + \ell_{\bar{\xi}}^2 q'_y}(k'_1+\xi b-q'_z, \omega_{n'_1 + n_2 - n_3})  \\
    && \times \, g_1 g_2 \sum_{q_x q_y} \sum_{k\omega_{n_1}} \mathcal{G}^0_{\xi}(k-\xi b+q'_z, \omega_{n_1}) \mathcal{G}^0_{\bar{\xi}}(k, \omega_{n_1 + n_2 - n_3}) \nn .
\end{eqnarray}
As this is a correction to the backward vertex, we conclude that the contribution to the beta function of $g_1$ arising from this channel is 
\begin{eqnarray}
    \left( \frac{dg_1}{dl} \right)_{P,F\cdot B} &=& g_1 g_2 \frac{1}{2\pi \ell_{A}^2} \frac{\lambda_P(l)}{1 + \delta v/2v_F} . \label{eq:beta-ee-P-FB}
\end{eqnarray}

\subsubsection{Cooper channel}

\begin{figure}[t]
    \centering
    \includegraphics[scale=0.6]{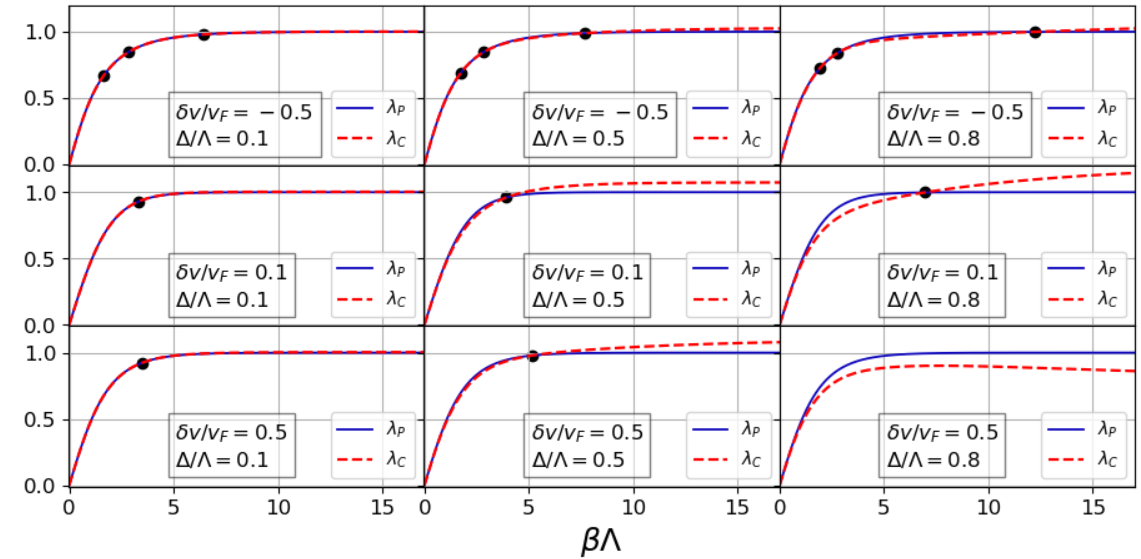}
    \caption{Plots of $\lambda_P(\nu_m=0,l)$ and $\lambda_C(\nu_m=0,l)$ as functions of $\beta \Lambda(l)$ for different values of $\gamma$ and $\Delta/\Lambda(l)$. The black dots corresponds to intersections between the two functions.}
    \label{fig:Landau-Cooper-loops}
\end{figure}

Let us now consider the vertex corrections arising from the Cooper channel. The contribution arising from forward scattering is the following
\begin{eqnarray}
    &&\frac{1}{2}\braket{\left(S^F_{ee}\right)_C^2} = 2\left( \frac{\pi v_F}{2 \beta \mathcal{V}} \right)^2 \sum_{\xi \textbf{q}} \sum_{X_{1,\xi} X_{2,\Bar{\xi}}} \sum_{k_1 k_2} \sum_{\left\{\omega_{n_i}\right\} }    \nonumber \\
    && \times e^{iq_x(X_{1,\xi} - X_{2,\bar{\xi}})} \Psi^{\dag}_{\xi X_{1,\xi}}(k_1, \omega_{n_1}) \Psi^{\dag}_{\Bar{\xi} X_{2,\Bar{\xi}}}(k_2, \omega_{n_2})  \nonumber \\
    && \times \Psi_{\Bar{\xi}, X_{2,\Bar{\xi}} - \ell_{\Bar{\xi}}^2 q_y}(k_2+q_z, \omega_{n_3})   \nonumber \\
    &&\times \Psi_{\xi, X_{1,\xi} + \ell_{\xi}^2 q_y}(k_1-q_z, \omega_{n_1+n_2 - n_3})\nn\\
     && \times  g_2^2  \sum_{q'_x q'_y} \sum_{k \omega_{n}} \mathcal{G}^0_{\xi}(k_1 + k_2 - k, \omega_{n_1 + n_2 - n}) \mathcal{G}^0_{\bar{\xi}}(k, \omega_{n}) \, .
\end{eqnarray}
The last line gives a correction to the forward coupling $g_2$. The sum over the pair of Green's functions in the last factor can be symmetrized in order to define the polarization insertion from the Cooper channel as follows (details in Appendix~\ref{App_Cooper}) 
\bea
    \Pi_C(\nu_m) &= &  \frac{1}{2}\sum_{k} \sum_{\omega_{n}}  \left[ \mathcal{G}^0_{+}\left(k, \omega_{n}\right) \right.\nn\\
    &&\left.\times\mathcal{G}^0_{-}\left(k_{+}+ k_{-}- k, \omega_{n_1 + n_2 - n}\right) \right. \nonumber \\
&&\left. +  \mathcal{G}^0_{+}\left(k_{+}+ k_{-}- k, \omega_{n_1 + n_2 - n}\right) \mathcal{G}^0_{-}\left(k, \omega_{n}\right) \right]\nn\\
&=& \frac{\beta L_z}{2\pi \hbar v_F} \cdot \frac{dl}{1+\delta v / 2v_F} \lambda_C(T,\delta v, \Delta, \nu_m,l),
\label{eq:polarization-cooper}
\eea
where we have defined
\bea
&&\lambda_C(T, \delta v, \Delta, \nu_m,l) 
    = \frac{1}{4}
\sum_{s=\pm 1} \frac{1}{1 + s\frac{\gamma}{1 + \gamma} \frac{\Delta}{\Lambda(l)}}\nn\\
&&\times \frac{1}{1- \frac{i s}{1+ s\frac{\gamma}{1+\gamma} \frac{\Delta}{\Lambda(l)}}  \frac{\hbar\nu_m}{(1+\gamma)\Delta}} \left[ \frac{1}{2}\tanh\left( \frac{\beta \gamma (\Lambda(l) + s\Delta)}{2}  \right)\right.\nn\\ 
&&\left.+ \frac{1}{2}\tanh\left( \frac{\beta (\Lambda(l) + s\gamma \Delta)}{2\gamma} \right) + \tanh\left( \frac{\beta \Lambda(l)}{2} \right)\right],
\label{eq:lambda_C}
\eea
and
\begin{equation}
    \Delta \equiv \hbar v_F \left[ k_{F_+}  + k_{F_-} \right] \,.
    \label{eq:Delta}
\end{equation}
In Fig.~\ref{fig:Landau-Cooper-loops}, we represent $\lambda_C(\nu_m=0)$ as function of the cutoff $\beta \Lambda(l)$ for different values of $\gamma$ and $\Delta/\Lambda(l)$. It is important to note that
\begin{equation}
    0 \leq \lambda_C(T, \delta v, \Delta, 0,l) < \frac{1}{1-\left( \frac{\gamma}{1 + \gamma} \frac{\Delta}{\Lambda(l)} \right)^2} 
\end{equation}
and
\begin{equation}
    \lambda_C(T, \delta v, \Delta, \nu_m,l) = \lambda_C(T, \delta v, -\Delta, \nu_m,l) \, .
\end{equation}
\onecolumngrid\
\twocolumngrid\

Thus, after setting $\omega_{n_1} + \omega_{n_2} = \nu_m$, we conclude that the contribution to the beta function of $g_2$ arising from this diagram is given by 
\begin{eqnarray}
    \left( \frac{dg_2}{dl} \right)_C &=& -g_2^2 \frac{1}{4\pi \ell_A^2} \frac{\lambda_C(l)}{1 + \delta v/2v_F} \, . \label{eq:beta-ee-C-F2}
\end{eqnarray}

Let us now calculate the contribution from backward scattering in the Cooper channel:  
\begin{eqnarray}
    &&\frac{1}{2}\braket{\left(S^B_{ee}\right)_C^2} =  2\left( \frac{\pi v_F}{2 \beta \mathcal{V}} \right)^2 \frac{\ell_G^4}{\ell_A^4} \sum_{\xi \textbf{q}'} \sum_{X_{1,\xi} X_{2,\Bar{\xi}}} \sum_{k_1 k_2} \sum_{\{\omega_{n}\}}    \nonumber \\
    && \times e^{i \frac{\ell_{\bar{\xi}}^2}{\ell_A^2} q_x (X_{1,\xi} - \varepsilon_{\xi}^2 X_{2,\bar{\xi}})} \Psi^{\dag}_{\xi X_{1,\xi}}(k_1, \omega_{n_1}) \Psi^{\dag}_{\Bar{\xi} X_{2,\Bar{\xi}}}(k_2, \omega_{n_2}) \nonumber \\
    && \times \Psi_{\Bar{\xi}, X_{2,\Bar{\xi}} - \ell_{\Bar{\xi}}^2 q'_y}(k_2+q'_z, \omega_{n'_3})  \nonumber \\
    && \times  \Psi_{\xi, X_{1,\xi} + \ell_{\xi}^2 q'_y}(k_1-q'_z, \omega_{n_1 + n_2 - n'_3}) \nn\\
    && \times g_1^2 \sum_{q_x q_y}  \sum_{k \omega_{n_3}} \mathcal{G}^0_{\xi}(k_2 + k_1 - k, \omega_{n_3}) \mathcal{G}^0_{\bar{\xi}}(k, \omega_{n_1 + n_2 - n_3}). 
\end{eqnarray}

Introducing the approximation $k_1 + k_2 \approx k_+ + k_-$, then setting $\nu_m = \omega_{n_1} + \omega_{n_2}$, followed by the change of variable $\omega_{n_3} \to \nu_m-\omega_{n_3}$, and further applying the same symmetrization as before, the Matsubara sum and the $k$-integral correspond exactly to the definition of $\Pi_C(\nu_m)$ given in Eq.~\eqref{eq:polarization-cooper} (and in Appendix~\ref{App_Cooper}). 

Thus, defining $\omega_{n_1} + \omega_{n_2} = \nu_m$, we conclude that the contribution to the beta function of $g_2$ from this channel is 
\begin{eqnarray}
    \left( \frac{dg_2}{dl} \right)_{C,B^2} &=& -g_1^2 \frac{\ell_G^2}{4\pi \ell_A^4 \sqrt{2 - \frac{\ell_G^4}{\ell_A^4}}} \frac{\lambda_C(l)}{1 + \delta v / 2v_F} \, . \label{eq:beta-ee-C-B2}
\end{eqnarray}

The last contribution from the Cooper channel corresponds to the product $g_1 g_2$ between forward and backward scattering couplings,  
\begin{eqnarray}
&&\braket{\left( S^F_{ee} S^B_{ee} \right)_C} = 4\left( \frac{\pi v_F}{2 \beta \mathcal{V}} \right)^2 \frac{\ell_G^2}{\ell_A^2} \sum_{\xi \textbf{q}'} \sum_{X_{1,\xi} X_{2,\Bar{\xi}}} \sum_{k_1 k_2} \sum_{\{ \omega_n \}}  \\
    && \times e^{i \frac{\ell_{\bar{\xi}}^2}{\ell_A^2} q'_x(X_{1,\xi} - \varepsilon_{\xi}^2 X_{2,\bar{\xi}})}\Psi^{\dag}_{\xi X_{1,\xi}}(k_1, \omega_{n_1}) \nonumber \\
    && \times \Psi^{\dag}_{\Bar{\xi} X_{2,\Bar{\xi}}}(k_2, \omega_{n_2})  \Psi_{\xi, \varepsilon_{\xi}^2 X_{2,\bar{\xi}} - \ell_{\xi}^2 q'_y}(k_2-\xi b+q'_z, \omega_{n_3}) \nonumber \\
    && \times \, \Psi_{\bar{\xi}, \varepsilon_{\bar{\xi}}^2 X_{1,\xi} + \ell_{\bar{\xi}}^2 q'_y}(k_1+\xi b-q'_z, \omega_{n_1 + n_2 - n_3})\nn\\
    &&\times g_1 g_2\sum_{q_x q_y}\sum_{k \omega_n} \mathcal{G}^0_{\xi}\left(k, \omega_{n} \right) \mathcal{G}^0_{\bar{\xi}}\left(k_1 + k_2 - k, \omega_{n_1 + n_2 - n}\right). \nn
\end{eqnarray}

Introducing the approximation $k_1 + k_2 \approx k_+ + k_-$, then setting $\nu_m = \omega_{n_1} + \omega_{n_2}$, and further applying the same symmetrization procedure as before, the Matsubara sum and the $k$-integral correspond precisely to the definition of $\Pi_C(\nu_m)$ given in Eq.~\eqref{eq:polarization-cooper} (and in Appendix~\ref{App_Cooper}). By finally setting $\omega_{n_1} + \omega_{n_2} = \nu_m$, we conclude that the contribution to the beta function of $g_1$ arising from this term is  
\begin{eqnarray}
    \left( \frac{dg_1}{dl} \right)_{C,F\cdot B} &=& -g_1 g_2 \frac{1}{2\pi  \ell_A^2} \frac{\lambda_C(l)}{1 + \delta v/2v_F} \, . \label{eq:beta-ee-C-FB}
\end{eqnarray}

\subsubsection{Electronic beta functions}

\begin{figure}[t]
    \centering
    \includegraphics[scale=0.55]{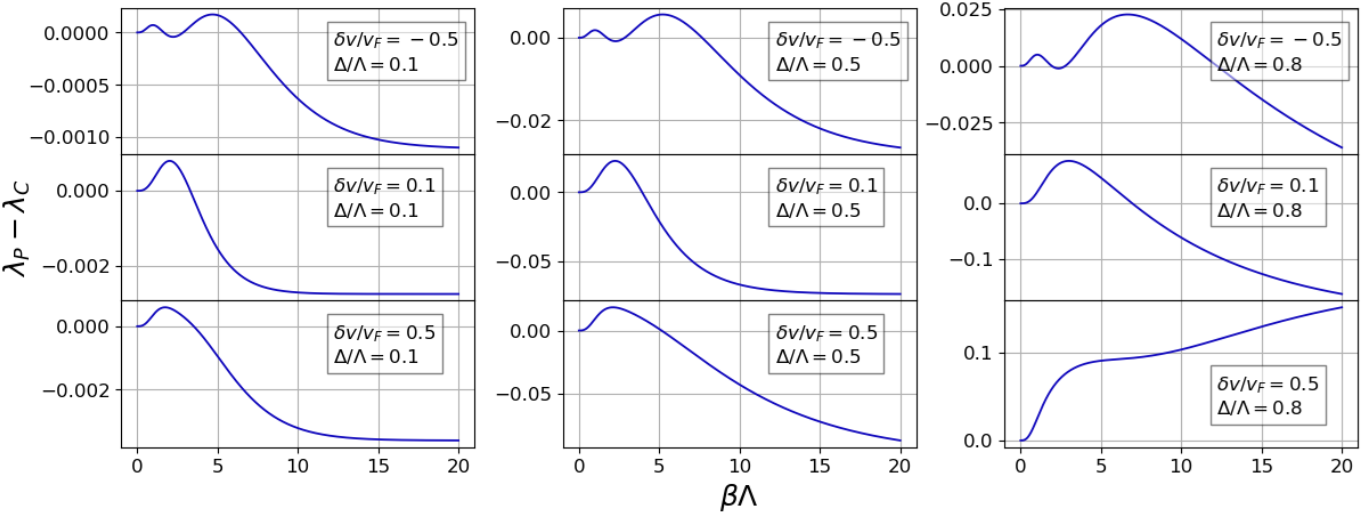}
    \caption{Plots of $\lambda_P - \lambda_C$ as a function of $\beta \Lambda(l)$ for different values of $\gamma$ and $\Delta /\Lambda(l)$.}
    \label{fig:loops-difference}
\end{figure}

Combining Eqs.~\eqref{eq:beta-ee-P-B2}, \eqref{eq:beta-ee-P-FB} and \eqref{eq:beta-ee-C-FB}, we obtain the total beta function and the corresponding flow for the forward coupling $g_1$. Similarly, by adding Eqs.~\eqref{eq:beta-ee-C-B2}, \eqref{eq:beta-ee-P-F2} and \eqref{eq:beta-ee-C-F2}, we obtain the total beta function and the corresponding flow for the backward coupling $g_2$.  In summary, we arrive at the system of equations that describes the renormalization flow of both parameters
\bea
    \frac{dg_1}{dl} &=& - \frac{g_1^2}{2}\alpha_G \lambda_P+  g_1 g_2\alpha_G \left(\frac{\ell_G}{\ell_A}\right)^2 \left(\lambda_P - \lambda_C\right).\\ \label{eq:beta-ee-forward}
    \frac{dg_2}{dl} &=& - \frac{g_1^2}{2}\alpha_G\frac{ \left(\frac{\ell_G}{\ell_A}\right)^4\lambda_C}{\sqrt{2 - \left(\frac{\ell_G}{\ell_A}\right)^4}} +  \frac{g_2^2}{2}\left(\frac{\ell_G}{\ell_A}\right)^2\alpha_G \left(\lambda_P - \lambda_C\right).\nn \label{eq:beta-ee-backward}
\eea
Here, we defined for brevity the coefficient
\bea
\alpha_G = \frac{1}{2\pi\ell_G^2\left( 1 + \frac{\delta v}{2 v_F} \right)}.
\eea
Note that this subsystem of differential equations is decoupled from the electron-phonon sector. In Fig.~\ref{fig:Landau-Cooper-loops}, we represent $\lambda_P$ and $\lambda_C$ as functions of the cutoff $\beta \Lambda(l)$, for different values of $\gamma$ and $\Delta / \Lambda(l)$. We illustrate, by black dots therein, the points at which both curves intersect, representing the conditions when one of the pairing channels dominates over the other. In addition, in Fig.~\ref{fig:loops-difference} we display the difference $\lambda_P - \lambda_C$ as a function of the same set of parameters.
\onecolumngrid\
\twocolumngrid\
\subsection{One-loop renormalization of the electron-phonon vertex}
Let us now analyze the renormalization effects over the electron-phonon coupling, as predicted by our model. We first consider the contribution to the backward channel, which is given by  
\begin{widetext}
\bea
&&\braket{S_{ep}S^B_{ee}} = - \left(\frac{\pi v_F}{\beta \mathcal{V}}\right)^{3/2} \frac{\ell_G^3}{\ell_A^3} \sum_{\xi} \sum_{\textbf{q}\, j}  \sum_{q'_x} \sum_{X_{1,\xi} X_{2,\bar{\xi}}} \sum_{k_1 k_2} \sum_{\omega_{n_1} } 
 g_1 z_{\xi \bar{\xi}} g_{\xi \bar{\xi},j} e^{i \frac{\ell_{\xi}^2}{\ell_A^2} q'_x X_{2,\bar{\xi}}} e^{i \frac{\ell_{\bar{\xi}}^2}{\ell_A^2} q_x \left( X_{1,\xi} - \varepsilon_{\xi}^2 X_{2,\bar{\xi}} \right)} \\
    &&\times\sum_{\omega_n \nu_m} \mathcal{G}^0_{\xi}(k_2 - \xi b + q_z, \omega_n + \nu_m) \mathcal{G}^0_{\bar{\xi}}(k_2, \omega_n)    \Psi^{\dag}_{\xi, X_{1,\xi}}(k_1, \omega_{n_1}) \Psi^{\dag}_{\bar{\xi}, \varepsilon_{\bar{\xi}}^2 X_{1,\xi} + \ell_{\bar{\xi}}^2 q_y}(k_1 + \xi b - q, \omega_{n_1} - \nu_{m}) \phi_{j}(q'_x, q_y, q_z, \nu_m)\nn.
\eea
\end{widetext}

The Matsubara sum and the $k_2$-integral, using the approximation $\xi b - q_z \approx 2k_F$, give the same expression presented in Eq.~\eqref{eq:peierls-symmetrization-1} (see Appendix~\ref{App_Peierls} for details). 
Thus, we conclude that the contribution to the beta function for the electron-phonon vertex part arising from this backward channel is
\begin{equation}
    \left[ \frac{d}{dl} \log \left( z_{\xi \bar{\xi}} \right) \right]_B = - g_1 \frac{1}{4\pi \ell_G^2} \frac{\lambda_P(l)}{1 + \delta v / 2 v_F} \, . \label{eq:beta-ep-backward}
\end{equation}

Similarly, for the forward channel, we have the contribution
\begin{widetext}
\bea
&&\braket{S_{ep}S^F_{ee}} = \sqrt{\frac{\pi v_F}{\beta \mathcal{V}}} \frac{\pi v_F}{2\beta \mathcal{V}} \frac{\ell_G}{\ell_A} \sum_{\xi} \sum_{\textbf{q}\, j}  \sum_{\textbf{q}'} \sum_{X_{1,\xi} k_1 } \sum_{\omega_{n_1} }  g_2 z_{\xi \bar{\xi},j} g_{\xi \bar{\xi}} e^{i \frac{\ell_{\bar{\xi}}^2}{\ell_A^2} q'_x X_{1,\xi}}  \Psi^{\dag}_{\xi, X_{1,\xi}}(k_1, \omega_{n_1})  \nonumber \\
    && \times \,  \Psi^{\dag}_{\bar{\xi}, \varepsilon_{\bar{\xi}}^2 X_{1,\xi} + \ell_{\bar{\xi}}^2 q'_y}(k_1 + \xi b - q'_z, \omega_{n_1} - \nu_{m}) \phi_{j}(\textbf{q}', \nu_m)
    \sum_{\omega_n \nu_m} \mathcal{G}^0_{\xi}(k_1 - q_z, \omega_n + \nu_m) \mathcal{G}^0_{\bar{\xi}}(k_1 + \xi b - (q_z+q'_z), \omega_n).
\eea
\end{widetext}
After introducing $k = k_1 - q_z$ to remove $q_z$, using the approximation $\xi b - q_z' \approx 2k_F$, and introducing the change of variable $\omega_n \to \omega_n - \nu_m$, we obtain
\bea
\sum_{k \omega_n} \mathcal{G}^0_{\xi}(k, \omega_n) \mathcal{G}^0_{\bar{\xi}}(k - 2k_F, \omega_n - \nu_m) 
    =\Pi_P(\nu_m).
\eea

Therefore, the contribution to the beta function of the electron-phonon coupling arising from the forward channel is given by
\begin{equation}
    \left[ \frac{d}{dl} \log \left( z_{\xi \bar{\xi}} \right) \right]_F = g_2 \frac{1}{4\pi \ell_A^2} \frac{\lambda_P(l)}{1 + \delta v / 2 v_F} \, ,   \label{eq:beta-ep-forward}
\end{equation}
where the magnetic length $\ell_A$ is defined in Eq.~\eqref{eq:length-scales-2}.

After adding the two contributions to the beta function from Eq.~\eqref{eq:beta-ep-backward} and Eq.~\eqref{eq:beta-ep-forward}, the full electron-phonon vertex flow is given by the equation
\begin{equation}
    \frac{d}{dl} \log \left( z_{\xi \bar{\xi}} \right) = \frac{\alpha_G}{2} \left( \left(\frac{\ell_G}{\ell_A}  \right)^2 g_2 - g_1  \right)\lambda_P(l).   \label{eq:betaflow-ep}
\end{equation}

\subsection{One-loop renormalization of the phonon propagator}

The electron-phonon interaction is expressed by the term defined by Eq.~\eqref{eq:S_ep_final} in the action. Then, we obtain
\bea
&&\frac{1}{2}\left\langle S_{ep}^2\right\rangle_{dl}=
-\frac{1}{4}\frac{v_F}{\beta L_z} \sum_{\xi}\sum_{\mathbf{q},j,\nu_m}\sum_{k}\sum_{\omega_n}z^2_{\xi\bar{\xi}}g^2_{\xi\bar{\xi},j} \frac{1}{\ell_G^2}\nn\\
    &&\times \mathcal{G}_{\xi}^0(k,\omega_n)\mathcal{G}_{\bar{\xi}}^0(k+\xi b-q_z,\omega_n+\nu_m) |\Phi_{j}(\mathbf{q},\nu_m)|^2\nn\\
    && =\sum_{\mathbf{q},j,\nu_m}\frac{\lambda_P(T, \delta v, \nu_m,l)}{2} dl\frac{z^2_{+-}g^2_{+-,j}|\Phi_{j}(\mathbf{q},\nu_m)|^2}{2\pi\hbar(1+\delta v / 2v_F)\ell_G^2}\nn\\
\eea
where we have used the definition of the magnetic lengths in Eq.~\eqref{eq:lchi}, and the result given in Eq.~\eqref{eq:I_P}. Then, to order $dl$ we have that the inverse of the renormalized phonon propagator is given by 
\bea
    &&\mathcal{D}^{-1}_{j}(\mathbf{q},\nu_m,l+dl)=\mathcal{D}^{-1}_{j}(\mathbf{q},\nu_m,l)\\
    &&-\frac{g^2_{+-,j}}{4\pi\ell_G^2(1+\delta v / 2v_F)}\,z^2_{+-}(l)\lambda_P(l) dl,\nn
\eea
and therefore we obtain the corresponding differential equation for the renormalization flow as follows
\begin{equation}
   \frac{d}{dl}\mathcal{D}^{-1}_j(\mathbf{q},\nu_m,l) =-\frac{\alpha_G }{2}g^2_{+-,j} \, \lambda_P(l)\,z^2_{+-}(l).
\end{equation}
The solution of this differential equation, subject to the initial condition $\mathcal{D}^{-1}_j(\mathbf{q},\nu_m,l=0)=\mathcal{D}_{j,0}^{-1}(\mathbf{q},\nu_m)$ after Eq.~\eqref{eq:D0}, is given by
\bea
\mathcal{D}^{-1}_j(\mathbf{q},\nu_m,l)&=&\nu_m^2+\nu_{0,j}^2(\mathbf{q})\left(1-\frac{\alpha_G g'^2_j\, }{2}\chi(l)\right),\label{eq:D(l)}
\eea
where we defined $g'_j=g_{+-,j}/\nu_{0,j}(\mathbf{q})$. Clearly, the expression in parentheses represents a renormalization of the phonon dispersion relation as a consequence of interactions. In this expression, we identify the CDW susceptibility for the electron system
\bea
\chi(l) = \int_0^l z^2_{+-}(l')\lambda_P(l')
dl',\label{eq:chi(l)}
\eea
and therefore the vanishing of the term in brackets in Eq.~\eqref{eq:D(l)} signals the emergence of the Peierls instability in these WSMs at high pseudo-magnetic fields, thus defining the corresponding scale for the critical temperature.
 
\section{Running Couplings}
In this section, we shall analyze the flow of the renormalized couplings in our model. We shall also identify the possible existence of fixed points. In the present work, we shall focus on the adiabatic limit $\nu_m \to 0$.
\subsection{Fixed Points in the fermion sector}
Let us start with the electronic sector of the running couplings, which is defined by the system of non-linear equations \eqref{eq:beta-ee-forward} and \eqref{eq:beta-ee-backward}. For technical reasons, we express it in matrix form
\begin{equation}
    \frac{d}{dl} \mbf{g} = \mbf{M}(\mbf{g}) \cdot \mbf{g}  \label{eq:beta-electronic-matrix}.
\end{equation}
Here, $\mbf{g} \equiv \left(g_1 , g_2\right)^T$ is the two-component vector for the couplings, while the matrix 
\begin{equation}
    \mbf{M}(\mbf{g}) \equiv 
    \begin{pmatrix}
        -a_1 g_1 & 2a_2 g_1 \\ -a_3 g_1 & a_2 g_2
    \end{pmatrix} ,   \label{eq:matrix-M-electronic-Callan-Symanzik}
\end{equation}
with
\begin{align}
    a_1 &= \frac{\alpha_G}{2} \lambda_P\nn  \\
    a_2 &= \frac{\alpha_G}{2}\left(\frac{\ell_G}{\ell_A}\right)^2 \left(\lambda_P - \lambda_C\right)\nn \\
    a_3 &= \frac{\alpha_G}{2} \frac{\left( \frac{\ell_G}{\ell_A} \right)^4}{\sqrt{2 - \ell_G^4/\ell_A^4}} \lambda_C\, .
\end{align}
For the system \eqref{eq:beta-electronic-matrix}, the fixed points $(g_1^* , g_2^*)$ are given by the stationary condition $d\mbf{g}/dl = 0$, \textit{i.e.}  
\begin{equation}
    \begin{pmatrix}
        -a_1 g_1^* & 2a_2 g_1^* \\ -a_3 g_1^* & a_2 g_2^*
    \end{pmatrix} 
    \begin{pmatrix}
        g_1^* \\ g_2^*
    \end{pmatrix} =
    \begin{pmatrix}
        0 \\ 0
    \end{pmatrix} \, .
\end{equation}
We clearly have the trivial solution $g_1^* = g_2^* = 0$, corresponding to the non-interacting Gaussian fixed point where the two couplings become irrelevant.

A second possible, non-trivial fixed point can thus be discovered when the matrix becomes singular 
\begin{equation}
    \det \begin{pmatrix}
        -a_1 g_1^* & 2a_2 g_1^* \\ -a_3 g_1^* & a_2 g_2^*
    \end{pmatrix}
    = 0 \, .
\end{equation}
The two possible cases are: (1) The trivial solution $g_1^*= g_2^* = 0$ already identified, and (2) the straight line  $g_2^* = 2(a_3/a_1) g_1^*$, \textit{i.e.}
\begin{equation}
    g_2^* =  \frac{2\left(\frac{\ell_G}{\ell_A}\right)^4}{\sqrt{2 - \ell_G^4/\ell_A^4}} \cdot \frac{\lambda_C}{\lambda_P}\, g_1^* \,.    \label{eq:fixed-straight-line-1}
\end{equation}
It is straightforward to check, by direct substitution onto the system~\eqref{eq:beta-electronic-matrix}, that this second case is only possible if the condition $a_1^2/4a_3 = a_2$ is fulfilled. In terms of the model parameters, this corresponds to the condition
\begin{equation}
    \frac{\lambda_C}{\lambda_P} \left( 1 - \frac{\lambda_C}{\lambda_P} \right) = \frac{1}{4} \frac{\ell_A^4}{\ell_G^4} \sqrt{2 \frac{\ell_A^4}{\ell_G^4} - 1} \, ,  \label{eq:condition-existence-fixed-straight-line-1}
\end{equation}
which relates the parameters $\delta v$ and $\Delta$ with $B_0/B_S$ and temperature $T$. Note that since $\lambda_C,\lambda_P\geq 0$, this equation only admits a solution when $\lambda_C < \lambda_P$.
In order to study the possible existence of this straight line of fixed points, let us analyze the order of magnitude of the ratio $\ell_A/\ell_G$. To this aim, we define
\begin{equation}
    \delta \equiv \frac{\ell_-^2 - \ell_+^2}{2}
\end{equation}
as a measure of the nodal asymmetry induced by the pseudo-magnetic field due to torsional strain,
such that
\begin{equation}
    \ell_{\pm}^2 = \ell_A^2 \mp \delta \, .
\end{equation}
Then, according to the definition in Eq.~\eqref{eq:length-scales-2},
\bea
    \ell_G^2 &=& \sqrt{\ell_+^2 \ell_-^2}  
    = \ell_A^2 \sqrt{1 - \left( \delta/\ell_A^2 \right)^2} \, . \label{eq:ell_G/ell_A-aux1}
\eea
The ratio in the last expression is
\begin{align}
    \frac{\delta}{\ell_A^2}  = \frac{\ell_-^2/\ell_+^2 - 1}{\ell_-^2/\ell_+^2 + 1}, \label{eq:delta/ell_A-aux}
\end{align}
and therefore we have
\begin{equation}
    \frac{\ell_+^2}{\ell_-^2} = \frac{|B_-|}{|B_+|} = \frac{1 - \varepsilon}{1 + \varepsilon} \, ,
\end{equation}
where we defined the ratio between the external magnetic field and the strain pseudo-magnetic field
\begin{equation}
    \varepsilon \equiv \left|\frac{B_0}{B_S}\right| \, .
\end{equation}
Then, substituting into Eq.~\eqref{eq:delta/ell_A-aux}, we obtain $\delta/\ell_A^2 = \varepsilon$.

Finally, from Eq.~\eqref{eq:ell_G/ell_A-aux1} and assuming $\varepsilon < 1$ as previously stated, we have
\begin{align}
    \frac{\ell_G^2}{\ell_A^2} = \sqrt{1 - \varepsilon^2} \approx 1 - \frac{1}{2}\varepsilon^2 = 1 - \frac{1}{2} \left( \frac{B_0}{B_S} \right)^2 \, .
\end{align}
Thus,
\begin{equation}
    \frac{\ell_A^2}{\ell_G^2} \approx 1 + \frac{1}{2}\varepsilon^2 = 1 + \frac{1}{2} \left( \frac{B_0}{B_S} \right)^2 \, .
\end{equation}

Using this final result, the condition in Eq.~\eqref{eq:condition-existence-fixed-straight-line-1} for the existence of the straight line of fixed points can be approximated as  
\begin{equation}
    \frac{\lambda_C}{\lambda_P} \left( 1 - \frac{\lambda_C}{\lambda_P} \right) = \frac{1}{4} \left( 1 + 2\frac{B_0^2}{B_S^2} \right) \, .
\end{equation}
This is a quadratic equation in the ratio $y = \lambda_C/\lambda_P > 0$,
\bea
y^2 - y + \frac{1}{4} \left( 1 + 2\frac{B_0^2}{B_S^2} \right) = 0,\nn
\eea
that does not possess any real solutions since its discriminant is negative.

Therefore, we conclude that the condition in Eq.~\eqref{eq:condition-existence-fixed-straight-line-1} is not fulfilled in the regimes of strain and magnetic field assumed in our model, and hence the existence of this non-trivial fixed point is ruled out.

\subsection{Fixed points in the electron-phonon sector}

After the renormalization flow of the electron-phonon vertex (see Eq.~\eqref{eq:betaflow-ep}), it is straightforward to conclude that it reaches a fixed point under the condition
\begin{equation}
    g_1^* =  \left(\frac{\ell_G}{\ell_A}\right)^2 g_2^* \, ,
    \label{eq:g1g2invariat}
\end{equation}
where this straight line thus represents an invariant submanifold of the $(g_1, g_2)$-manifold. We note that even though the beta function can change sign across this line, the renormalization factor $z_{\xi \bar{\xi}}(l)$ cannot, since the solution to Eq.~\eqref{eq:betaflow-ep} is exponential and given the initial condition $z_{\xi \bar{\xi}}(l=0)=1$, it follows that
\begin{equation}
    z_{\xi \bar{\xi}}(l) = e^{  \frac{\alpha_G}{2}\int_0^l dl'  \lambda_P(l')\left( \left( \frac{\ell_G}{\ell_A} \right)^2g_2(l') -  g_1(l') \right)  } > 0 \, .
\end{equation}
Therefore, since $\lambda_P(l)>0$, we have that $z_{\xi \bar{\xi}}$ is marginally irrelevant in the submanifold $g_1 >  \left(\frac{\ell_G}{\ell_A}\right)^2 g_2$, while it becomes relevant in the submanifold $g_1 <  \left(\frac{\ell_G}{\ell_A}\right)^2 g_2$. It is interesting to note that either in the absence of strain but with an external magnetic field present, or conversely in the presence of strain solely, $\ell_G = \ell_A$ and hence the invariant submanifold reduces to the straight line $g_1^* = g_2^*$. In contrast, when both effects are present, the asymmetry factor $\left(\ell_G / \ell_A\right)^2 = \sqrt{1 - \varepsilon^2}$ controls the slope of such a straight line. Interestingly, if $\varepsilon = \left|B_0/B_S\right|\rightarrow 1$, the invariant manifold reduces to $g_1^* = 0$, and sufficiently close to this particular limit the electron-phonon coupling $z_{\xi \bar{\xi}}$ will be relevant for $g_1 < 0$, and marginally irrelevant for $g_1 > 0$.

Based on the general analysis leading to the renormalization of the phonon frequency, as presented in Eq.~\eqref{eq:D(l)}, it depends on the susceptibility $\chi(l)$ determined by Eq.~\eqref{eq:chi(l)} as an integral of the electron-phonon coupling $z_{+-,j}(l)$. Therefore, in the submanifold $g_1 \le  \left(\frac{\ell_G}{\ell_A}\right)^2 g_2$ where this coupling is relevant, there will always be a Peierls instability at some finite critical temperature provided $\alpha_G g_j'^2 > 0$. In the following sections, we shall explore this possibility and how the combination of strain and external magnetic field through the parameter $\varepsilon$ determines the corresponding critical behavior. We shall consider $\varepsilon < 1$, and hence the invariant submanifold in Eq.~\eqref{eq:g1g2invariat} becomes $g_1^* = \left(1-\varepsilon^2/2\right) g_2^*$.

\section{Renormalization Group analysis of the adiabatic Peierls instability}
\label{sec_Peierls_instab}
In what follows, we shall restrict ourselves to the adiabatic regime $\nu_m = 0$. We shall also assume that the external magnetic field is much weaker than the pseudo-magnetic field due to strain, such that the ratio $\varepsilon = |B_0/B_S|\ll 1$. Therefore, the RG system of equations reduces to (up to $O(\varepsilon^4)$)
\bea
&&\frac{d g_1}{dl} = -\frac{g_1^2}{2}\alpha_G\lambda_P + g_1g_2\alpha_G\left(1 - \frac{\varepsilon^2}{2}\right)\left( \lambda_P - \lambda_C \right)\nn\\
&&\frac{d g_2}{dl} = -\frac{g_1^2}{2}\alpha_G\left( 1 - \frac{3}{2}\varepsilon^2 \right)\lambda_C + \frac{g_2^2}{2}\alpha_G\left(1 - \frac{\varepsilon^2}{2}\right)\left( \lambda_P - \lambda_C \right)\nn\\
&&\frac{d\log z_{\xi,\bar{\xi}}}{dl} = \frac{\alpha_G}{2}\lambda_P
\left(\left( 1 - \frac{\varepsilon^2}{2} \right)g_2 - g_1 \right).
\label{eq:NRGflow_Epsilon}
\eea
We note that this system, in the limit $\varepsilon\rightarrow 0$, is equivalent to the one obtained in Ref.~\cite{Kundu_2022}. Therefore, our analysis will represent a natural extension towards the asymmetric nodal condition induced by the combination of strain and magnetic field, when $\varepsilon \ll 1$.
\subsection{Adiabatic regime with mirror symmetry $\Delta = 0$}
Let us first assume that, in the absence of magnetic field and strain, the two Weyl nodes are related by a mirror plane. Under this condition, the parameters $\Delta = 0$ and $\delta v = 0$. If the pseudo-magnetic field arising from the combination of strain and external magnetic field is oriented along the z-direction, as we assume in our theoretical setup, then the mirror symmetry will be preserved. Under this condition, as can be directly verified by comparing Eq.~\eqref{eq:lambda_C} and Eq.~\eqref{eq:lambda_P}, we have $\lambda_P = \lambda_C$, and the RG flow Eq.~\eqref{eq:NRGflow_Epsilon} simplifies to
\bea
\frac{d g_1}{dl} = -\frac{g_1^2}{2}\alpha_G\lambda_P,\,\,\,\,\,\frac{d g_2}{dl} = \left( 1 - \frac{3}{2}\varepsilon^2 \right)\frac{d g_1}{dl}.
\eea
Therefore, the combination
\bea
\gamma_{G}\equiv\alpha_G \left[g_2(l) - \left( 1 - \frac{3}{2}\varepsilon^2 \right)g_1(l)\right]
\label{eq:g2_symm}
\eea
remains a marginal and non-universal constant. Moreover, solving for $g_1(l)$ we have (for $\lambda_P\sim 1$ at low temperatures)
\bea
g_1(l) = \frac{g_1(0)}{1 + \frac{\alpha_G}{2}g_1(0)\int_0^l\lambda_P(l')dl'} \sim \frac{g_1(0)}{1 + \frac{\alpha_G}{2}g_1(0)l},\nn\\
\label{eq:g1_symm}
\eea
thus verifying that the repulsive backward scattering flows towards its marginally irrelevant fixed point $g_1\rightarrow g_1^* = 0$ already identified in the previous section.

After substituting Eq.~\eqref{eq:g2_symm} into the third equation in the system~\eqref{eq:NRGflow_Epsilon}, we notice that the electron-phonon vertex flows according to
\bea
\frac{d\log z_{\xi,\bar{\xi}}}{dl} = \frac{\gamma_G}{2}\left( 1 - \frac{\varepsilon^2}{2} \right)\lambda_P - \varepsilon^2\alpha_G\lambda_P g_1(l).
\eea
Integrating this equation combined with Eq.~\eqref{eq:g1_symm}, and taking into account that $\lambda_P\sim 1$ up to the temperature dependent cutoff $l < l_T$, with
\be
l_T = \log\left(\frac{\Lambda_0}{k_B T}  \right),
\ee
we obtain
\bea
z_{\xi,\bar{\xi}}(l) &=& e^{\frac{\gamma_G}{2}\left( 1 - \frac{\varepsilon^2}{2} \right)\int_0^l\lambda_P(l')dl'-\varepsilon^2\alpha_G \int_{0}^l\lambda_P(l')g_1(l')dl'}\nn\\
&=& e^{\frac{\gamma_{\varepsilon}}{2}l}, 
\eea
were we defined the non universal exponent
\be
\gamma_{\varepsilon} \equiv \gamma_G\left( 1 - \frac{\varepsilon^2}{2} \right) - \varepsilon^2 g_1(0).
\label{eq:gamma_epsilon}
\ee
The Peierls instability arises when the renormalized phonon frequency (in the adiabatic limit $\nu_m=0$) in Eq.~\eqref{eq:D(l)} vanishes, and thus the critical temperature $T_{c,0}^j$ is defined after the condition
\bea
1 - \frac{\alpha_G g_j'^2}{2}\chi(l_{T_{c,0}^j}) = 0.
\label{eq:Tccondsymm}
\eea
Integrating Eq.~\eqref{eq:chi(l)} we obtain the susceptibility $\chi(l_{T_{c,0}^j})$, and then solving Eq.~\eqref{eq:Tccondsymm}, we find the critical temperature for the $j$-phonon mode
\bea
T_{c,0}^j = \frac{\Lambda_0}{k_B}\left(  \frac{\alpha_G g_{j}'^2}{2\gamma_{\varepsilon} + \alpha_G g_{j}'^2}\right)^{1/\gamma_{\varepsilon}},
\label{eq:TPerierls1}
\eea
with the Peierls instability determined by the phonon mode with the highest critical temperature.
As clearly seen in Eq.~\eqref{eq:gamma_epsilon}, the exponent $\gamma_{\varepsilon}$ that controls the critical temperature is determined by a competition between the Peierls and Cooper channels. This conclusion agrees with Ref.~\cite{Kundu_2022}, both in the limit $\varepsilon  \rightarrow 0$, as well as in the combined presence of strain and an external magnetic field $\varepsilon > 0$ considered here, but disagrees with some previous mean-field studies~\cite{Zhang_2016,Qin_2020} that ignored the contribution from the Cooper channel. As mentioned in the previous section, for $\left(1 - \varepsilon^2/2\right)g_2>g_1$ and provided $\alpha_G g_j'^2 > 0$, the Peierls instability will arise at a finite critical temperature. Moreover, in the limit $\gamma_{\varepsilon}\rightarrow 0^+$, after Eq.~\eqref{eq:TPerierls1} the critical temperature reduces to a BCS-like expression $T_{c,0}^j \rightarrow \frac{\Lambda_0}{k_B}\exp\left[-2/\left( \alpha_G g_j'^2 \right) \right]$. On the other hand, in the submanifold $\left( 1 - \varepsilon^2/2 \right)g_2 < g_1$, it is still possible to develop a Peierls instability, provided the electron-phonon coupling exceeds the threshold $\alpha_G g_j'^2 > -2\gamma_{\varepsilon}$. This second condition arises if the Thomas Fermi screening vector $q_{TF} > {\rm{max}}\left\{ k_F,\ell_A \right\}$ (notice that $\ell_G \le \ell_A$), and hence the tendency to suppress the Peierls instability in this region of parameter space is consistent with the quasi-1D behavior of the system in the presence of a strong pseudomagnetic field~\cite{Kundu_2022}.

\subsection{Adiabatic regime without mirror symmetry $\Delta > 0$}
Let us now consider the case when the mirror symmetry between the Weyl nodes of opposite chirality is already broken in the absence of a pseudomagnetic field, i.e. $\Delta >0$ and $\delta v > 0$.
We notice that when $\Delta > 0$, the RG flows can be separated into two different regimes, as follows: 
(1) For $l < l_{\Delta} = \log(\Lambda_0/\Delta)$, we have $\lambda_P \sim \lambda_C \sim 1$. In this first case, the flow equations are still determined by the system Eq.~\eqref{eq:D(l)}. Therefore, the corresponding expression for the Peierls critical temperature $T_{c,\Delta}^j$ remains the same as in the mirror symmetric case Eq.~\eqref{eq:TPerierls1}, i.e. $T_{c,\Delta}^j = T_{c,0}^j$, provided $k_B T_{c,0}^j > \Delta$.
\\
(2) For $l \ge l_{\Delta}$ and $\Delta \ge k_B T_{c,0}^j$ defined in Eq.~\eqref{eq:TPerierls1}, we have $\lambda_C \rightarrow 0$ (due to the upper cutoff defined by $\Delta$ in Eq.~\eqref{eq:lambda_C}), while $\lambda_P \sim 1$. Therefore, in this regime the flow equations reduce to
\bea
\frac{d g_1}{d l} &=& -\frac{g_1^2}{2}\alpha_G + g_1 g_2 \alpha_G \left( 1 - \frac{\varepsilon^2}{2} \right)\nn\\
\frac{d g_2}{d l} &=& \frac{g_2^2}{2}\alpha_G \left( 1 - \frac{\varepsilon^2}{2}\right)\nn\\
\frac{d\log z_{+-}}{dl} &=& \frac{\alpha_G}{2}\left[ \left( 1 - \frac{\varepsilon^2}{2}\right)g_2 - g_1 \right]
\label{eq:NRGflowDelta}
\eea
Combining the first two equations, we obtain
\bea
\frac{d}{dl}\left[ \left( 1 - \frac{\varepsilon^2}{2}\right)g_2 - g_1 \right]
= \frac{\alpha_G}{2}\lambda_P\left[ \left( 1 - \frac{\varepsilon^2}{2}\right)g_2 - g_1 \right]^2,\nn\\
\eea
which can be integrated for $l > l_{\Delta}$, to obtain
\bea
\left[ \left( 1 - \frac{\varepsilon^2}{2}\right)g_2 - g_1 \right](l) = \frac{\gamma_{\Delta,\varepsilon}/\alpha_G}{1 - \frac{\gamma_{\Delta,\varepsilon}}{2}\left( l - l_{\Delta} \right) },
\label{eq:g2g1RGflow}
\eea
where we defined
\bea
\gamma_{\Delta,\varepsilon} \equiv \alpha_G \left[ \left( 1 - \frac{\varepsilon^2}{2}\right)g_2(\Delta) - g_1(\Delta) \right].
\eea
It is clear from Eq.~\eqref{eq:g2g1RGflow} that the combination of couplings reaches a singularity at a characteristic temperature defined by $l_{T_0^*} = \log\left( \Lambda_0/k_B T_0^* \right) > l_{\Delta}$, defined by the pole
\bea
1 - \frac{\gamma_{\Delta,\varepsilon}}{2}\left( l_{T_0^*} - l_{\Delta} \right) = 0.
\eea
Solving this equation, we find the transition temperature
\bea
T_0^* = \frac{\Delta}{k_B}e^{-\frac{2}{\gamma_{\Delta,\varepsilon}}},
\label{eq:T0*}
\eea
Now, substituting Eq.~\eqref{eq:g2g1RGflow} into the third flow equation in Eq.~\eqref{eq:NRGflowDelta}, and integrating up to $l \ge l_{\Delta}$, we obtain
\bea
z_{+-}(l) = \frac{z_{+-}(\Delta)}{1 - \frac{\gamma_{\Delta,\varepsilon}}{2}\left( l - l_{\Delta}\right)},
\eea
thus clearly showing that the electron-phonon vertex develops a pole singularity precisely at $T_0^{*}$, that is then interpreted as the characteristic temperature for the onset of an instability against the CDW formation.

Finally, we return to the analysis of the Peierls temperature for this case $\Delta > 0$, by calculating explicitly the susceptibility for $l > l_{\Delta}$,
\bea
\chi(l) - \chi(\Delta) &=& \int_{l_{\Delta}}^{l}z_{+-}^2(l')dl'\nn\\
&=& \frac{(l - l_{\Delta})z_{+-}^2(\Delta)}{1-\frac{\gamma_{\Delta,\varepsilon}}{2}(l - l_{\Delta})}.
\eea
As in the former case, the Peierls temperature is set by the scale $l_{T_\Delta} = \log(\Lambda_0/k_B T_{c,\Delta}^j)$ when the renormalized phonon frequency vanishes, i.e.
\bea
1 - \frac{\alpha_G}{2}g_j'^2\chi(l_{T_\Delta}) = 0.
\eea
Solving the equation above, we obtain the explicit expression for the Peierls critical temperature
\bea
T_{c,\Delta}^j = \frac{\Delta}{k_B}e^{-\frac{2 - \alpha_G g_j'^2\chi(\Delta)}{\gamma_{\Delta,\varepsilon}\left[ 1 - \frac{\alpha_G}{2}g_j'^2\chi(\Delta) \right] + \alpha_G g_j'^2 z_{+-}^2(\Delta)}}.
\label{eq:TDelta1}
\eea
(3) The expressions above are valid provided $\Delta \ge k_B T_{c,0}^j$. However, in the opposite regime where $k_B T_{c,0}^j > \Delta$, then the Cooper loop remains bounded at $\lambda_C \sim 1$, and the corresponding expression for the Peierls critical temperature remains as in the symmetric case, given by Eq.~\eqref{eq:TPerierls1}.\\\\
(4) Finally, we have the regime where $\Delta > \Lambda_0$. In this case, the lower integration limit for $l$ must be set to zero instead of $l_{\Delta}$ in the above equations. This is equivalent to replace $l_{\Delta} \rightarrow 0$, $\gamma_{\Delta}\rightarrow \gamma_{\varepsilon}$, and $\chi(\Delta)\rightarrow 0$, such that the corresponding expression for the Peierls critical temperature in this case reduces to the expression
\bea
T_{c,\Delta}^j = \frac{\Lambda_0}{k_B}
e^{-\frac{2}{\gamma_{\varepsilon} + \alpha_G g_{j}^{'2}}}
\label{eq:TDelta2}
\eea
Interestingly, these results show that when the mirror symmetry between the two Weyl nodes of opposite chirality is broken (i.e. $\Delta > 0$, $\delta v >0$), the Peierls instability develops even in the limit of vanishing electron-phonon interaction $g'^2_j\rightarrow 0$, as clearly seen from Eq.~\eqref{eq:TDelta1} and Eq.~\eqref{eq:TDelta2}, in contrast with the mirror symmetric case Eq.~\eqref{eq:TPerierls1}. This effect is a consequence of the competition between the Cooper and Peierls channels~\cite{Kundu_2022}. In particular, when the mirror symmetry is broken the Cooper channel is suppressed, leaving room for a purely electronic instability in the vanishing limit of the electron-phonon coupling. Nevertheless, a finite value of the later enhances the instability, and hence a lattice instability characterized by $T_0^*$ in Eq.~\eqref{eq:T0*} precludes the purely electronic instability.

\section{Discussion and Conclusions}

In this work, we developed a theoretical model to describe the combined effects of mechanical strain and an external magnetic field $\mathbf{B}_0 = \hat{z}B_0$ over electron-phonon interactions in a type I Weyl semimetal with a single pair of nodes. The effects of mechanical strain on electronic degrees of freedom are represented by a gauge field whose curl determines a pseudomagnetic field $\mathbf{B}_S = \hat{z}B_S$. Therefore, our theory leads to the presence of an effective pseudo-magnetic field that couples differently at each Weyl node ($B_{\xi} = B_S + \xi B_0$), thus breaking the nodal symmetry. In this sense, our theoretical model represents an extension to the renormalization group analysis in Ref.~\cite{Kundu_2022}, by incorporating the effects of strain in the corresponding RG flow equations~\eqref{eq:NRGflow_Epsilon} by means of the parameter $\varepsilon = |B_0/B_S|$ representing the ratio between the external magnetic field and the strain field, respectively. We obtained the RG flow equations for the couplings in the very strong pseudomagnetic field regime, within the LLL subspace. Moreover, we explored the regime where $\varepsilon < 1$, and its consequences on the onset of the CDW and Peierls instability in the adiabatic limit. 

Our results show that the Peierls instability and the corresponding critical temperature, both for the case of mirror symmetry and in the absence of it, are clearly affected by the presence of strain. Concretely, the critical exponents $\gamma_{\varepsilon} = \gamma_G\left( 1 - \varepsilon^2/2 \right) - \varepsilon^2 g_1(0)$ and $\gamma_{\Delta,\varepsilon} = \alpha_G\left[ \left( 1 - \varepsilon^2/2 \right)g_2(\Delta) - g_1(\Delta)\right]$, that control the critical Peierls temperature $T_{c,0}^j$ with ($\Delta = 0$) and without ($\Delta\ne 0$) mirror symmetry $T_{c,\Delta}^j$, respectively, are clearly modified in the presence of the combination of strain and external magnetic field, by the ratio $\varepsilon = |B_0/B_S|$. In the limit $\varepsilon\rightarrow 0$, the nodal symmetry with respect to the pseudo-magnetic field is recovered, and our expressions for these exponents as well as the corresponding Peierls critical temperature reduce exactly to those presented in Ref.~\cite{Kundu_2022} for the adiabatic limit. In that sense, our results represent a natural extension of that previous analysis when the nodal asymmetry imposed by the combination of magnetic field and mechanical strain is included. Therefore, by comparison, we notice that $\gamma_{\varepsilon} \le \gamma_G$, and therefore we conclude that the critical temperature in the mirror symmetry case $\Delta=0$ increases for $\varepsilon > 0$ with respect to the limit $\varepsilon = 0$. For the absence of mirror symmetry $\Delta \ne 0$, we have $\gamma_{\Delta,\varepsilon} \le \gamma_{\Delta,0}$ when $\varepsilon > 0$, and hence in this second case the Peierls critical temperature decreases for $\varepsilon > 0$ compared to the limit $\varepsilon = 0$. Therefore, a main conclusion in our work is that the onset of the Peierls instability, both in the mirror symmetric $\Delta = 0$ and asymmetric $\Delta \ne 0$ cases, can be controlled by a combination of mechanical strain and an external magnetic field in the context of this minimal model for a type I Weyl semimetal, as a consequence of the nodal asymmetry in the effective pseudo-magnetic field that emerges from this particular configuration. In agreement with previous studies~\cite{Kundu_2022}, we observe that the role of the Cooper channel (controlled by $\lambda_C$), in addition to the Peierls channel (controlled by $\lambda_P$), is fundamental for the renormalization group flows that lead to the Peierls instability and the subsequent determination of the critical temperature. Remarkably, our theoretical results open up the possibility for magneto-strain engineering of the electronic properties in this topological materials.

Even though we solved the RG equations and studied the Peierls instability in the adiabatic limit, we are currently extending our analysis to include the role of non-adiabatic effects, as well as the role of the higher LL's, and these results will be communicated in a forthcoming publication.

\section{Acknowledgements}
E.M. acknowledges financial support from ANID Fondecyt Grant No 1230440. D.A.B. was supported by the DGAPA-UNAM Posdoctoral Program. F.J.P. was founded by the ANID Beca Doctorado Nacional 2022 Grant No. 21221919.

%

\appendix
\section{The Landau level basis for a pseudomagnetic field due to torsional strain and an external magnetic field}
\label{App_Landau_basis}
The low energy, effective Hamiltonian that incorporates both torsional strain and an external magnetic field via minimal coupling is
\begin{align}
     \hat{H}_{\xi}=\xi b_0 \sigma_0 + \xi v_{\xi} \bs{\sigma}\cdot\left[\mbf{p}+e\mbf{A}_{\xi}(x)-\hbar\mathbf{K}_{\xi} \right]. \label{eq:A_hamiltonian_2}
\end{align}

Here $\xi=\pm$ is the chirality index of each Weyl node, and $v_{\xi}$ the corresponding Fermi velocity. $b_0$ is the shift with respect to the zero energy, $\sigma_{i}$ with $i=1,2,3$ are the Pauli matrices, and $\sigma_0$ is the identity matrix. We assume that the Weyl nodes are parallel to the direction $\mathbf{\hat{z}}$ of the total pseudomagnetic field, such that the location of each Weyl node in the momentum space is $\mathbf{K}_{\xi}=k_{\xi}\mbf{\hat{z}}$.

Now, we redefine the momentum operator 
\begin{align}
    \mathbf{p}'_{\xi}=\mbf{p}-\hbar\mathbf{K}_{\xi}=\frac{\hbar}{i}\left(\frac{\pd}{\pd x},\frac{\pd}{\pd y}, \frac{\pd}{\pd z}- i k_{\xi}\right).\label{eq:p_prime_coordinates}
\end{align}
We look for the solution to the eigenvalue problem
\begin{equation}
    \left\lbrace\xi b_0 \sigma_0 + \xi v_{\xi} \bs{\sigma}\cdot\left[\mbf{p}'_{\xi}+e\mbf{A}_{\xi}(x) \right]\right\rbrace\Psi(\mbf{x})=E\,\Psi(\mbf{x}),\label{eq:eigen_eqn_1}
\end{equation}
where $\Psi$ is a two-component spinor of the form
\begin{equation}
    \Psi(\mbf{x})=\begin{pmatrix}
        \psi_1(x,y,z) \\ \psi_2(x,y,z)
    \end{pmatrix}.
\end{equation}
Let us define $E=\xi E'$. Then, rearranging, we get 
\begin{equation}
   \left\lbrace  v_{\xi} \bs{\sigma}\cdot\left[\mbf{p}'_{\xi} +e\mbf{A}_{\xi}(x)\right]-\left(E'- b_0\right)\sigma_0\right\rbrace\Psi=0.\label{eq:eigen_eqn_2}
\end{equation}

Notice that, in the chosen gauge, the Hamiltonian is translationally invariant in $y$ and $z$ directions. Therefore, separation of variables leads to the form
\begin{align}
    \Psi(\mbf{x})=\frac{1}{\sqrt{L_y L_z}}e^{-ik_{2}y+i(k+k_{\xi})z}\begin{pmatrix}
        \Phi_1(x)\\ \Phi_2(x)
    \end{pmatrix}. \label{eq:Psi_ansatz}
\end{align}
Let us define the Landau's magnetic length 
\begin{align}
    \ell_{\xi}^2 = \frac{\sign( B_{\xi}) \hbar}{e B_{\xi}},
\end{align}
and the guiding center for each node
\begin{align}
    X_{\xi}=\frac{\hbar k_2}{eB_{\xi}},
\end{align}
such that
\begin{equation}
    k_2=\frac{X_{\xi}}{\ell_{\xi}^2}\sign( B_{\xi}).
\end{equation}
Inserting the spinor in Eq.~\eqref{eq:Psi_ansatz} into Eq.~\eqref{eq:eigen_eqn_2}, we obtain
\bea
     &&\left[ - i  \frac{d}{dx} \sigma_1+  \sign(B_\xi)\frac{x-X_{\xi}}{\ell_{\xi}^2} \sigma_2+k\sigma_3\right.\nn\\
     &&\left.- \frac{ (E'-b_0)}{\hbar v_{\xi}}\sigma_0 \right] \begin{pmatrix}
        \Phi_1(x)\\ 
        \Phi_2(x)
    \end{pmatrix} =0.
\eea
Expanding the product, this leads to the system of equations
\begin{widetext}
\begin{align}
    \left( k-\frac{ (E'-b_0)}{\hbar v_{\xi}} \right)\Phi_1(x)-i\left(\frac{d}{dx}+\sign(B_\xi)\frac{x-X_{\xi}}{\ell_{\xi}^2} \right) \Phi_2(x)&=0, \label{eq:eqn1}\\
    -i\left(\frac{d}{dx}-\sign(B_\xi)\frac{x-X_{\xi}}{\ell_{\xi}^2} \right) \Phi_1(x)-\left( k+\frac{ (E'-b_0)}{\hbar v_{\xi}} \right)\Phi_2(x)&=0 \label{eq:eqn2}.
\end{align}
\end{widetext}
From Eq.~\eqref{eq:eqn1}, we have
\begin{equation}
   \Phi_1(x)= \frac{i\left(\frac{d}{dx}+\sign(B_\xi)\frac{x-X_{\xi}}{\ell_{\xi}^2} \right) \Phi_2(x)}{ \left( k-\frac{ (E'-b_0)}{\hbar v_{\xi}} \right)}, \label{eq:phi_1_phi_2}
\end{equation}
and after replacing in Eq.~\eqref{eq:eqn2} we get
\bea
      &&\frac{d^2\Phi_2(x)}{dx^2}+\frac{\sign(B_\xi)}{\ell_{\xi}^2}\Phi_2(x)-\frac{(x-X_{\xi})^2}{\ell_{\xi}^4}\Phi_2(x)\nn\\
      &&- \left( k^2-\frac{ (E'-b_0)^2}{\hbar^2 v^2_{\xi}} \right)\Phi_2(x)=0.
\eea
Introducing the change of variable $s=\frac{x' - X_{\xi}}{\ell_{\xi}}$, and the substitution
\begin{align}
    \Phi_2(s)=e^{-s^2/2}v(s),
\end{align}
we obtain
\bea
    &&\left\{ \frac{d^2}{ds} - 2s \frac{d}{ds} + \ell_{\xi}^2 \left[ \frac{(E' - b_0)^2}{\hbar^2 v_{\xi}^2} - k^2\right.\right.\nn\\
    &&\left.\left.+\frac{1-\sign(B_\xi)}{\ell_\xi^2} \right] \right\} v(s) = 0. \label{eq:diff-u(s)-incomplete_2}
\eea
Since the functions $v(s)$ must be normalizable, they must be in $L^2$. The only solutions to the above equation that fulfill this requirement are the Hermite polynomials, provided that the following condition is satisfied
\begin{equation}
     \ell_{\xi}^2 \left[ \frac{(E' - b_0)^2}{\hbar^2 v_{\xi}^2} - k^2 +\frac{1-\sign(B_\xi)}{\ell_\xi^2} \right] = 2n_\xi,   \label{eq:dispersion-relation}
\end{equation}
with $n_\xi=0,1,2,\cdots$. Bearing in mind that $\sign(B_\xi)=\pm 1$, we can rearrange the terms in the above equation as follows
\begin{equation}
     \ell_{\xi}^2 \left[ \frac{(E' - b_0)^2}{\hbar^2 v_{\xi}^2} - k^2 \right] = 2n_\xi-1+\sign(B_\xi)\equiv 2n,   \label{eq:dispersion-relation_1}
\end{equation}
with $n$ an integer. Then, the energy spectrum is (remembering that $E' = \xi E$)
\begin{equation}  \label{eq:A_eigenvalues_2}
  E_{\lb \xi n k } =\left\lbrace\begin{array}{cc}
      \xi b_0 +\xi\lb \epsilon_{\xi nk},& n\neq0  \\
      \xi b_0+\xi \hbar v_\xi k, & n=0
  \end{array}\right.\; ,
\end{equation}
where 
\begin{equation}
    \epsilon_{\xi nk} = \hbar v_{\xi} \sqrt{k^2 + \frac{2n}{\ell_{\xi}^2}},
\end{equation}
and $\lb=\pm 1$ is the band index. The equation \eqref{eq:diff-u(s)-incomplete_2} reduces to
\begin{equation}
    \left\{ \frac{d^2}{ds} - 2s \frac{d}{ds} + 2n \right\} v(s) = 0 \, ,
\end{equation}
whose solutions are the Hermite polynomials $H_{n}(s)$. Using the orthogonality relation of the Hermite polynomials, i.e.,  
\begin{align}
    \int_{-\infty}^{\infty} H_{n}(s)H_{m}(s)e^{-s^2}\,ds =\sqrt{\pi}2^{n}n! \delta_{nm},   \label{eq:orthogonality-hermite}
\end{align}
we define the normalized orthogonal functions
\begin{align}
    \varphi_{n}(x)=\frac{(-1)^n}{\sqrt{2^{n}n!}} \left( \frac{1}{\pi \ell_{\xi}^2} \right)^{1/4} e^{-x^2/2\ell_{\xi}^2}\, H_{n}\left( \frac{x}{\ell_{\xi}} \right),\label{eq:A_def_phi_n}
\end{align}
where 
\begin{equation}
    \int_{-\infty}^{\infty}  dx \, \varphi_n(x) \varphi_m(x) = \delta_{nm}.
\end{equation}

Then, we can write the solution of the differential equation as
\begin{equation}
    \Phi_2(x)=N \varphi_{n}(x-X_{\xi}),
\end{equation}
where $N$ is a normalization coefficient to be determined. From Eq.~\eqref{eq:phi_1_phi_2} we have
\begin{align}
     \Phi_1(x) &= \frac{i\left(\frac{d}{dx}+\frac{x-X_{\xi}}{\ell_{\xi}^2} \right) \Phi_2(x)}{ \left( k-\frac{ (E'-b_0)}{\hbar v_{\xi}} \right)}\\
     &= \frac{i\, N}{k-\frac{\lb\epsilon_{\xi nk}}{\hbar v_{\xi}}} \left(\frac{d}{dx}+\frac{x-X_{\xi}}{\ell_{\xi}^2} \right) \varphi_n\left( x - X_{\xi} \right). \nn
\end{align}
Applying the identity $H'_n(x)=2nH_{n-1}(x)$, we obtain
\begin{align}
    \Phi_1(x) 
    &= \frac{i\, N}{\frac{\lb\epsilon_{\xi nk}}{\hbar v_{\xi}}-k} \frac{\sqrt{2n}}{\ell_{\xi}} \varphi_{n-1}(x - X_{\xi}) ,
\end{align}
where we have used de definition \eqref{eq:def_phi_n}. Note that, using equation \eqref{eq:dispersion-relation},
\begin{align}
    \frac{1}{\frac{\lb\epsilon_{\xi nk}}{\hbar v_{\xi}}-k} \frac{\sqrt{2n}}{\ell_{\xi}}
    &= \frac{1}{\frac{\lb\epsilon_{\xi nk}}{\hbar v_{\xi}}-k} \sqrt{\left( \frac{\epsilon_{\xi nk}}{\hbar v_{\xi}} - k \right) \left( \frac{\epsilon_{\xi nk}}{\hbar v_{\xi}} + k \right)} .
\end{align}
By considering separately the cases $\lb=+1$ and $\lb=-1$, we can see that the result can be cast in the simpler form
\begin{equation}
  \lb \sqrt{\frac{\frac{\lb \epsilon_{\xi nk}}{\hbar v_{\xi} k} + 1}{\frac{\lb \epsilon_{\xi nk}}{\hbar v_{\xi} k} - 1}} .
\end{equation}
Then,
\begin{align}
    \Phi_1(x) = i N \lb \sqrt{\frac{\frac{\lb \epsilon_{\xi nk}}{\hbar v_{\xi} k} + 1}{\frac{\lb \epsilon_{\xi nk}}{\hbar v_{\xi} k} - 1}} \varphi_{n-1}(x - X_{\xi}).
\end{align}
Normalizing the eigenspinors $\begin{pmatrix} \Phi_1(x)\\ \Phi_2(x) \end{pmatrix}$, we have
\begin{align}
    N = \frac{1}{\sqrt{2}} \sqrt{1 - \frac{\hbar v_{\xi} k}{\lb \epsilon_{\xi nk}}} .
\end{align}
Hence,
\begin{align}
    \Phi_1(x) &=  \frac{ i\lb}{\sqrt{2}} \sqrt{1 + \frac{\hbar v_{\xi} k}{\lb \epsilon_{\xi nk}}} \varphi_{n-1}(x - X_{\xi}) , \\
    \Phi_2(x) &= \frac{1}{\sqrt{2}} \sqrt{1 - \frac{\hbar v_{\xi} k}{\lb \epsilon_{\xi nk}}} \varphi_{n}(x - X_{\xi}) .
\end{align}

We can introduce the coefficients: 
\begin{align}
    u_{\xi\lb nk}= \frac{i\lb}{\sqrt{2}}\sqrt{1+\frac{\hbar v_{\xi} k}{\lb \epsilon_{\xi nk}}}, \label{eq:def_u}
\end{align}
and 
\begin{align}
    v_{\xi\lb nk}=\frac{1}{\sqrt{2}}\sqrt{1-\frac{\hbar v_{\xi} k}{\lb \epsilon_{\xi nk}}}, \label{eq:def_v}
\end{align}
such that the final expression for each spinor eigenstate is
\begin{align}
    \Psi(\mbf{x}) = \frac{e^{- i\left(X_{\xi}/\ell_{\xi}^2\right)y + i(k+k_{\xi})z}}{\sqrt{L_y L_z}}\begin{pmatrix}
        u_{\xi \lb nk}\,\varphi_{n-1}(x-X_{\xi})  \\  v_{\xi \lb nk}\,\varphi_{n}(x-X_{\xi})
    \end{pmatrix}. \label{eq:A_spinor_2}
\end{align}
We can see that
\bea
    u_{-\xi nk}^{*}u_{\xi nk}+v_{-\xi nk}^{*}v_{\xi nk}&=&-\frac{1}{2}\sqrt{1-\left(\frac{\hbar v_{\xi} k}{\epsilon_{\xi nk}}\right)^2}\\
    &+&\frac{1}{2}\sqrt{1-\left(\frac{\hbar v_{\xi} k}{\epsilon_{\xi nk}}\right)^2}=0\nn
\eea
as expected, states with different band index are mutually orthogonal.

For the case $n=0$, we must consider $\lb=+1$ and $\lb=-1$ separately. For the case $\lb=+1$, from Eq.\eqref{eq:def_u} and Eq.\eqref{eq:def_v} we get
\begin{align}
    u_{+\xi 0k} = \frac{i\lb}{\sqrt{2}}, \hspace{0.5cm}  v_{+\xi 0k} =0
\end{align}
Then, the corresponding eigenspinor is
\begin{align}
    \Psi(\mbf{x})_{n=0,\lb=+1} = \frac{i\,e^{- i\left(X_{\xi}/\ell_{\xi}^2\right)y + i(k+k_{\xi})z}}{\sqrt{ 2 L_y L_z}}\begin{pmatrix}
        \varphi_{-1}(x-X_{\xi})  \\  0
    \end{pmatrix}. \label{eq:A_spinor_3}
\end{align}
For $\lb=-1$, the corresponding eigenspinor is
\begin{align}
    \Psi(\mbf{x})_{n=0,\lb=-1} = \frac{e^{- i\left(X_{\xi}/\ell_{\xi}^2\right)y + i(k+k_{\xi})z}}{\sqrt{2 L_y L_z}}\begin{pmatrix}
        0  \\  \varphi_{0}(x-X_{\xi})
    \end{pmatrix}. \label{eq:A_spinor_4}
\end{align}
However, the function $\varphi_{-1}(x)\sim H_{-1}(x)$ is not normalizable, and therefore only the case $(n=0,\lambda=-1)$ exists.
\section{Evaluation of Matrix elements}
\label{App_Landau_Matrix}
The interaction matrix elements are proportional to the following inner product:
\begin{equation}
    \int d^3x \; \Psi^{\dag}_{\lambda \xi nX_{\xi}k}(\textbf{x}) e^{i\textbf{q} \cdot \textbf{x}} \Psi_{\lambda' \xi' n'X'_{\xi'}k'}(\textbf{x}) .  \label{eq:braket-interaction-matrix-element}
\end{equation}
In order to simplify the algebraic manipulations, let us define the separation of variables $\Psi_{\alpha}(\textbf{x}) \equiv (1/L_z)e^{-i(k+k_{\xi})z}\mathbf{h}_{\alpha}(\textbf{x}_{\perp})$. Therefore, the matrix element Eq.~\eqref{eq:braket-interaction-matrix-element} is
\bea
&&\int d^3x \; \Psi^{\dag}_{\lambda \xi nX_{\xi}k}(\textbf{x}) e^{i\textbf{q} \cdot \textbf{x}} \Psi_{\lambda' \xi' n'X'_{\xi'}k'}(\textbf{x}) \nn\\
    &&=\frac{1}{L_z} \int dz \, e^{i \left[ (k' + k_{\xi'}) - (k + k_{\xi}) + q_z \right]z}\nn\\
    &&\times\int d^2x_{\perp} \,  \mathbf{h}^{\dag}_{\lambda \xi nX_{\xi}k}(\textbf{x}_{\perp}) e^{i\textbf{q}_{\perp} \cdot \textbf{x}_{\perp}} \mathbf{h}_{\lambda' \xi' n'X'_{\xi'}k'}(\textbf{x}_{\perp})  \nonumber \\
    &&=\, \delta_{k' + k_{\xi'} , (k + k_{\xi}) - q_z}\nn\\
    &&\times \int d^2x_{\perp}
    \mathbf{h}^{\dag}_{\lambda \xi nX_{\xi}k}(\textbf{x}_{\perp}) e^{i\textbf{q}_{\perp} \cdot \textbf{x}_{\perp}} \mathbf{h}_{\lambda' \xi' n'X'_{\xi'}k'}(\textbf{x}_{\perp}).
\eea
Substituting the spinor components in the explicit form $\mathbf{h}_{\lambda \xi nX_{\xi}k}(\textbf{x}_{\perp}) \equiv \left( u_{\lambda \xi nk} h_{\xi (n-1)X_{\xi}}(\textbf{x}_{\perp}) \, , v_{\lambda \xi nk} h_{nX_{\xi}\xi}(\textbf{x}_{\perp}) \right)^{T}$, we have for the in the $(x,y)$-integral:
\begin{widetext}
\begin{align}
    & \int d^2x_{\perp} \,  \mathbf{h}^{\dag}_{\lambda \xi nX_{\xi}k}(\textbf{x}_{\perp}) e^{i\textbf{q}_{\perp} \cdot \textbf{x}_{\perp}} \mathbf{h}_{\lambda' \xi' n'X'_{\xi'}k'}(\textbf{x}_{\perp})  \\
    =& \int d^2 x_{\perp}  \left[ u^*_{\lambda \xi nk}u_{\lambda' \xi' n'k'} h^*_{\xi (n-1)X_{\xi}}(\textbf{x}_{\perp}) h_{(n'-1)X'_{\xi'} \xi'}(\textbf{x}_{\perp})  + v^*_{\lambda \xi nk}v_{\lambda' \xi' n'k'} h^*_{nX_{\xi}\xi}(\textbf{x}_{\perp}) h_{n'X'_{\xi'}\xi'}(\textbf{x}_{\perp}) \right] e^{i\textbf{q}_{\perp} \cdot \textbf{x}_{\perp}} . \nonumber
\end{align}
\end{widetext}
Let us consider the first integral expression
\begin{widetext}
\begin{align}
    & \hspace{5mm} \int d^2 x_{\perp}  h^*_{nX_{\xi}\xi}(\textbf{x}_{\perp}) h_{n'X'_{\xi'}\xi'}(\textbf{x}_{\perp}) e^{i\textbf{q}_{\perp} \cdot \textbf{x}_{\perp}} = \frac{1}{L_y} \int dy \, e^{iX_{\xi}y/\ell_{\xi}^2} e^{-iX'_{\xi'}y/\ell_{\xi'}^2} e^{iq_y y} \int dx \, \varphi^*_{n}(x - X_{\xi}) \varphi_{n'}(x - X'_{\xi'}) e^{iq_x x} \nonumber  \\
    &= \delta_{X'_{\xi'}/\ell_{\xi'}^2, X_{\xi}/\ell_{\xi}^2 + q_y} \frac{1}{\sqrt{\pi \ell_{\xi} \ell_{\xi'}}} \frac{1}{\sqrt{2^n n!}} \frac{1}{\sqrt{2^{n'} n'!}} \, \int dx \, H_{n}\left( \frac{x - X_{\xi}}{\ell_{\xi}} \right) H_{n'}\left( \frac{x - X'_{\xi'}}{\ell_{\xi'}} \right) e^{-\frac{(x - X_{\xi})^2}{2\ell_{\xi}^2}-\frac{(x - X'_{\xi'})^2}{2\ell_{\xi'}^2} +iq_x x}  . \label{eq:integral-91}
\end{align}
\end{widetext}
Completing the square in the argument of the exponential, we have
\bea
    && \frac{(x - X_{\xi})^2}{2\ell_{\xi}^2}+\frac{(x - X'_{\xi'})^2}{2\ell_{\xi'}^2} - iq_x x  \\
    &&= \, \frac{\ell_{\xi}^2 + \ell_{\xi'}^2}{2\ell_{\xi}^2 \ell_{\xi'}^2} \left( x - \frac{\ell_{\xi'}^2 X_{\xi} + \ell_{\xi}^2 X'_{\xi'}}{\ell_{\xi}^2 + \ell_{\xi'}^2} - iq_x \frac{\ell_{\xi}^2 \ell_{\xi'}^2}{\ell_{\xi}^2 + \ell_{\xi'}^2} \right)^2    \nonumber \\
     &&+ \frac{1}{2} \frac{\left(X_{\xi} - X'_{\xi'}\right)^2}{\ell_{\xi}^2 + \ell_{\xi'}^2}-iq_x \frac{\ell_{\xi}^2 X'_{\xi'} + \ell_{\xi'}^2 X_{\xi}}{\ell_{\xi}^2 + \ell_{\xi'}^2} + \frac{q_x^2}{2} \frac{\ell_{\xi}^2 \ell_{\xi'}^2}{\ell_{\xi}^2 + \ell_{\xi'}^2}.\nn
\eea
By further defining the magnetic length scales
\begin{equation}
    \ell^A_{\xi \xi'} \equiv \sqrt{\frac{\ell_{\xi}^2 + \ell_{\xi'}^2}{2}} \hspace{4mm} \text{and} \hspace{4mm} \ell^G_{\xi \xi'} \equiv \sqrt{\ell_{\xi} \ell_{\xi'}},   \label{eq:length-scales}
\end{equation}
corresponding to the arithmetic and geometric averages, respectively, followed by the change of integration variable
\begin{equation}
    s = x - \frac{1}{2\left( \ell^A_{\xi \xi'} \right)^2} \left( \ell_{\xi'}^2 X_{\xi} + \ell_{\xi}^2 X'_{\xi'} + i \left( \ell^G_{\xi \xi'} \right)^4 q_x \right),
\end{equation}
the integral \eqref{eq:integral-91} becomes
\begin{widetext}
\bea
\int d^2 x_{\perp}  h^*_{nX_{\xi}\xi}(\textbf{x}_{\perp}) h_{n'X'_{\xi'}\xi'}(\textbf{x}_{\perp}) e^{i\textbf{q}_{\perp} \cdot \textbf{x}_{\perp}} =
    &\delta_{X'_{\xi'}/\ell_{\xi'}^2, X_{\xi}/\ell_{\xi}^2 + q_y} \; \frac{\ell^G_{\xi \xi'}}{\ell^A_{\xi \xi'}} \frac{1}{\sqrt{\pi}} \frac{1}{\sqrt{2^n n!}} \frac{1}{\sqrt{2^{n'} n'!}} e^{-\Tilde{\alpha}_{\xi \xi'}(q_x)} \times \nonumber \\
    &\times \int ds \, H_n \left( \frac{\ell_{\xi'}}{\ell^A_{\xi \xi'}} s + \Tilde{\beta}_{\xi \xi'}(q_x) \right) H_{n'} \left( \frac{\ell_{\xi}}{\ell^A_{\xi \xi'}} s + \Tilde{\beta}_{\xi' \xi}(q_x) \right) e^{-s^2}.
\eea
\end{widetext}
Here, we defined the parameters
\bea
    &&\Tilde{\alpha}_{\xi \xi'}(q_x) \equiv \frac{1}{4\left( \ell^A_{\xi \xi'} \right)^2} \left[ \left( X_{\xi} - X'_{\xi'} \right)^2 + \left( \ell^G_{\xi \xi'} \right)^4 q_x^2 \right]\nn\\
    &&- \frac{iq_x}{2\left( \ell^A_{\xi \xi'} \right)^2} \left( \ell_{\xi'}^2 X_{\xi} + \ell_{\xi}^2 X'_{\xi'} \right)   \label{eq:parameters-1} \\
    && \Tilde{\beta}_{\xi \xi'}(q_x) \equiv \frac{\ell_{\xi'}}{2\left( \ell^A_{\xi \xi'} \right)^2} \left[ \left( \frac{\ell_{\xi'}}{\ell_{\xi}} - 2\left( \frac{ \ell^A_{\xi \xi'} }{\ell^G_{\xi \xi'}} \right)^2 \right) X_{\xi}\right.\nn\\
    &&\left.+ \frac{\ell_{\xi}}{\ell_{\xi'}}X'_{\xi'} + i\ell^G_{\xi \xi'}q_x \right]  \, ,\label{eq:parameters-2}
\eea

Note that, due to the presence of the Kronecker delta, we can replace ${X'}_{\xi'} = X_{\xi} \ell_{\xi'}^2 / \ell_{\xi}^2 + \ell_{\xi'}^2 q_y$ in the definitions of the parameters $\Tilde{\alpha}_{\xi \xi'}(q_x)$ and $\Tilde{\beta}_{\xi \xi'}(q_x)$.

Thus, we seek for a solution to the generic integral
\begin{equation}
    I_{nn'}(a,a',b,b') = \int_{-\infty}^{\infty} dx \, e^{-x^2} H_n(ax+b) H_{n'}(a'x + b') \, .  \label{eq:integral-hermite}
\end{equation}
By using the identity
\begin{equation}
    H_n(x + y) = \sum_{k=0}^{n} {n \choose k} H_k(x) \left( 2y \right)^{n-k} \, ,
\end{equation}
the integral reduces to (omitting the arguments in $I_{nn'}$ for brevity)
\begin{widetext}
\begin{align}
    I_{nn'} &= \int_{-\infty}^{\infty} dx \, e^{-x^2} \left[ \sum_{k=0}^{n} {n \choose k} \left( 2ax \right)^{n-k} H_k(b) \right] \left[ \sum_{k'=0}^{n'} {n' \choose k'} \left( 2a'x \right)^{n'-k'} H_{k'}(b') \right] \nonumber \\
    &= \sum_{k=0}^{n} {n \choose k} a^{n-k} H_k(b) \sum_{k'=0}^{n'} {n' \choose k'} 2^{n+n'-k-k'} {a'}^{n'-k'} H_{k'}(b')  \underbrace{ \int_{-\infty}^{\infty} dx \, e^{-x^2} x^{n+n'-k-k'} }_{\Tilde{A}_{nn'kk'}} .
\end{align}
\end{widetext}
Note that the integral $\Tilde{A}$ is $0$ when $(n+n'-k-k')$ is odd. Then, for the more general case, we obtain the result
\begin{align}
    \Tilde{A}_{nn'kk'} &= \int_{-\infty}^{\infty} dx \, e^{-x^2} x^{n+n'-k-k'} \nonumber \\
    &= \delta_{n+n'-k-k',2p} \, 2 \int_{0}^{\infty} dx \, e^{-x^2} \left(x^2 \right)^p \nonumber \\
    &= \delta_{n+n'-k-k',2p} \, \int_{0}^{\infty} d\Bar{x} \, e^{-\Bar{x}} \, \Bar{x}^{p - \frac{1}{2}} \nonumber \\
    &= \delta_{n+n'-k-k',2p} \, \Gamma\left( p + \frac{1}{2} \right) \nonumber \\
    &= \delta_{n+n'-k-k',2p} \, \frac{\sqrt{\pi}(2p)!}{4^p p!} \nonumber \\
    &= \frac{1}{2}\left[ 1 + (-1)^{n+n'-k-k'} \right] \frac{\sqrt{\pi}(n+n'-k-k')!}{2^{n+n'-k-k'} \left( \frac{n+n'-k-k'}{2} \right)!} \, .
\end{align}
Substituting this expression into Eq.~\eqref{eq:integral-hermite}, we obtain
\begin{widetext}
\begin{align}
    I_{nn'} &= \sqrt{\pi} \sum_{k=0}^{n} {n \choose k} a^{n-k} H_k(b) \sum_{k'=0}^{n'} \delta_{n+n'-k-k',2p} {n' \choose k'} \frac{(2p)!}{p!} {a'}^{n'-k'} H_{k'}(b') \nonumber \\
    &= \sqrt{\pi} \sum_{k=0}^{n} {n \choose k} a^{n-k} H_k(b) \sum_{p=\frac{n-k}{2}}^{\frac{n+n'-k}{2}} {n' \choose n+n'-k-2p} \frac{(2p)!}{ p!} {a'}^{2p+k-n} H_{n+n'-k-2p}(b') \nonumber \\
    &= \sqrt{\pi} \sum_{k=0}^{n} {n \choose k} a^{n-k} H_k(b) \sum_{k'=0}^{n'} \frac{1 + (-1)^{n+n'-k-k'}}{2} {n' \choose k'} \frac{(n+n'-k-k')!}{\left( \frac{n+n'-k-k'}{2} \right)!} {a'}^{n'-k'} H_{k'}(b'). \label{eq:I_nm}
\end{align}
\end{widetext}
Finally, the inner product defined in Eq.~\eqref{eq:braket-interaction-matrix-element} is then given by 
\bea
    &&\int d^3x \; \Psi^{\dag}_{\lambda \xi nX_{\xi}k}(\textbf{x}) e^{i\textbf{q} \cdot \textbf{x}} \Psi_{\lambda' \xi' n'X'_{\xi'}k'}(\textbf{x}) \nonumber \\
    &&=  \; \delta_{k' + k_{\xi'} , (k + k_{\xi}) - q_z} \delta_{X'_{\xi'}/\ell_{\xi'}^2, X_{\xi}/\ell_{\xi}^2 + q_y} \left[ u^*_{\lambda \xi nk}u_{\lambda' \xi' n'k'}\right.\nn\\
    &&\left.F_{n-1,n'-1}(\textbf{q}_{\perp}) + v^*_{\lambda \xi nk}v_{\lambda' \xi' n'k'} F_{nn'}(\textbf{q}_{\perp}) \right],  \label{eq:electron-phonon-coupling}
\eea
where we defined
\bea
    &&F_{nn'}(\textbf{q}_{\perp}) = \frac{\ell^G_{\xi \xi'}}{\ell^A_{\xi \xi'}} \frac{1}{\sqrt{\pi}} \frac{1}{\sqrt{2^n n!}} \frac{1}{\sqrt{2^{n'} n'!}} e^{-\Tilde{\alpha}} \nn\\ 
    &&\times I_{nn'} \left(\frac{\ell_{\xi'}}{\ell^A_{\xi \xi'}}, \frac{\ell_{\xi}}{\ell^A_{\xi \xi'}}, \Tilde{\beta}_{\xi \xi'}(q_x), \Tilde{\beta}_{\xi' \xi}(q_x) \right)  \label{eq:Form-Factors-F_nm}
\eea
and $I_{nn'}$ is defined by Eq.~\eqref{eq:I_nm}.

As a particular case of interest, we consider the eigenspinor associated to the chiral Landau level $n=0$, which is given by Eq.~\eqref{eq:A_def_phi_n}. Substituting $u_{\lambda \xi 0 k} = 0$ and $v_{\lambda \xi 0 k} = 1$ for this case, we have
\begin{equation}
    \Psi_{n=0}(\mbf{x}) = \frac{(\pi \ell^2_{\xi})^{-1/4}}{\sqrt{L_y L_z}}e^{-(x - X_{\xi})^2/2\ell^2_{\xi}  - i\left(X_{\xi}/\ell_{\xi}^2\right)y + i(k+k_{\xi})z}\begin{pmatrix}
        0  \\ 1
    \end{pmatrix} \label{eq:chiral-eigenspinor}
\end{equation}

For the chiral Landau level, it is straightforward to verify that the integral in Eq.~\eqref{eq:I_nm} reduces to $I_{00} = \sqrt{\pi}$. Then, the corresponding form factor in Eq.~\eqref{eq:Form-Factors-F_nm} reduces to  
\bea
    &&F_{00}(\textbf{q}_{\perp}) = \frac{\ell^G_{\xi \xi'}}{\ell^A_{\xi \xi'}} \, e^{\frac{iq_x}{2\left(\ell^A_{\xi \xi'}\right)^2} \left( \ell_{\xi'}^2 X_{\xi} + \ell_{\xi}^2 X'_{\xi'} \right)}\nn\\
    &&\times e^{-\frac{1}{4\left(\ell^A_{\xi \xi'}\right)^2} \left[ \left( X_{\xi} - X'_{\xi'} \right)^2 + \left( \ell^G_{\xi \xi'} \right)^4 q_x^2 \right]} .
\eea
Thus, the inner product Eq.~\eqref{eq:electron-phonon-coupling} for the chiral Landau level gives
\bea
    &&\int d^3x \; \Psi^{\dag}_{\xi 0 X_{\xi}k}(\textbf{x}) e^{i\textbf{q} \cdot \textbf{x}} \Psi_{\xi' 0 X'_{\xi'}k'}(\textbf{x})\nn\\ 
    &&= \; \delta_{k' + k_{\xi'} , (k + k_{\xi}) - q_z} \, \delta_{X'_{\xi'}/\ell_{\xi'}^2, X_{\xi}/\ell_{\xi}^2 + q_y} \, F_{00}(\textbf{q}_{\perp}) , \label{eq:matrix-element-chiral-level-01}
\eea
where we omitted the band index $\lambda$ since it is fixed for $n=0$.

This expression will enter in the couplings when we calculate the vertex and propagator correction in the quantum limit (\textit{i.e.} Fermi energy only intersects the chiral Landau level and it is sufficiently far from the excited bands).

When the strain field $B_S$ is turned off, the magnetic length loses its nodal dependence $\ell_{\xi} \to \ell_{B}$, so $\ell^A_{\xi \xi'}, \ell^G_{\xi \xi'} \to \ell_B$. Then, those two matrix elements reduce to the corresponding expressions given in Ref.~\cite{Kundu_2022}.

Evaluating the matrix element expressed in Eq.~\eqref{eq:matrix-element-chiral-level-01} for the backward tunneling, and expressing it in terms of the magnetic length $\ell_{\xi}$ and the ratio
\begin{equation}
\varepsilon_{\xi} \equiv \ell_{\xi}/\ell_{\Bar{\xi}} \, ,
\end{equation}
where $\Bar{\xi} \equiv -\xi$, we obtain   
\begin{eqnarray}
    &&\mathcal{M}_{\xi X_{\xi} k}^{\Bar{\xi} X'_{\Bar{\xi}} k'}(\textbf{q})\equiv \int d^3x \; \Psi^{\dag}_{\xi 0X_{\xi}k}(\textbf{x}) e^{i\textbf{q} \cdot \textbf{x}} \Psi_{\bar{\xi} 0X'_{\xi'}k'}(\textbf{x})  \nonumber \\
    &&= \frac{\ell_G}{\ell_A} \delta_{k'+k_{\Bar{\xi}}, k+k_{\xi}-q_z} \delta_{X'_{\Bar{\xi}}/\ell^2_{\Bar{\xi}}, X_{\xi}/\ell^2_{\xi}+q_y}  \nonumber \\
    && \times e^{-\frac{1}{4} \frac{\ell_G^4}{\ell_A^2} q_x^2} e^{-\frac{1}{4} \frac{\ell_{\bar{\xi}}^4}{\ell_A^2} q_y^2} e^{\frac{i}{2} \frac{\ell_G^4}{\ell_A^2} q_x q_y}  e^{i\frac{\ell_{\bar{\xi}}^2}{\ell_A^2}q_x X_{\xi}}  \nn\\
    &&\times e^{-\frac{1}{4}(\varepsilon_{\xi}^2 - 1)^2 \frac{X_{\xi}^2}{\ell_A^2}} e^{\frac{1}{2} \left( \varepsilon_{\xi}^2 - 1 \right) \frac{\ell_{\bar{\xi}}^2}{\ell_A^2} q_y X_{\xi}} \, .
\end{eqnarray}
Given that in the strong magnetic field limit, the ratio $\varepsilon_{\xi}\sim 1$, for simplicity we neglected the last two exponential factors, as presented in the main text.

On the other hand, the matrix element corresponding to the forward tunneling reduces to
\begin{eqnarray}
    &&\mathcal{M}_{\xi X_{\xi} k}^{\xi X'_{\xi} k'}(\textbf{q}) = \delta_{k', k-q_z} \delta_{X'_{\xi}, X_{\xi} + \ell^2_{\xi}q_y}\nn\\
    &&\times e^{-\frac{1}{4} \ell_{\xi}^2 q_x^2} e^{-\frac{1}{4} \ell_{\xi}^2 q_y^2} e^{\frac{i}{2} \ell_{\xi}^2 q_x q_y} e^{i q_x X_{\xi}}  \, .
\end{eqnarray}
\section{The Wilson renormalization group procedure}
\label{app_rengen}
\subsection{First order contributions}
\begin{figure}
    \centering
    \begin{subfigure}{0.3\textwidth}
        \includegraphics[scale=0.6]{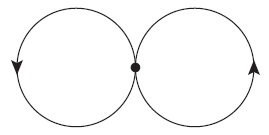}
        \caption{}
        \label{subfig:vacuum-bubbles}
    \end{subfigure}
    \begin{subfigure}{0.3\textwidth}
        \includegraphics[scale=0.6]{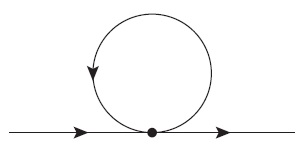}
        \caption{}
        \label{subfig:chemical-potential-correction-1loop}
    \end{subfigure}
    \begin{subfigure}{0.3\textwidth}
        \includegraphics[scale=0.6]{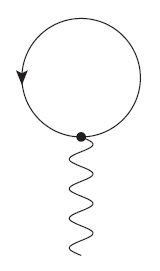}
        \caption{}
        \label{subfig:tadpole-phonon}
    \end{subfigure}
    \caption{First order corrections to the effective action, corresponding to: (a) First term in Eq.~\eqref{C_1ord-See-4bar}, (b) Second term in Eq.~\eqref{C_1ord-See-4bar}, and (c) Eq.~\eqref{C_1ord-Sep}.}
    \label{fig:diagrams-1loop-corrections}
\end{figure}

We shall start with the first order contribution to the effective action,
\begin{equation}
    \braket{S_I} = \braket{S_{ee}} + \braket{S_{ep}} \,.
\end{equation}
For this purpose, we shall apply Wick's contraction for the two-point correlation function
\bea
    \braket{\Bar{\Psi}_{\alpha}(\omega_{n}) \Bar{\Psi}^{\dag}_{\alpha'}(\omega_{n'})} = \mathcal{G}^0_{\alpha}(\omega_n) \delta_{\alpha \alpha'} \delta_{\omega_n \omega_{n'}},
\eea
and Wick's theorem to factorize higher-order correlation functions, such as the 4-point function
\bea
&&\braket{\Bar{\Psi}^{\dag}_{\alpha_1}(\omega_{n_1}) \Bar{\Psi}^{\dag}_{\alpha_2}(\omega_{n_2}) \Bar{\Psi}_{\alpha_3}(\omega_{n_3}) \Bar{\Psi}_{\alpha_4}(\omega_{n_4})}\nn\\
&&=\mathcal{G}^0_{\alpha_1}(\omega_{n_1}) \mathcal{G}^0_{\alpha_2}(\omega_{n_2}) \left[ \delta_{\alpha_1 \alpha_4} \delta_{\alpha_2 \alpha_3} \delta_{\omega_{n_1} \omega_{n_4}} \delta_{\omega_{n_2} \omega_{n_3}} \right. \nonumber \\
&& \left. \; \; \; - \, \delta_{\alpha_1 \alpha_3} \delta_{\alpha_2 \alpha_4} \delta_{\omega_{n_1} \omega_{n_3}} \delta_{\omega_{n_2} \omega_{n_4}} \right]  \, .
\eea
By performing the corresponding contractions, we obtain
\bea
&&\langle S_{ee}\rangle = \frac{2\pi v_F}{\beta \mathcal{V}} \SumInt_{\alpha_1 \alpha_2} \sum_{\omega_{n_1} \omega_{n_2}} \sum_{\textbf{q}} g_{\alpha_1 \alpha_2}^{\alpha_2 \alpha_1}(\textbf{q}) \mathcal{G}^0_{\alpha_1}(\omega_{n_1}) \mathcal{G}^0_{\alpha_2}(\omega_{n_2})\nn\\
&&+\frac{\pi v_F}{\mathcal{\beta V}} \SumInt_{\alpha_1 \alpha_2 \alpha_4} \sum_{\omega_{n} \omega_{n_1}} \sum_{\textbf{q}} g_{\alpha_1 \alpha_2}^{\alpha_1 \alpha_4}(\textbf{q}) \mathcal{G}^0_{\alpha_1}(\omega_{n_1}) \Psi^{\dag}_{\alpha_2}(\omega_{n}) \Psi_{\alpha_4}(\omega_{n})\nn\\
\label{C_1ord-See-4bar}
\eea
where the first term represents a constant and thus irrelevant contribution to the effective action (see Fig.~\ref{subfig:vacuum-bubbles}), while the second represents a constant overall shift in the chemical potential (see Fig.~\ref{subfig:chemical-potential-correction-1loop}).

The next first order contribution is given by 
\bea
    \braket{S_{ep}} 
    = -\sqrt{\frac{\pi v_F}{\beta \mathcal{V}}} \SumInt_{\alpha} \sum_{\omega_n} \sum_{\textbf{q}\, j} g_{\alpha \alpha , j}^{ep}(\textbf{q}) \mathcal{G}^0_{\alpha}(\omega_n) \phi_{j}(\textbf{q},0)\nn\\ \label{C_1ord-Sep}
\eea
This provides a new term to the effective action, which is linear in the phonon field. In diagrammatic language, it is represented by a tadpole, as shown in Fig.~\ref{subfig:tadpole-phonon}.
\subsection{Second order contributions}
\begin{figure}[b]
    \centering
    \begin{subfigure}{0.3\textwidth}
    \includegraphics[scale=0.5]{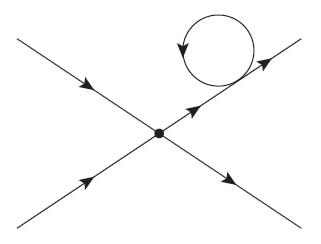}
    \caption{}
    \label{subfig:coulomb-vertex-1dressed-line}
    \end{subfigure}
    \hspace{1cm}
    \begin{subfigure}{0.3\textwidth}
\includegraphics[scale=0.4]{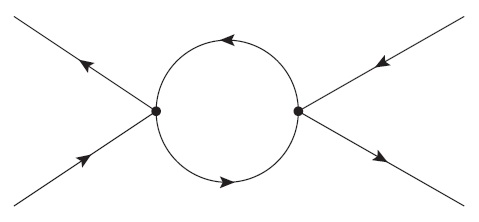}
\caption{}
    \label{subfig:electron-electron-vertex-correction}
    \end{subfigure}
    \caption{Diagram for the first term in Eq.~\eqref{eq:C_SeeSee}.}
\end{figure}

\begin{figure}[b]
    \centering
    \begin{subfigure}{0.3\textwidth}
        \includegraphics[scale=0.5]{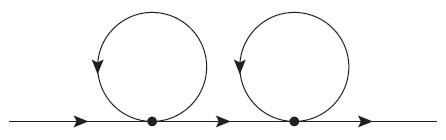}
        \caption{}
        \label{subfig:separable-correction-fermion-propagator}
    \end{subfigure}
    \hspace{1cm}
    \begin{subfigure}{0.3\textwidth}
        \includegraphics[scale=0.5]{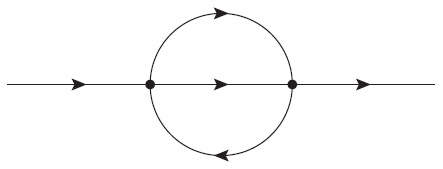}
        \caption{}
        \label{subfig:fermion-propagator-correction}
    \end{subfigure}
    \hspace{1cm}
    \begin{subfigure}{0.3\textwidth}
        \includegraphics[scale=0.5]{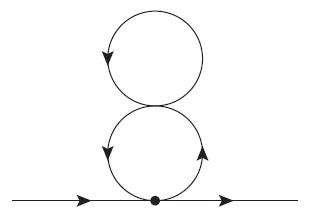}
        \caption{}
    \label{subfig:chemical-potential-correction-2loops}
    \end{subfigure}
    \caption{Second order corrections to the fermion propagator, from Eq.~\eqref{eq:C_SeeSee}.}
    \label{fig:diagrams-2loop-corrections-fermion-propagator}
\end{figure}

Let us now calculate the second order contribution to the effective action, defined as follows
\bea
    \frac{1}{2} \left( \braket{S_I^2} - \braket{S_I}^2 \right) &=& \frac{1}{2} \left( \braket{S_{ee}^2} + \braket{S_{ep}^2}\right.\nn\\
    &&\left.+ 2\braket{S_{ee} S_{ep}} - \braket{S_I}^2 \right) \, .
\eea
For $\langle S_{ee}^2\rangle$, we obtain
\begin{widetext}
\bea
&&\langle S_{ee}^2\rangle =
 2\left( \frac{\pi v_F}{\beta \mathcal{V}} \right)^2 \SumInt_{\{\alpha\}} \sum_{\{\omega_{n}\}} \sum_{\textbf{q}, \textbf{p}}\Bigg[ g_{\alpha_1 \alpha_2}^{\alpha_3 \alpha_6}(\textbf{q}) g_{\alpha_5 \alpha_6}^{\alpha_5 \alpha_4}(\textbf{p}) \mathcal{G}^0_{\alpha_5}(\omega_{n_5}) \mathcal{G}^0_{\alpha_6}(\omega_{n_1 + n_2 - n_3} )  \Psi^{\dag}_{\alpha_1}(\omega_{n_1}) \Psi^{\dag}_{\alpha_2}(\omega_{n_2})\Psi_{\alpha_3}(\omega_{n_3}) \Psi_{\alpha_4}(\omega_{n_1 + n_2 - n_3})\nn\\
 &&+ \Bigg\{ g_{\alpha_1 \alpha_2}^{\alpha_6 \alpha_4}(\textbf{q}) g_{\alpha_5 \alpha_6}^{\alpha_1 \alpha_2}(\textbf{p}) \mathcal{G}^0_{\alpha_1}(\omega_{n_1}) \mathcal{G}^0_{\alpha_2}(\omega_{n_2}) \mathcal{G}^0_{\alpha_6}(\omega_{n_1 + n_2 - n_4})+  2 g_{\alpha_1 \alpha_2}^{\alpha_2 \alpha_4}(\textbf{q}) g_{\alpha_5 \alpha_6}^{\alpha_6 \alpha_1}(\textbf{p}) \mathcal{G}^0_{\alpha_1}(\omega_{n_4}) \mathcal{G}^0_{\alpha_2}(\omega_{n_2}) \mathcal{G}^0_{\alpha_6}(\omega_{n_6}) \Bigg\} \Psi^{\dag}_{\alpha_5}(\omega_{n_4}) \Psi_{\alpha_4}(\omega_{n_4})\nn\\  
    &&+ 2 g_{\alpha_5 \alpha_6}^{\alpha_3 \alpha_4}(\textbf{q}) g_{\alpha_1 \alpha_2}^{\alpha_6 \alpha_5}(\textbf{p}) \mathcal{G}^0_{\alpha_5}(\omega_{n_5}) \mathcal{G}^0_{\alpha_6}(\omega_{n_1 + n_2 - n_5})   \Psi^{\dag}_{\alpha_1}(\omega_{n_1}) \Psi^{\dag}_{\alpha_2}(\omega_{n_2}) \Psi_{\alpha_3}(\omega_{n_3}) \Psi_{\alpha_4}(\omega_{n_1 + n_2 - n_3}) \nn\\
    &&+  \Bigg\{ g_{\alpha_1 \alpha_3}^{\alpha_2 \alpha_3}(\textbf{q}) g_{\alpha_4 \alpha_5}^{\alpha_4 \alpha_5}(\textbf{p}) \mathcal{G}^0_{\alpha_3}(\omega_{n_3}) \mathcal{G}^0_{\alpha_4}(\omega_{n_4}) \mathcal{G}^0_{\alpha_5}(\omega_{n_5})  +  2g_{\alpha_3 \alpha_1}^{\alpha_5 \alpha_2}(\textbf{q}) g_{\alpha_4 \alpha_3}^{\alpha_5 \alpha_4}(\textbf{p}) \mathcal{G}^0_{\alpha_3}(\omega_{n_3}) \mathcal{G}^0_{\alpha_4}(\omega_{n_4}) \mathcal{G}^0_{\alpha_5}(\omega_{n_3}) \Bigg\} \Psi^{\dag}_{\alpha_1}(\omega_{n_1}) \Psi_{\alpha_2}(\omega_{n_1})\Bigg]
    \, . \
    \label{eq:C_SeeSee}
\eea
\end{widetext}
Here, the first term is represented by the diagram in Fig.~\ref{subfig:coulomb-vertex-1dressed-line}, which is a correction to the external fermion line.
The second group of terms represents a correction to the fermion propagator, corresponding to the sum of the diagrams in Fig.~\ref{subfig:separable-correction-fermion-propagator} and Fig.~\ref{subfig:fermion-propagator-correction}.
The third contribution represents a vertex correction, as represented by the diagram in Fig.~\ref{subfig:electron-electron-vertex-correction}. Finally, the fourth term, in principle, gives a correction to the fermion propagator. However, this is the sum of two contributions: One disconnected diagram, corresponding to the product of the first order diagrams in Fig.~\ref{subfig:vacuum-bubbles} and Fig.~\ref{subfig:chemical-potential-correction-1loop}, which is then canceled by the term $\braket{S_I}^2$; and a connected diagram, as shown in Fig.~\ref{subfig:chemical-potential-correction-2loops}.

In summary, the second order corrections to the fermion propagator arising from the $\langle S_{ee}^2 \rangle$ term are presented in Fig.~\ref{fig:diagrams-2loop-corrections-fermion-propagator}, where the diagram in Fig.~\ref{subfig:separable-correction-fermion-propagator} is a reducible one.

\begin{figure}[t]
    \centering
\includegraphics[scale=0.6]{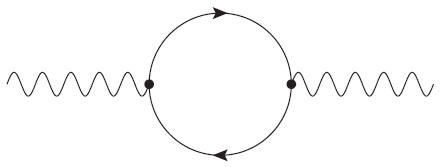}
    \caption{Diagram for the correction to the phonon propagator given in the second term of Eq.~\eqref{C_2nd-ord-Sep-2bar-2bar}.}
    \label{fig:phonon-propagator-correction}
\end{figure}

For the $\langle S_{ep}^2 \rangle$ contribution, we obtain
\begin{widetext}
\bea
\langle S_{ep}^2\rangle
&=& \frac{\pi v_F}{\beta \mathcal{V}} \sum_{\textbf{q} \textbf{p}} \sum_{ji} \SumInt_{\alpha_1 \alpha_2} \left[ \sum_{\omega_{n_1} \omega_{n_2}} g^{ep}_{\alpha_1 \alpha_1,j}(\textbf{q}) g^{ep}_{\alpha_2 \alpha_2,i}(\textbf{p}) \mathcal{G}^0_{\alpha_1}(\omega_{n_1}) \mathcal{G}^0_{\alpha_2}(\omega_{n_2}) \phi_j(\textbf{q},0) \phi_i(\textbf{p},0) \right. \nn\\
    && \left. - \sum_{\nu_{m} \omega_n} g^{ep}_{\alpha_1 \alpha_2,j}(\textbf{q}) g^{ep}_{\alpha_2 \alpha_1,i}(\textbf{p}) \mathcal{G}^0_{\alpha_1}(\omega_{n}) \mathcal{G}^0_{\alpha_2}(\omega_{n} - \nu_{m}) \phi_j(\textbf{q},\nu_{m}) \phi_i(\textbf{p},-\nu_{m}) \right]  \, . \label{C_2nd-ord-Sep-2bar-2bar}
\eea
\end{widetext}
This contribution is composed of two terms. The first is a disconnected diagram, corresponding to the square of the tadpole Fig.~\ref{subfig:tadpole-phonon}. The second corresponds to the diagram given in Fig.~\ref{fig:phonon-propagator-correction}, which represents a correction to the phonon propagator.
\begin{figure}[t]
\centering
    \begin{subfigure}{0.3\textwidth}
    \includegraphics[scale=0.5]{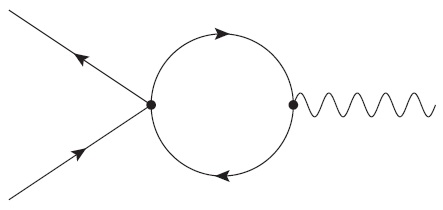}
    \caption{}
    \label{subfig:electron-phonon-vertex-correction}
\end{subfigure}
\begin{subfigure}{0.3\textwidth}
    \includegraphics[scale=0.5]{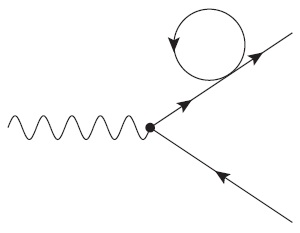}
    \caption{}
    \label{subfig:electron-phonon-vertex-1dressed-line}
\end{subfigure}
\caption{Diagrams for the correction to the electron-phonon vertex in Eq.~\eqref{eq:C_2ord-SeeSep-2bar-2bar}}
\end{figure}
Finally, for $\langle S_{ee}S_{ep}\rangle$ we obtain the expression
\begin{widetext}
\bea
\langle S_{ee}S_{ep}\rangle &=&    \left( \frac{\pi v_F}{\beta \mathcal{V}} \right)^{\frac{3}{2}} \SumInt_{\{\alpha\}} \sum_{\textbf{q} \textbf{p} j} \left[ \sum_{\{\omega_{n_i}\}} g_{\alpha_1 \alpha_3}^{\alpha_3 \alpha_2}(\textbf{q}) g^{ep}_{\alpha_4 \alpha_4, j}(\textbf{p}) \mathcal{G}^0_{\alpha_3}(\omega_{n_2}) \mathcal{G}^0_{\alpha_4}(\omega_{n_3}) \phi_j(\textbf{p},0) \Psi^{\dag}_{\alpha_1}(\omega_{n_1}) \Psi_{\alpha_2}(\omega_{n_1}) \right. \nn\\
    && \left. - \sum_{\{\omega_{n_i}\} } \sum_{\nu_m}g_{\alpha_1 \alpha_3}^{\alpha_4 \alpha_2}(\textbf{q}) g^{ep}_{\alpha_4 \alpha_3, j}(\textbf{p}) \mathcal{G}^0_{\alpha_3}(\omega_{n_2}) \mathcal{G}^0_{\alpha_4}(\omega_{n_2} - \nu_{m}) \phi_j(\textbf{p},\nu_{m}) \Psi^{\dag}_{\alpha_1}(\omega_{n_1}) \Psi_{\alpha_2}(\omega_{n_1} - \nu_{m}) \right.\nn\\
&&\left.+2 \sum_{\{\omega_{n_i}\} } \sum_{\nu_m} g_{\alpha_4 \alpha_3}^{\alpha_4 \alpha_2}(\textbf{q}) g^{ep}_{\alpha_1 \alpha_3 , j}(\textbf{p}) \mathcal{G}^0_{\alpha_3}(\omega_{n_1} - \nu_m) \mathcal{G}^0_{\alpha_4}(\omega_{n_2}) \phi_j(\textbf{p},\nu_{m}) \Psi^{\dag}_{\alpha_1}(\omega_{n_1}) \Psi_{\alpha_2}(\omega_{n_1} - \nu_{m})\right]  
    \label{eq:C_2ord-SeeSep-2bar-2bar}
\eea
\end{widetext}
The first term here corresponds to a disconnected diagram formed of the product between the first-order diagrams Fig. \ref{subfig:chemical-potential-correction-1loop} and Fig. \ref{subfig:tadpole-phonon}. The second term is the relevant correction to the electron-phonon vertex, represented by the diagram in Fig. \ref{subfig:electron-phonon-vertex-correction}. Finally, the last term gives a correction to the electron propagator, as shown in Fig. \ref{subfig:electron-phonon-vertex-1dressed-line}.
\section{Explicit renormalization in the high pseudo-magnetic field regime}

Along this section, we shall apply the general one-loop renormalization scheme discussed before, to obtain explicit results in the regime of very strong pseudo-magnetic fields at both nodes. This physical condition is in principle easier to achieve in the presence of strain, since the pseudo-magnetic fields related to strain are predicted to be much stronger~\cite{Cortijo_2015,Cortijo_2016,Arjona_Vozmediano_PRB2018,nature_1} than the actual magnetic fields that can be imposed under normal laboratory conditions $B_0/B_S \ll 1$, and hence we have $\ell_{\xi}^2 \sim B_S^{-1} \ll k_F^{-1}$. In this regime, the expansion over Landau levels is highly dominated (due to the presence of a large gap) by the lowest Landau level. 

We first need to evaluate the matrix elements according to Eq.~\eqref{eq:electron-phonon-matrix-elements} for the lowest Landau level, with $\alpha = \left(\lambda = -1,\xi,n = 0,X_{\xi},k\right)$. For backward tunneling, 
where $\Bar{\xi} \equiv -\xi$, $\alpha' = (\lambda' = -1,\Bar{\xi},n' = 0,X'_{\Bar{\xi}},k')$, we obtain (see Appendix~\ref{App_Landau_Matrix}) 
\begin{eqnarray}
&&\mathcal{M}_{\alpha}^{\alpha'}(\textbf{q}) 
    = \frac{\ell_G}{\ell_A} \delta_{k'+k_{\Bar{\xi}}, k+k_{\xi}-q_z} \delta_{X'_{\Bar{\xi}}/\ell^2_{\Bar{\xi}}, X_{\xi}/\ell^2_{\xi}+q_y}  \nonumber \\
    && \times e^{-\frac{1}{4} \frac{\ell_G^4}{\ell_A^2} q_x^2} e^{-\frac{1}{4} \frac{\ell_{\bar{\xi}}^4}{\ell_A^2} q_y^2} e^{\frac{i}{2} \frac{\ell_G^4}{\ell_A^2} q_x q_y}  e^{i\frac{\ell_{\bar{\xi}}^2}{\ell_A^2}q_x X_{\xi}},  \label{eq:D_M_backward}
\end{eqnarray}
where for notational convenience we defined the magnetic length scales
\begin{equation}
    \ell_A \equiv  \sqrt{\frac{\ell_{+}^2 + \ell_{-}^2}{2}} \hspace{4mm} \text{and} \hspace{4mm} \ell_G \equiv  \sqrt{\ell_{+} \ell_{-}} \, .   \label{eq:D_length-scales-2}
\end{equation}

On the other hand, for forward tunneling (where $\xi' = \xi$) the corresponding matrix element reduces to the expression (see Appendix~\ref{App_Landau_Matrix}) 
\begin{eqnarray}
    &&\mathcal{M}_{\alpha}^{\alpha'}(\textbf{q}) = \delta_{k', k-q_z} \delta_{X'_{\xi}, X_{\xi} + \ell^2_{\xi} q_y}\nn\\ 
    &&\times e^{-\frac{1}{4} \ell_{\xi}^2 (q_x^2 + q_y^2)} e^{\frac{i}{2} \ell_{\xi}^2 q_x q_y} e^{i q_x X_{\xi} }   \, .  \label{eq:D_M_forward}
\end{eqnarray}

We shall consider only the LLL ($n=0$, $\lambda=-1$). Furthermore, since the couplings have a very compact support in momentum space due to gaussian exponentials along the transverse directions $q_x$ and $q_y$, as seen in Eqs.~\eqref{eq:M_backward} and \eqref{eq:M_forward}, we apply the approximation $\exp\left( -q_i^2 \mathcal{L}^2_{i}/2 \right) f(q_i\mathcal{L}_i) \approx f(0)$ for $|q_i|\mathcal{L}_i<1$, and $\exp\left( -q_i^2 \mathcal{L}^2_{i}/2 \right) f(q_i\mathcal{L}_i) \approx 0$ for $|q_i|\mathcal{L}_i>1$, with $i = \{x,y\}$ and $\mathcal{L}_i$ the inverse of the corresponding standard deviation. Using the deltas to sum over $X_{2,\Bar{\xi}}$ and $k_2$, and defining
\begin{equation}
    \xi b \equiv k_{\xi} - k_{\bar{\xi}} \, ,
\end{equation}
with $b \equiv k_+ - k_- >0$, we obtain the simplified form of the electron-phonon component of the action: 
\begin{eqnarray}
    S_{ep} &=& \sqrt{\frac{\pi v_F}{\beta \mathcal{V}}} \frac{\ell_G}{\ell_A} \sum_{\xi} \sum_{\omega_n \nu_m} \sum_{\textbf{q}, j} \sum_{X_{1,\xi} } \sum_{k_1} \nonumber   \\
    && \times z_{\xi\bar{\xi}}\,g_{\xi \Bar{\xi},j} e^{i \frac{\ell_{\bar{\xi}}^2}{\ell_A^2} q_x X_{1,\xi}}  \Psi^{\dag}_{\xi ,X_{1,\xi}}(k_1, \omega_n + \nu_m) \nn\\ &&\times\Psi_{\Bar{\xi}, \varepsilon_{\Bar{\xi}}^2 X_{1,\xi} + \ell_{\Bar{\xi}}^2 q_y}(k_1 + \xi b - q_z, \omega_n)\Phi_j(\textbf{q},\nu_m), \,  \label{eq:D_S_ep_final}
\end{eqnarray}
where we included the vertex renormalization factor $z_{\xi\bar{\xi}} = \sqrt{z_{\xi}z_{\bar{\xi}}}$ arising from the field strength renormalization $\Psi_{\xi X_{\xi}}\rightarrow \sqrt{z_{\xi}}\Psi_{\xi X_{\xi}}$ and $\Psi_{\bar{\xi} X_{\bar{\xi}}}\rightarrow \sqrt{z_{\bar{\xi}}}\Psi_{\bar{\xi} X_{\bar{\xi}}}$, respectively.
Let us now consider the electron-electron interaction. For this purpose, we shall distinguish the short-wavelength part of the Coulomb interaction, involved in backward scattering, from the long-wavelength part, which is mainly involved in forward scattering.

Let us start from the backward scattering contribution by defining the distance in momentum space between both nodal momenta,
\begin{equation}
    2k_F \equiv k_+ - k_- \, .
\end{equation}
We further define the backward Coulomb coupling $g_1$ (at the bare level) as an average over the sphere $S$ of radius $\ell_A/\ell_G^2$, as this length scale defines the support in $q_x$ of the gaussian exponentials in Eq.~\eqref{eq:D_M_backward},
\bea
g_1 &=& \frac{1}{\pi \hbar v_F} \frac{\int_{S} d^2q_{\perp} \, V(2k_F \mbf{\hat{z}} + \mbf{q}_{\perp})}{\int_{S} d^2q_{\perp}}\nn\\
&=& \frac{e^2 \ell_G^4}{\pi \hbar v_F \ell_A^2 \epsilon_0 \epsilon_{\infty}} \log \left( 1 + \frac{\ell_A^2}{4k_F^2 \ell_G^4} \right).\label{eq:D_g1-definition}
\eea

The "backward" contribution of the electron-electron interaction to the action is thus given by the expression 
\begin{widetext}
\begin{eqnarray}
    S_{ee}^{B} 
    &=& g_1 \frac{\pi v_F}{2 \beta \mathcal{V}} \frac{\ell_G^2}{\ell_A^2} \sum_{\xi \textbf{q}} \sum_{\{ \omega_{n} \}} \sum_{X_{1,\xi} X_{2,\Bar{\xi}}} \sum_{k_1 k_2} e^{i \frac{\ell_{\bar{\xi}}^2}{\ell_A^2} q_x \left( X_{1,\xi} - \varepsilon_{\xi}^2 X_{2,\Bar{\xi}} \right)}  \Psi^{\dag}_{\xi X_{1,\xi}}(k_1, \omega_{n_1}) \Psi^{\dag}_{\Bar{\xi} X_{2,\Bar{\xi}}}(k_2, \omega_{n_2})\nonumber \\
    && \times \, \Psi_{\xi, \varepsilon_{\xi}^2 X_{2,\Bar{\xi}} - \ell_{\xi}^2 q_y}(k_2-\xi b+q_z, \omega_{n_3})   \Psi_{\Bar{\xi}, \varepsilon_{\Bar{\xi}}^2 X_{1,\xi} + \ell_{\Bar{\xi}}^2 q_y}(k_1+\xi b-q_z, \omega_{n_1 + n_2 - n_3} ) \, .
\end{eqnarray}
\end{widetext}

Performing a similar analysis for the long-wavelength component of the Coulomb interaction, named forward scattering, the corresponding contribution to the action is
\begin{eqnarray}
&&S_{ee}^{F} 
= g_2 \frac{\pi v_F}{2 \beta \mathcal{V}} \sum_{\xi \textbf{q}} \sum_{\{ \omega_{n} \}} \sum_{X_{1,\xi} X_{2,\Bar{\xi}}} \sum_{k_1 k_2}  e^{i q_x \left( X_{1,\xi} - X_{2,\Bar{\xi}} \right)}  \nonumber \\
    && \times \Psi^{\dag}_{\xi X_{1,\xi}}(k_1, \omega_{n_1})  \Psi^{\dag}_{\Bar{\xi} X_{2,\Bar{\xi}}}(k_2, \omega_{n_2}) \Psi_{\Bar{\xi}, X_{2,\Bar{\xi}} - \ell_{\Bar{\xi}}^2 q_y}(k_2 +q_z, \omega_{n_3}) \nonumber \\
&& \times  \Psi_{\xi, X_{1,\xi} + \ell_{\xi}^2 q_y}(k_1-q_z, \omega_{n_1 + n_2 - n_3} ) \, , 
\end{eqnarray}
where the forward (long-range) Coulomb coupling $g_2$ is defined by
\begin{equation}
    \lim_{q\to 0} \frac{V(\mbf{q})}{\epsilon(\mbf{q},0)} \equiv g_2 \pi \hbar v_F \, , \label{eq:D_g2-definition}
\end{equation}
with 
\begin{equation}
    \epsilon(\mbf{q}, \nu_m) = 1 - V(\mbf{q})\Pi_{\text{RPA}}(\mbf{q},\nu_m)
\end{equation}
being the dielectric function in the random phase approximation (RPA), and 
\bea
    \Pi_{\text{RPA}}(\mbf{q},\nu_m) &=& \frac{1}{\mathcal{V}} \SumInt_{\alpha \alpha'} \left| \int d^3x \; \Psi^{\dag}_{\alpha}(\textbf{x}) e^{i\textbf{q} \cdot \textbf{x}} \Psi_{\alpha'}(\textbf{x}) \right|^2\nn\\ 
    &&\times\frac{f_0(E_{\alpha}) - f_0(E_{\alpha'})}{E_{\alpha} - E_{\alpha'} + \hbar \nu_m + i\eta} \label{eq:D_polarization-definition}
\eea
is the electronic polarization function in RPA, $f_0(E) = \left( \exp(\beta (E - \mu)) + 1 \right)^{-1}$ is the Fermi distribution function, and $\eta \to 0^+$ is a regulator. In the high pseudo-magnetic field regime, we can approximate our calculations by restricting ourselves to the chiral Landau level, by considering $\xi = \xi'$ and $n=n'=0$ in the sets of indices $\alpha$ and $\alpha'$.

As explained in detail in Appendix~\ref{app_RPA}, we obtain
\begin{equation}
    g_2 = 4\pi \ell_G^2 \, \frac{1 + \frac{\delta v}{v_F}}{2\frac{\ell_A^2}{\ell_G^2} + \frac{\delta v}{v_F}\varepsilon_-}  \, .
\end{equation}

\subsection{One-loop renormalization of the electron-electron vertex}

The electron-electron internodal vertex will introduce corrections for the backward $g_1$ and forward $g_2$ couplings arising from 2 different channels. The Peierls channel ($\sim\Bar{\Psi}_+^{\dag} \Psi_-^{\dag} \Bar{\Psi}_- \Psi_+ + \Psi_+^{\dag} \Bar{\Psi}_-^{\dag} \Psi_- \Bar{\Psi}_+$) represents the pairing of one electron and one hole with opposite chiralities, while the Cooper channel ($\sim\Bar{\Psi}_+^{\dag} \Bar{\Psi}_-^{\dag} \Psi_- \Psi_+ + \Psi_+^{\dag} \Psi_-^{\dag} \Bar{\Psi}_- \Bar{\Psi}_+$) represents the pairing of two electrons (or two holes) with opposite chiralities. In addition, the Landau channel ($\sim\Bar{\Psi}_+^{\dag} \Psi_-^{\dag} \Psi_- \Bar{\Psi}_+ + \Psi_+^{\dag} \Bar{\Psi}_-^{\dag} \Bar{\Psi}_- \Psi_+$) only contributes to intranodal scattering and therefore does not contribute to the renormalization of the $g_1$ and $g_2$ couplings.

\subsubsection{Peierls channel}

Let us first analyze the vertex corrections arising from the Peierls channel. Applying the general procedure described in Section~\ref{field_theory_ren}, and restricting ourselves to the LLL subspace, the contribution from forward scattering is given by 
\begin{widetext}
\bea
    &&\frac{1}{2}\braket{\left(S^F_{ee}\right)_P^2} = 2 g_2^2 \left( \frac{\pi v_F}{2 \beta \mathcal{V}} \right)^2 \sum_{\xi \textbf{q}} \sum_{X'_{1,\xi} X_{2,\Bar{\xi}}} \sum_{k_1 k'_2} \sum_{q_x q_y} \sum_{\omega_{n'_1} \omega_{n_2} \omega_{n'_3}} e^{iq_x(X'_{1,\xi} - X_{2,\bar{\xi}})}  \Psi^{\dag}_{\xi X'_{1,\xi}}(k'_1, \omega_{n'_1}) \Psi^{\dag}_{\Bar{\xi} X_{2,\Bar{\xi}}}(k_2, \omega_{n_2})   \\
    && \times \Psi_{\Bar{\xi}, X_{2,\Bar{\xi}} - \ell_{\Bar{\xi}}^2 q_y}(k_2+q_z, \omega_{n'_3}) \Psi_{\xi, X'_{1,\xi} + \ell_{\xi}^2 q_y}(k'_1-q_z, \omega_{n'_1 + n_2- n'_3})    \sum_{q'_x q'_y} \sum_{k \omega_{n_1}} \mathcal{G}^0_{\xi}(k + k'_1 - k_2 - q_z, \omega_{n_1}) \mathcal{G}^0_{\bar{\xi}}(k, \omega_{n_1 - n'_1 + n'_3}) \nn ,  \label{eq:D_correction-F_P^2-aux1}
\eea
\end{widetext}
where the factor of 2 arises as the multiplicity factor of the third term in Eq.~\eqref{eq:C_SeeSee}.

In what follows, we shall evaluate the sum over $q'_x$ and $q'_y$ in the continuum limit  
\begin{equation}
    \sum_{q'_x q'_y} \rightarrow \frac{L_x L_y}{(2\pi)^2} \int\int_{-1/\ell_A}^{1/\ell_A} dq'_x dq'_y = \frac{L_x L_y}{\pi^2 \ell_A^2}  \, , \label{eq:D_sum_qx_qy_forward}
\end{equation}
where we used the length scale defined in the gaussian exponential in Eq.~\eqref{eq:D_M_forward}. as a compact support in momentum space.

Since we are computing the action within the lowest Landau level subspace, after Eq.~\eqref{eq:A_eigenvalues_2} the corresponding energy eigenvalue (for $n=0$) is $\epsilon_{n=0,\xi}=\xi\hbar v_F (k - k_{F_{\xi}})$, and hence the corresponding explicit expression for the free Fermion Green's functions is
\begin{equation}
    \mathcal{G}^0_{\xi}\left(k, \omega_{n}\right) = \frac{1}{i\hbar \omega_n - \xi \hbar v_{\xi} (k - k_{F_{\xi}})} \, .
\end{equation}
We can approximate the relative difference between external momenta in Eq.~\eqref{eq:D_correction-F_P^2-aux1} as $k'_1 - k_2 \approx k_{\xi} - k_{\bar{\xi}} \equiv \xi b$, since each momenta in this expression are very close to their respective Weyl node.  Also, introducing $\nu_m = \omega_{n'_1} - \omega_{n'_3}$, the Matsubara sums and the $k$-integrals can be expressed as (for $\xi b - q_z  \approx 2 k_{F}$)
\bea 
&&\sum_{k} \sum_{\omega_{n}} \mathcal{G}^0_{\xi}\left(k+\xi b-q_{z}, \omega_{n}\right) \mathcal{G}^0_{\bar{\xi}}\left(k, \omega_{n}-\nu_{m}\right) \nonumber \\
&&=\sum_{k} \sum_{\omega_{n}} \mathcal{G}^0_{\xi}\left(k+2 k_{F}, \omega_{n}\right) \mathcal{G}^0_{\bar{\xi}}\left(k, \omega_{n}-\nu_{m}\right) \nonumber \\
&&=\sum_{k} \sum_{\omega_{n}} \mathcal{G}^0_{\xi}\left(k+2 k_{F}, \omega_{n}+\nu_{m}\right) \mathcal{G}^0_{\bar{\xi}}\left(k, \omega_{n}\right), \label{eq:D_peierls-symmetrization-1}
\eea
where in the second step we performed the frequency shift $\omega_n \to \omega_n + \nu_m$.
This operation can be symmetrized, in order to define the polarization insertion $\Pi_P(\nu_m)$ for the Peierls channel (details in Appendix~\ref{App_Peierls})
\bea
    \Pi_P(\nu_m) &\equiv& \frac{1}{2} \sum_{k} \sum_{\omega_{n}}\left[\mathcal{G}^0_{+}\left(k, \omega_{n}\right) \mathcal{G}^0_{-}\left(k - 2 k_{F}, \omega_{n} - \nu_m\right)\right.\nn\\
    &&\left.+ \mathcal{G}^0_{+}\left(k+2 k_{F}, \omega_{n} + \nu_m \right) \mathcal{G}^0_{-}\left(k, \omega_{n}\right) \right]\nn\\
    &=&-\frac{\beta L_z}{2 \pi \hbar v_F} \cdot \frac{dl}{1+ \delta v/2v_F} \lambda_P(T,\delta v. \nu_m,l) , \label{eq:D_I_P}
\eea
Here, we defined
\bea
   &&\lambda_P(T,\delta v, \nu_m,l) \equiv \frac{1}{1-i\frac{\hbar \nu_m}{(1+\gamma)\Lambda(l)}} \frac{1}{4} \left[ 2\tanh\left( \frac{\beta \Lambda(l)}{2} \right)\right.\nn\\
    &&\left.+ \tanh\left( \frac{\beta \Lambda(l)}{2} \gamma \right) + \tanh\left( \frac{\beta \Lambda(l)}{2\gamma}  \right) \right] \, . \label{eq:D_lambda_P}
\eea

By substituting Eq.~\eqref{eq:D_sum_qx_qy_forward} and Eq.~\eqref{eq:D_I_P} into Eq.~\eqref{eq:D_correction-F_P^2-aux1}, and defining $\nu_m = \omega_{n'_1} - \omega_{n'_3}$, we conclude that the contribution to the beta function from this diagram is   
\begin{eqnarray}
    \left( \frac{dg_2}{dl} \right)_P = g_2^2 \frac{1}{2\pi^2\ell_A^2} \frac{\lambda_P(l)}{1 + \delta v/2v_F}.  \label{eq:D_beta-ee-P-F2}
\end{eqnarray}

Let us now consider the contribution to the Peierls channel arising from backward scattering.  Following the general procedure described in Section~\ref{field_theory_ren} and restricting ourselves to the LLL subspace, we obtain 
\begin{eqnarray}
    &&\frac{1}{2}\braket{\left(S^B_{ee}\right)_P^2} = -2 g_1^2 \left( \frac{\pi v_F}{2 \beta \mathcal{V}} \right)^2 \frac{\ell_G^4}{\ell_A^4} \sum_{\xi \textbf{q}} \sum_{k'_1 k_2} \sum_{\{\omega_{n}\}}    \nonumber \\
    && \times  \sum_{X'_{1,\xi} X_{2,\Bar{\xi}}}\sum_{X'_{2,\bar{\xi}}} e^{i \frac{\ell_{\xi}^2}{\ell_A^2} q_x(X'_{2,\bar{\xi}} - X_{2,\bar{\xi}})} \sum_{q'_x} e^{i \frac{\ell_{\bar{\xi}}^2}{\ell_A^2} q'_x(X'_{1,\xi} - \varepsilon_{\xi}^2 X'_{2,\bar{\xi}})}    \nonumber \\
    && \times  \sum_{k'_2 \omega_{n'_1}} \mathcal{G}^0_{\xi}(k'_2 - \xi b + q_z, \omega_{n'_1}) \mathcal{G}^0_{\bar{\xi}}(k'_2, \omega_{n'_1+ n_2 - n_3})  \nonumber \\
    && \times
    \Psi^{\dag}_{\xi X'_{1,\xi}}(k'_1, \omega_{n_1})
    \Psi^{\dag}_{\Bar{\xi} X_{2,\Bar{\xi}}}(k_2, \omega_{n_2})\nonumber\\
    &&\times\Psi_{\xi, \varepsilon_{\xi}^2 X_{2,\Bar{\xi}} - \ell_{\xi}^2 q_y}(k_2-\xi b+q_z, \omega_{n_3})  \nonumber \\
    && \times  \Psi_{\bar{\xi}, \varepsilon_{\bar{\xi}}^2 X'_{1,\xi} + \ell_{\bar{\xi}}^2 q_y}(k'_1+ \xi b -q_z, \omega_{n_1 + n_2 - n_3}) .
    \label{eq:D_See2Peierls}
\end{eqnarray}

Applying the approximation $\xi b - q_z \approx 2k_F$, and after the same symmetrization procedure leading to Eq.~\eqref{eq:D_peierls-symmetrization-1}, we obtain in this case
\be
    \sum_{k'_2 \omega_{n'_1}} \mathcal{G}^0_{\xi}(k'_2 - 2k_F, \omega_{n'_1} - \nu_m) \mathcal{G}^0_{\bar{\xi}}(k'_2, \omega_{n'_1}) 
    =\, \Pi_P(\nu_m).  \label{eq:D_peierls-polarization-aux2}
\ee

The sum over $q'_x$ is calculated in the continuum limit as follows 
\begin{align}
\sum_{q'_x} e^{i \frac{\ell_{\bar{\xi}}^2}{\ell_A^2} q'_x(X'_{1,\xi} - \varepsilon_{\xi}^2 X'_{2,\bar{\xi}})} &= \frac{L_x}{2\pi} \int_{-\infty}^{+\infty} dq'_x e^{i \frac{\ell_{\bar{\xi}}^2}{\ell_A^2} q'_x(X'_{1,\xi} - \varepsilon_{\xi}^2 X'_{2,\bar{\xi}})} \nonumber \\
&= L_{x} \frac{\ell_A^2}{\ell_{\bar{\xi}}^2} \frac{1}{\varepsilon_{\xi}^2} \delta\left(X'_{2,\bar{\xi}} - \varepsilon_{\xi}^{2} X'_{1,\xi} \right).
\end{align}
Substituting this result into the subsequent $X_{\bar{\xi}}$-integral, we obtain
\bea
&&\frac{L_y}{2\pi \ell_{\bar{\xi}}^2} \int_{-\pi \ell_{\bar{\xi}}/L_y}^{\pi \ell_{\bar{\xi}}/L_y} dX'_{2,\bar{\xi}} e^{i \frac{\ell_{\xi}^2}{\ell_A^2} q_{x} \left(X'_{2,\xi} - X_{2,\bar{\xi}}\right)}\nn\\ 
&&\times L_{x} \frac{\ell_A^2}{\ell_{\bar{\xi}}^2} \frac{1}{\varepsilon_{\xi}^2} \delta\left(X'_{2,\bar{\xi}} - \varepsilon_{\xi}^{2} X'_{1,\xi} \right)  \nonumber \\
&&= \frac{L_x L_y \ell_A^2}{2\pi \ell_G^4} e^{i \frac{\ell_{\bar{\xi}}^2}{\ell_A^2} q_{x} \left(X'_{1,\xi} - \varepsilon_{\xi}^{2} X_{2,\bar{\xi}}\right)} \, .
\eea

Thus, upon substituting into Eq.\eqref{eq:D_See2Peierls}, 
the contribution to the beta function for the $g_1$ coupling is  
\begin{eqnarray}
    \left( \frac{dg_1}{dl} \right)_{P,B^2} &=& -g_1^2 \frac{1}{4\pi \ell_G^2} \frac{\lambda_P(l)}{1 + \delta v/2v_F} \, , \label{eq:D_beta-ee-P-B2}
\end{eqnarray}
where we applied the definitions of the magnetic lengths defined in Eq.~\eqref{eq:D_length-scales-2}.

The last contribution arising from the Peierls channel corresponds to the product between backward and forward couplings $g_1 g_2$, given by 
\begin{eqnarray}
    &&\braket{\left( S^F_{ee} S^B_{ee} \right)_P} = 4\left( \frac{\pi v_F}{2 \beta \mathcal{V}} \right)^2 \frac{\ell_G^2}{\ell_A^2} \sum_{\xi \textbf{q}'} \sum_{X'_{1,\xi} X_{2,\Bar{\xi}}} \sum_{k'_1 k_2}\sum_{\{ \omega_n \}}   \nonumber \\
    && \times  e^{i \frac{\ell_{\bar{\xi}}^2}{\ell_A^2} q'_x(X'_{1,\xi} - \varepsilon_{\xi}^2 X_{2,\bar{\xi}})} \Psi^{\dag}_{\xi X'_{1,\xi}}(k'_1, \omega_{n'_1}) \Psi^{\dag}_{\Bar{\xi} X_{2,\Bar{\xi}}}(k_2, \omega_{n_2})  \nonumber \\
    &&\times \Psi_{\xi, X_{1,\xi} + \ell_{\xi}^2 q_y}(k-\xi b+q'_z, \omega_{n_3}) \nn\\
    && \times \, \Psi_{\bar{\xi}, \varepsilon_{\bar{\xi}}^2 X'_{1,\xi} + \ell_{\bar{\xi}}^2 q'_y}(k'_1+\xi b-q'_z, \omega_{n'_1 + n_2 - n_3})  \\
    && \times \, g_1 g_2 \sum_{q_x q_y} \sum_{k\omega_{n_1}} \mathcal{G}^0_{\xi}(k-\xi b+q'_z, \omega_{n_1}) \mathcal{G}^0_{\bar{\xi}}(k, \omega_{n_1 + n_2 - n_3}) \nn .
\end{eqnarray}

As this is a correction to the backward vertex, the sum over $q_x$ and $q_y$ uses the forward length scales, as we have already calculated in Eq.~\eqref{eq:D_sum_qx_qy_forward}. Using the approximation $\xi b - q'_z \approx 2k_F$ and doing the changes $\omega_{n_2} - \omega_{n_3} = \nu_m$ and $\omega_{n_1} \to \omega_{n_1} - \nu_m$, the Matsubara sum and the $k$-integral are exactly the same as we obtained in Eq.~\eqref{eq:D_peierls-polarization-aux2} for the previous diagram, so it reduces to the polarization $\Pi_P(\nu_m)$. 
Thus, we conclude that the contribution to the beta function of $g_1$ arising from this diagram is 
\begin{eqnarray}
    \left( \frac{dg_1}{dl} \right)_{P,F\cdot B} &=& g_1 g_2 \frac{1}{\pi^2 \ell_{A}^2} \frac{\lambda_P(l)}{1 + \delta v/2v_F} . \label{eq:D_beta-ee-P-FB}
\end{eqnarray}


\subsubsection{Cooper channel}

Let us now consider the vertex corrections arising from the Cooper channel. The contribution arising from forward scattering is the following
\begin{eqnarray}
    &&\frac{1}{2}\braket{\left(S^F_{ee}\right)_C^2} = 2\left( \frac{\pi v_F}{2 \beta \mathcal{V}} \right)^2 \sum_{\xi \textbf{q}} \sum_{X_{1,\xi} X_{2,\Bar{\xi}}} \sum_{k_1 k_2} \sum_{\left\{\omega_{n_i}\right\} }    \nonumber \\
    && \times e^{iq_x(X_{1,\xi} - X_{2,\bar{\xi}})} \Psi^{\dag}_{\xi X_{1,\xi}}(k_1, \omega_{n_1}) \Psi^{\dag}_{\Bar{\xi} X_{2,\Bar{\xi}}}(k_2, \omega_{n_2})  \nonumber \\
    && \times \Psi_{\Bar{\xi}, X_{2,\Bar{\xi}} - \ell_{\Bar{\xi}}^2 q_y}(k_2+q_z, \omega_{n_3})   \nonumber \\
    &&\times \Psi_{\xi, X_{1,\xi} + \ell_{\xi}^2 q_y}(k_1-q_z, \omega_{n_1+n_2 - n_3})\nn\\
     && \times  g_2^2  \sum_{q'_x q'_y} \sum_{k \omega_{n}} \mathcal{G}^0_{\xi}(k_1 + k_2 - k, \omega_{n_1 + n_2 - n}) \mathcal{G}^0_{\bar{\xi}}(k, \omega_{n}) \, .
\end{eqnarray}
The last line gives a correction to the forward coupling $g_2$. For the sum over $q'_x$ and $q'_y$, we have the same result as in Eq.~\eqref{eq:D_sum_qx_qy_forward}. This expression can be symmetrized in order to define the polarization insertion from the Cooper channel as follows (details in Appendix~\ref{App_Cooper}) 
\bea
    \Pi_C(\nu_m) &= &  \frac{1}{2}\sum_{k} \sum_{\omega_{n}}  \left[ \mathcal{G}^0_{+}\left(k, \omega_{n}\right) \right.\nn\\
    &&\left.\times\mathcal{G}^0_{-}\left(k_{+}+ k_{-}- k, \omega_{n_1 + n_2 - n}\right) \right. \nonumber \\
&&\left. +  \mathcal{G}^0_{+}\left(k_{+}+ k_{-}- k, \omega_{n_1 + n_2 - n}\right) \mathcal{G}^0_{-}\left(k, \omega_{n}\right) \right]\nn\\
&=& \frac{\beta L_z}{2\pi \hbar v_F} \cdot \frac{dl}{1+\delta v / 2v_F} \lambda_C(T,\delta v, \Delta, \nu_m,l),
\label{eq:D_polarization-cooper}
\eea
where we have defined
\bea
&&\lambda_C(T, \delta v, \Delta, \nu_m,l) 
    = \frac{1}{4}
\sum_{s=\pm 1} \frac{1}{1 + s\frac{\gamma}{1 + \gamma} \frac{\Delta(l)}{\Lambda}}\nn\\
&&\times \frac{1}{1- \frac{i s}{1+ s\frac{\gamma}{1+\gamma} \frac{\Delta}{\Lambda(l)}}  \frac{\hbar\nu_m}{(1+\gamma)\Delta}} \left[ \frac{1}{2}\tanh\left( \frac{\beta \gamma (\Lambda(l) + s\Delta)}{2}  \right)\right.\nn\\ 
&&\left.+ \frac{1}{2}\tanh\left( \frac{\beta (\Lambda(l) + s\gamma \Delta)}{2\gamma} \right) + \tanh\left( \frac{\beta \Lambda(l)}{2} \right)\right],
\eea
and
\begin{equation}
    \Delta \equiv \hbar v_F \left[ (k_{F_+} - k_+) + (k_{F_-} - k_-) \right] \,.
\end{equation}
Thus, after setting $\omega_{n_1} + \omega_{n_2} = \nu_m$, we conclude that the contribution to the beta function of $g_2$ arising from this diagram is given by 
\begin{eqnarray}
    \left( \frac{dg_2}{dl} \right)_C &=& -g_2^2 \frac{1}{2\pi^2 \ell_A^2} \frac{\lambda_C}{1 + \delta v/2v_F} \, . \label{eq:D_beta-ee-C-F2}
\end{eqnarray}
Let us now calculate the backward scattering contribution in the Cooper channel:  
\begin{eqnarray}
    &&\frac{1}{2}\braket{\left(S^B_{ee}\right)_C^2} =  2\left( \frac{\pi v_F}{2 \beta \mathcal{V}} \right)^2 \frac{\ell_G^4}{\ell_A^4} \sum_{\xi \textbf{q}'} \sum_{X_{1,\xi} X_{2,\Bar{\xi}}} \sum_{k_1 k_2} \sum_{\{\omega_{n}\}}    \nonumber \\
    && \times e^{i \frac{\ell_{\bar{\xi}}^2}{\ell_A^2} q_x (X_{1,\xi} - \varepsilon_{\xi}^2 X_{2,\bar{\xi}})} \Psi^{\dag}_{\xi X_{1,\xi}}(k_1, \omega_{n_1}) \Psi^{\dag}_{\Bar{\xi} X_{2,\Bar{\xi}}}(k_2, \omega_{n_2}) \nonumber \\
    && \times \Psi_{\Bar{\xi}, X_{2,\Bar{\xi}} - \ell_{\Bar{\xi}}^2 q'_y}(k_2+q'_z, \omega_{n'_3})  \nonumber \\
    && \times  \Psi_{\xi, X_{1,\xi} + \ell_{\xi}^2 q'_y}(k_1-q'_z, \omega_{n_1 + n_2 - n'_3}) \\
    && \times g_1^2 \sum_{q_x q_y}  \sum_{k \omega_{n_3}} \mathcal{G}^0_{\xi}(k_2 + k_1 - k, \omega_{n_3}) \mathcal{G}^0_{\bar{\xi}}(k, \omega_{n_1 + n_2 - n_3}). \nonumber
\end{eqnarray}

Introducing the approximation $k_1 + k_2 \approx k_+ + k_-$, then setting $\nu_m = \omega_{n_1} + \omega_{n_2}$, followed by the change of variable $\omega_{n_3} \to \nu_m-\omega_{n_3}$, and further applying the same symmetrization as before, the Matsubara sum and the $k$-integral correspond exactly to the definition of $\Pi_C(\nu_m)$ given in Eq.~\eqref{eq:D_polarization-cooper} (and in Appendix~\ref{App_Cooper}). On the other hand, the sum over $q_x$ and $q_y$ is computed in the continuum limit using the momentum cutoff for the backward vertex, as follows 
\begin{align}
\sum_{q_x q_y} &= \frac{L_x L_y}{(2\pi)^2} \int_{- \ell_A/\ell_G^2}^{\ell_A/\ell_G^2} dq_x \int_{-\sqrt{2}\ell_A/\sqrt{\ell_+^4 + \ell_-^4}}^{ \sqrt{2}\ell_A/\sqrt{\ell_+^4 + \ell_-^4}} dq_y \nn \\
&= \frac{\sqrt{2}L_x L_y}{\pi^2 \sqrt{\ell_+^4 + \ell_-^4}} \frac{\ell_A^2}{\ell_G^2} \nn \\
&= \frac{L_x L_y}{\pi^2 \ell_G^2} \frac{1}{\sqrt{2 - \frac{\ell_G^4}{\ell_A^4}}} .
\end{align}
where the integration limits are defined by the magnetic length-scale that determines the compact support in momentum space of the gaussian exponential in Eq.~\eqref{eq:D_M_backward}.

Thus, defining $\omega_{n_1} + \omega_{n_2} = \nu_m$, we conclude that the contribution to the beta function of $g_2$ from this diagram is 
\begin{eqnarray}
    \left( \frac{dg_2}{dl} \right)_{C,B^2} &=& -g_1^2 \frac{\ell_G^2}{2\pi^2 \ell_A^4 \sqrt{2 - \frac{\ell_G^4}{\ell_A^4}}} \frac{\lambda_C(l)}{1 + \delta v / 2v_F} \, . \label{eq:D_beta-ee-C-B2}
\end{eqnarray}
The last contribution from the Cooper channel corresponds to the product between the forward and backward scattering couplings $g_1 g_2$,  
\begin{eqnarray}
&&\braket{\left( S^F_{ee} S^B_{ee} \right)_C} = 4\left( \frac{\pi v_F}{2 \beta \mathcal{V}} \right)^2 \frac{\ell_G^2}{\ell_A^2} \sum_{\xi \textbf{q}'} \sum_{X_{1,\xi} X_{2,\Bar{\xi}}} \sum_{k_1 k_2} \sum_{\{ \omega_n \}}  \\
    && \times e^{i \frac{\ell_{\bar{\xi}}^2}{\ell_A^2} q'_x(X_{1,\xi} - \varepsilon_{\xi}^2 X_{2,\bar{\xi}})}\Psi^{\dag}_{\xi X_{1,\xi}}(k_1, \omega_{n_1}) \nonumber \\
    && \times \Psi^{\dag}_{\Bar{\xi} X_{2,\Bar{\xi}}}(k_2, \omega_{n_2})  \Psi_{\xi, \varepsilon_{\xi}^2 X_{2,\bar{\xi}} - \ell_{\xi}^2 q'_y}(k_2-\xi b+q'_z, \omega_{n_3}) \nonumber \\
    && \times \, \Psi_{\bar{\xi}, \varepsilon_{\bar{\xi}}^2 X_{1,\xi} + \ell_{\bar{\xi}}^2 q'_y}(k_1+\xi b-q'_z, \omega_{n_1 + n_2 - n_3})\nn\\
    &&\times g_1 g_2\sum_{q_x q_y}\sum_{k \omega_n} \mathcal{G}^0_{\xi}\left(k, \omega_{n} \right) \mathcal{G}^0_{\bar{\xi}}\left(k_1 + k_2 - k, \omega_{n_1 + n_2 - n}\right). \nn
\end{eqnarray}

Introducing the approximation $k_1 + k_2 \approx k_+ + k_-$, then setting $\nu_m = \omega_{n_1} + \omega_{n_2}$, and further applying the same symmetrization procedure as before, the Matsubara sum and the $k$-integral correspond precisely to the definition of $\Pi_C(\nu_m)$ given in Eq.~\eqref{eq:D_polarization-cooper} (and in Appendix~\ref{App_Cooper}). As this is a correction to the backward vertex $g_1$, the sum over $q_x$ and $q_y$ is obtained from Eq.~\eqref{eq:D_sum_qx_qy_forward}. 
By finally setting $\omega_{n_1} + \omega_{n_2} = \nu_m$, we conclude that the contribution to the beta function of $g_1$ arising from this term is  
\begin{eqnarray}
    \left( \frac{dg_1}{dl} \right)_{C,F\cdot B} &=& -g_1 g_2 \frac{1}{\pi^2  \ell_A^2} \frac{\lambda_C(l)}{1 + \delta v/2v_F} \, . \label{eq:D_beta-ee-C-FB}
\end{eqnarray}

\subsubsection{Electronic beta functions}

Combining Eqs.~\eqref{eq:D_beta-ee-P-B2}, \eqref{eq:D_beta-ee-P-FB} and \eqref{eq:D_beta-ee-C-FB}, we obtain the total beta function and the corresponding flow for the forward coupling,  
\bea
    \frac{dg_1}{dl} = - \frac{g_1^2}{4\pi \ell_G^2} \frac{\lambda_P}{1 + \delta v / 2 v_F}+  \frac{g_1 g_2}{\pi^2 \ell_A^2} \frac{\lambda_P - \lambda_C}{1 + \delta v / 2 v_F}. \label{eq:D_beta-ee-forward}
\eea
Similarly, by adding Eqs.~\eqref{eq:D_beta-ee-C-B2}, \eqref{eq:D_beta-ee-P-F2} and \eqref{eq:D_beta-ee-C-F2}, we obtain the total beta function and the corresponding flow for the backward coupling:  
\bea
    \frac{dg_2}{dl} &=& - \frac{g_1^2\ell_G^2}{2\pi^2 \ell_A^4 \sqrt{2 - \frac{\ell_G^4}{\ell_A^4}}} \frac{\lambda_C}{1 + \delta v / 2 v_F}\\
    &&+  \frac{g_2^2}{2\pi^2 \ell_A^2} \frac{\lambda_P - \lambda_C}{1 + \delta v / 2 v_F} \nn . \label{eq:D_beta-ee-backward}
\eea
\subsection{One-loop renormalization of the electron-phonon vertex}
Let us now analyze the renormalization effects over the electron-phonon coupling, as predicted by our model. We first consider the contribution to the backward channel, which is given by 
\begin{widetext}
\bea
&&\braket{S_{ep}S^B_{ee}} = - \left(\frac{\pi v_F}{\beta \mathcal{V}}\right)^{3/2} \frac{\ell_G^3}{\ell_A^3} \sum_{\xi} \sum_{\textbf{q}\, j}  \sum_{q'_x} \sum_{X_{1,\xi} X_{2,\bar{\xi}}} \sum_{k_1 k_2} \sum_{\omega_{n_1} } 
 g_1 z_{\xi \bar{\xi}} g_{\xi \bar{\xi},j} e^{i \frac{\ell_{\xi}^2}{\ell_A^2} q'_x X_{2,\bar{\xi}}} e^{i \frac{\ell_{\bar{\xi}}^2}{\ell_A^2} q_x \left( X_{1,\xi} - \varepsilon_{\xi}^2 X_{2,\bar{\xi}} \right)} \\
    &&\times\sum_{\omega_n \nu_m} \mathcal{G}^0_{\xi}(k_2 - \xi b + q_z, \omega_n + \nu_m) \mathcal{G}^0_{\bar{\xi}}(k_2, \omega_n)    \Psi^{\dag}_{\xi, X_{1,\xi}}(k_1, \omega_{n_1}) \Psi^{\dag}_{\bar{\xi}, \varepsilon_{\bar{\xi}}^2 X_{1,\xi} + \ell_{\bar{\xi}}^2 q_y}(k_1 + \xi b - q, \omega_{n_1} - \nu_{m}) \phi_{j}(q'_x, q_y, q_z, \nu_m)\nn.
\eea
\end{widetext}

The Matsubara sum and the $k_2$-integral, using the approximation $\xi b - q_z \approx 2k_F$, give the same expression presented in Eq.~\eqref{eq:peierls-symmetrization-1} (see Appendix~\ref{App_Peierls} for details). Hence, applying the same symmetrization procedure, we obtain the result
\bea
\sum_{k_2 \omega_n} \mathcal{G}^0_{\xi}(k_2 + 2k_F, \omega_n + \nu_m) \mathcal{G}^0_{\bar{\xi}}(k_2, \omega_n)  
    = \Pi_P(\nu_m).
\eea
On the other hand, the sum over $q'_x$ and $X_{2,\bar{\xi}}$ is calculated in the continuum limit, as before, to obtain the result
\bea
    &&\sum_{q'_x X_{2,\bar{\xi}}} e^{i \frac{\ell_{\xi}^2}{\ell_A^2} X_{2,\bar{\xi}} (q'_x - q_x)} =  \frac{L_x L_y}{2\pi} \frac{\ell_A^2}{\ell_G^4}. \nn 
\eea
Thus, we conclude that the contribution to the beta function for the electron-phonon vertex arising from this backward channel is given by the expression
\begin{equation}
    \left[ \frac{d}{dl} \log \left( z_{\xi \bar{\xi}} \right) \right]_B = - g_1 \frac{1}{8\pi \ell_G^2} \frac{\lambda_P(l)}{1 + \delta v / 2 v_F} \, . \label{eq:D_beta-ep-backward}
\end{equation}

Similarly, for the forward channel, we have the contribution  
\begin{widetext}
\bea
&&\braket{S_{ep}S^F_{ee}} = \sqrt{\frac{\pi v_F}{\beta \mathcal{V}}} \frac{\pi v_F}{2\beta \mathcal{V}} \frac{\ell_G}{\ell_A} \sum_{\xi} \sum_{\textbf{q}\, j}  \sum_{\textbf{q}'} \sum_{X_{1,\xi} k_1 } \sum_{\omega_{n_1} }  g_2 z_{\xi \bar{\xi}} g_{\xi \bar{\xi},j} e^{i \frac{\ell_{\bar{\xi}}^2}{\ell_A^2} q'_x X_{1,\xi}}  \Psi^{\dag}_{\xi, X_{1,\xi}}(k_1, \omega_{n_1})  \nonumber \\
    && \times \,  \Psi^{\dag}_{\bar{\xi}, \varepsilon_{\bar{\xi}}^2 X_{1,\xi} + \ell_{\bar{\xi}}^2 q'_y}(k_1 + \xi b - q'_z, \omega_{n_1} - \nu_{m}) \phi_{j}(\textbf{q}', \nu_m)
    \sum_{\omega_n \nu_m} \mathcal{G}^0_{\xi}(k_1 - q_z, \omega_n + \nu_m) \mathcal{G}^0_{\bar{\xi}}(k_1 + \xi b - (q_z+q'_z), \omega_n).
\eea
\end{widetext}
After introducing $k = k_1 - q_z$ to remove $q_z$, using the approximation $\xi b - q_z' \approx 2k_F$, and introducing the change of variable $\omega_n \to \omega_n - \nu_m$, we obtain
\bea
\sum_{k \omega_n} \mathcal{G}^0_{\xi}(k, \omega_n) \mathcal{G}^0_{\bar{\xi}}(k - 2k_F, \omega_n - \nu_m) 
    =\Pi_P(\nu_m).
\eea
The sum over $q_x$ and $q_y$ is calculated in the continuum limit for the forward scattering, leading to Eq.~\eqref{eq:D_sum_qx_qy_forward}.

Therefore, the contribution to the beta function of the electron-phonon vertex arising from the forward channel is given by
\begin{equation}
    \left[ \frac{d}{dl} \log \left( z_{\xi \bar{\xi}} \right) \right]_F = g_2 \frac{1}{4\pi^2 \ell_A^2} \frac{\lambda_P(l)}{1 + \delta v / 2 v_F} \, ,   \label{eq:D_beta-ep-forward}
\end{equation}
where the magnetic length $\ell_A$ is defined in Eq.~\eqref{eq:D_length-scales-2}.

After adding the two contributions to the beta function from Eq.~\eqref{eq:D_beta-ep-backward} and Eq.~\eqref{eq:D_beta-ep-forward}, the full electron-phonon vertex renormalization flow is given by the equation
\begin{equation}
    \frac{d}{dl} \log \left( z_{\xi \bar{\xi}} \right) = \frac{1}{4\pi^2 \ell_A^2} \frac{\lambda_P(l)}{1 + \delta v / 2 v_F} \left( g_2 - g_1\frac{\pi}{2} \frac{\ell_A^2}{\ell_G^2} \right).   \label{eq:beta-ep}
\end{equation}

\subsection{One-loop renormalization of the phonon propagator}

The electron-phonon interaction is expressed by the term defined by Eq.~\eqref{eq:D_S_ep_final} in the action. The sum over $X_{1,\xi}$ is performed in the continuum limit, as follows  
\bea
&&\sum_{X_{1,\xi}}e^{i\frac{\ell_{\bar{\xi}}^2}{\ell_A^2}(q'_x+q_x)X_{1,\xi}}=\frac{L_y}{2\pi \ell_{\xi}^2}\int_{-\infty}^{\infty}dX_{1,\xi}e^{i\frac{\ell_{\bar{\xi}}^2}{\ell_A^2}(q'_x+q_x)X_{1,\xi}}\notag\\
    &&= \frac{L_y \ell_A^2}{2\pi\ell_G^4} (2\pi)\delta(q'_x+q_x).
\eea
Replacing this result, we obtain
\bea
&&\frac{1}{2}\left\langle S_{ep}^2\right\rangle_{dl}=
-\frac{1}{4}\frac{v_F}{\beta L_z} \sum_{\xi}\sum_{\mathbf{q},j,\nu_m}\sum_{k}\sum_{\omega_n}z^2_{\xi\bar{\xi}}g^2_{\xi\bar{\xi},j} \frac{1}{\ell_G^2}\nn\\
    &&\times \mathcal{G}_{\xi}^0(k,\omega_n)\mathcal{G}_{\bar{\xi}}^0(k+\xi b-q_z,\omega_n+\nu_m) |\Phi_{j}(\mathbf{q},\nu_m)|^2\nn\\
    && =\sum_{\mathbf{q},j,\nu_m}\frac{\lambda_P(T, \delta v, \nu_m) dl}{2} \frac{z^2_{+-}g^2_{+-,j}}{2\pi\hbar(1+\delta v / 2v_F)\ell_G^2}|\Phi_{j}(\mathbf{q},\nu_m)|^2\nn\\
\eea
where we have used the definition of the chiral-dependent magnetic lengths in Eq.~\eqref{eq:lchi}, and the result given in Eq.~\eqref{eq:I_P}. Then, to order $dl$ the renormalized inverse phonon propagator is
\bea
    &&\mathcal{D}^{-1}_{j}(\mathbf{q},\nu_m,l+dl)=\mathcal{D}^{-1}_{j}(\mathbf{q},\nu_m,l)\\
    &&-\frac{g^2_{+-,j}}{4\pi(1+\delta v / 2v_F)\ell_G^2}\,z^2_{+-}(l)\lambda_P(l) dl,\nn
\eea
and therefore we obtain the corresponding renormalization flow differential equation,
\begin{equation}
   \frac{d}{dl}\mathcal{D}^{-1}_j(\mathbf{q},\nu_m,l) =-\frac{g^2_{+-,j} \, \lambda_P(l)}{4\pi(1+\delta v / 2v_F)\ell_G^2}\,z^2_{+-}(l).
\end{equation}
The solution of this differential equation, subject to the initial condition $\mathcal{D}^{-1}_j(\mathbf{q},\nu_m,l=0)=\mathcal{D}_{j,0}^{-1}(\mathbf{q},\nu_m)$ after Eq.~\eqref{eq:D0}, is given by
\bea
&&\mathcal{D}^{-1}_j(\mathbf{q},\nu_m,l)=\mathcal{D}_{j,0}^{-1}(\mathbf{q},\nu_m)\\ 
&&-\frac{g^2_{+-,j}\, }{4\pi(1+\delta v / 2v_F)\ell_G^2}\int_0^l z^2_{+-}(l')\lambda_P dl'\nn\\
&&=\nu_m^2+\nu_{0,j}^2(\mathbf{q})-\frac{g^2_{+-,j}\, }{4\pi(1+\delta v / 2v_F) \ell_G^2}\int_0^l z^2_{+-}(l')\lambda_P(l') dl'.\nn
\eea
By defining $g'_j=g_{+-,j}/\nu_{0,j}(\mathbf{q})$, we can finally write
\bea
\mathcal{D}^{-1}_j(\mathbf{q},\nu_m,l)&=&\nu_m^2+\nu_{0,j}^2(\mathbf{q})\left(1-\frac{g'^2_j\, }{4\pi \ell_G^2(1+\delta v/2v_F)}\right.\nn\\
&&\left.\times\int_0^l z^2_{+-}(l')\lambda_P dl'\right).\label{eq:D_D(l)}
\eea
\section{The polarization function in the RPA}
\label{app_RPA}
As defined in the main text, the dielectric function in the random phase approximation (RPA) is given by the expression
\bea
    \Pi_{\text{RPA}}(\mbf{q},\nu_m) &=& \frac{1}{\mathcal{V}} \SumInt_{\alpha \alpha'} \left| \int d^3x \; \Psi^{\dag}_{\alpha}(\textbf{x}) e^{i\textbf{q} \cdot \textbf{x}} \Psi_{\alpha'}(\textbf{x}) \right|^2\nn\\ 
    &&\times\frac{f_0(E_{\alpha}) - f_0(E_{\alpha'})}{E_{\alpha} - E_{\alpha'} + \hbar \nu_m + i\eta}, \label{eq:E_polarization-definition}
\eea
where $f_0(E) = \left( \exp(\beta (E - \mu)) + 1 \right)^{-1}$ is the Fermi distribution function and $\eta \to 0^+$ is an infinitesimal regulator. In the high pseudo-magnetic field regime, we can approximate our calculations by restricting ourselves to the chiral Landau level, by considering $\xi = \xi'$ and $n=n'=0$ in the sets of indices $\alpha$ and $\alpha'$.
Using the result \eqref{eq:matrix-element-chiral-level-01},
\bea
    &&\left| \int d^3x \; \Psi^{\dag}_{\xi 0 X_{\xi} k}(\textbf{x}) e^{i\textbf{q} \cdot \textbf{x}} \Psi_{\xi 0 X'_{\xi} k'}(\textbf{x}) \right|^2\nn\\ 
    &&= \delta_{k',k-q_z} \delta_{X'_{\xi}, X_{\xi} + \ell_{\xi}^2 q_y} e^{-\frac{1}{2}\ell_{\xi}^2 q_{\perp}^2} \, .
\eea
Evaluating this in \eqref{eq:E_polarization-definition}, using the deltas and taking immediately the static limit $\nu_m \to 0$ (and $\eta \to 0$),
\begin{widetext}
\bea
    \Pi_{\text{RPA}}(\mbf{q},0) &=& \frac{1}{\mathcal{V}}\sum_{\xi} \sum_{X_{\xi}} \sum_k e^{-\frac{1}{2}\ell_{\xi}^2 q_{\perp}^2} \frac{\theta\left( -E_{\xi 0 k} \right) - \theta\left( -E_{\xi 0 (k-q_z)} \right)}{E_{\xi 0 k} - E_{\xi 0 (k-q_z)}}  \nonumber \\
    &=& \frac{1}{\mathcal{V}}\sum_{\xi} \left( \frac{L_y}{2\pi \ell_{\xi}^2} \right) \int_{-L_y/2}^{L_y/2} dX_{\xi} \left( \frac{L_z}{2\pi} \right) \int_{-\infty}^{\infty} dk e^{-\frac{1}{2}\ell_{\xi}^2 q_{\perp}^2} \frac{\theta\left( -\xi b_0 - \xi \hbar v_{\xi}k \right) - \theta\left( -\xi b_0 - \xi \hbar v_{\xi}(k - q_z) \right)}{\xi \hbar v_{\xi} q_z}  \nonumber \\
    &=& \frac{L_y}{(2\pi)^2 L_x} \sum_{\xi} \frac{e^{-\frac{1}{2}\ell_{\xi}^2 q_{\perp}^2}}{\hbar v_{\xi} q_z \ell_{\xi}^2} \int_{-\infty}^{\infty} dk \left[ \theta\left( -\xi b_0 - \xi \hbar v_{\xi}k \right) - \theta\left( -\xi b_0 - \xi \hbar v_{\xi}(k - q_z) \right) \right] \, . \label{eq:polarizacion-01}
\eea
\end{widetext}
For the $k$-integral we have
\bea
    &&\int_{-\infty}^{\infty} dk \left[ \theta\left( -\xi b_0 - \xi \hbar v_{\xi}k \right) - \theta\left( -\xi b_0 - \xi \hbar v_{\xi}(k - q_z) \right) \right] \nonumber \\
    &&= \, \xi \int_{-\infty}^{\infty} dk \left[ \theta\left( - b_0 - \hbar v_{\xi}k \right) - \theta\left( - b_0 - \hbar v_{\xi}(k - q_z) \right) \right] \nonumber \\
    &&= \, - \xi \int_{-b_0/\hbar v_{\xi}}^{q_z - b_0/\hbar v_{\xi}} dk  \nonumber \\
    &&= \, -\xi q_z \, ,
\eea
where we used the identity $\theta(x) = 1 - \theta(-x)$ to obtain the second expression from the first one. Evaluating in Eq.~\eqref{eq:polarizacion-01},
\bea
    &&\Pi_{\text{RPA}}(\mbf{q},0) = -\frac{L_y}{(2\pi)^2 L_x} \sum_{\xi} \frac{e^{-\frac{1}{2}\ell_{\xi}^2 q_{\perp}^2}}{\hbar v_{\xi} \ell_{\xi}^2}  \\
    &&= -\frac{L_y}{(2\pi)^2 L_x} \left( \frac{e^{-\frac{1}{2}\ell_{+}^2 q_{\perp}^2}}{\hbar v_{+} \ell_{+}^2} + \frac{e^{-\frac{1}{2}\ell_{-}^2 q_{\perp}^2}}{\hbar v_{-} \ell_{-}^2} \right) \nonumber \\
    &&= -\frac{L_y}{L_x} \frac{1}{(2\pi)^2 \hbar v_F \ell_+^2} \left( e^{-\frac{1}{2}\ell_{+}^2 q_{\perp}^2} + \frac{\varepsilon_{+}^2}{1 + \delta v / v_F} e^{-\frac{1}{2}\ell_{-}^2 q_{\perp}^2}\right)\nn .
\eea
Based on this result, we can obtain the limit of the long-range Coulomb interaction as follows
\begin{align}
    \lim_{q\to 0} \frac{V(\mbf{q})}{\epsilon(\mbf{q},0)} &= \lim_{q\to 0} \frac{V(\mbf{q})}{1 - V(\mbf{q})\Pi_{\text{RPA}}(\mbf{q},0)} \nonumber \\
    &= \lim_{q\to 0} \frac{e^2/\epsilon_0 \epsilon_{\infty}}{q_z - (e^2/\epsilon_0 \epsilon_{\infty})\Pi_{\text{RPA}}(\mbf{q},0)} \nonumber \\
    &= \frac{-1}{\Pi_{\text{RPA}}(0,0)} \nonumber \\
    &= \frac{L_x}{L_y} (2\pi)^2 \hbar v_F \ell_+^2 \left(1 + \frac{\varepsilon_{+}^2}{1 + \delta v /v_F} \right)^{-1} \nonumber \\
    &= \frac{L_x}{L_y} (2\pi)^2 \hbar v_F \ell_+^2 \frac{1 + \delta v /v_F}{1 + \varepsilon_{+}^2 + \delta v /v_F} \nonumber \\
    &= \frac{L_x}{L_y} (2\pi)^2 \hbar v_F \ell_G^2 \frac{1 + \delta v /v_F}{2 \ell_A^2/\ell_G^2 + \varepsilon_- \cdot \delta v / v_F} \, .
\end{align}
From the definition \eqref{eq:D_g2-definition} and taking the thermodynamic limit ($L_x / L_y \to 1$),
\begin{equation}
    g_2 = 4\pi \ell_G^2 \, \frac{1 + \frac{\delta v}{v_F}}{2\frac{\ell_A^2}{\ell_G^2} + \varepsilon_- \cdot \frac{\delta v}{v_F}} \, .
\end{equation}

We can alternatively express Eq.~\eqref{eq:D_g2-definition}, by introducing the Thomas-Fermi screening wave vector $q_{TF}$, such that
\begin{equation}
    g_2 \pi \hbar v_F = \frac{e^2}{\epsilon_0 \epsilon_{\infty}q^2_{TF}} \, ,
\end{equation}
where
\begin{equation}
    q^2_{TF} = \frac{e^2}{\epsilon_0 \epsilon_{\infty}} \nu(E_F)
\end{equation}
and
\begin{equation}
    \nu(E_F) = \frac{L_y}{L_x} \cdot \frac{1}{4\pi^2 \hbar v_F \ell_G^2} \cdot \frac{2\frac{\ell_A^2}{\ell_G^2} + \varepsilon_- \cdot \frac{\delta v}{v_F}}{1 + \frac{\delta v}{v_F}}
\end{equation}
is the density of states at the Fermi level.
\section{Polarization insertion in the Peierls channel}
\label{App_Peierls}
As explained in the main text, the bare Fermion propagators in the Matsubara space, for the lowest Landau level, are given by the expression
\begin{equation}
    \mathcal{G}^0_{\xi}\left(k, \omega_{n}\right) = \frac{1}{i\hbar \omega_n - \xi \hbar v_{\xi} (k - k_{F_{\xi}})} \, ,
\end{equation}
where $k$ and $k_{F_{\xi}}$ are measured from an arbitrary origin in $k$-space. We can approximate the relative difference between external fermionic momenta in \eqref{eq:correction-F_P^2-aux1} as $k'_1 - k_2 \approx k_{\xi} - k_{\bar{\xi}} \equiv \xi b$ because each momentum is very close to a different respective node. Note that $b>0$ when we choose a reference system in which $k_+ > k_-$, but $b<0$ in the other case. Also, introducing $\nu_m = \omega'_{n_1} - \omega'_{n_3}$, the Matsubara sum and the $k$-integral can be expressed as (for $\xi b - q_z  \approx 2 k_{F}$)
\bea 
&&\sum_{k} \sum_{\omega_{n}} \mathcal{G}^0_{\xi}\left(k+\xi b-q_{z}, \omega_{n}\right) \mathcal{G}^0_{\bar{\xi}}\left(k, \omega_{n}-\nu_{m}\right)\nonumber \\
&&=\sum_{k} \sum_{\omega_{n}} \mathcal{G}^0_{\xi}\left(k+2 k_{F}, \omega_{n}\right) \mathcal{G}^0_{\bar{\xi}}\left(k, \omega_{n}-\nu_{m}\right) \nonumber \\
&&=\sum_{k} \sum_{\omega_{n}} \mathcal{G}^0_{\xi}\left(k+2 k_{F}, \omega_{n}+\nu_{m}\right) \mathcal{G}^0_{\bar{\xi}}\left(k, \omega_{n}\right) \, , \label{eq:F_peierls-symmetrization-1}
\eea
where in the second step, we shifted $\omega_n \to \omega_n + \nu_m$. 
Generating an equivalent expression but changing $\omega_{n} \rightarrow \omega_{n}-\nu_{m}$ on that term, we can add them to obtain the equivalent combination
\bea
    &&\frac{1}{2} \sum_{k} \sum_{\omega_{n}}\left[\mathcal{G}^0_{\xi}\left(k+2 k_{F}, \omega_{n}+\nu_{m}\right) \mathcal{G}^0_{\bar{\xi}}\left(k, \omega_{n}\right)\right.\nn\\ 
    &&\left.+ \mathcal{G}^0_{\xi}\left(k+2 k_{F}, \omega_{n}\right) \mathcal{G}^0_{\bar{\xi}}\left(k, \omega_{n}-\nu_{m}\right)\right] .
\eea
Applying the node translation $k \to k - 2 k_{F}$,
and explicitly evaluating at the nodes $\xi=+1$ and $\xi=-1$, we obtain the expression for the polarization insertion $\Pi_P(\nu_m)$ in the Peierls channel
\bea
&&\Pi_P(\nu_m) \equiv\frac{1}{2} \sum_{k} \sum_{\omega_{n}}\left[\mathcal{G}^0_{+}\left(k+2 k_{F}, \omega_{n} + \nu_m \right) \mathcal{G}^0_{-}\left(k, \omega_{n}\right)\right.\nn\\
&&\left.+ \mathcal{G}^0_{+}\left(k, \omega_{n}\right) \mathcal{G}^0_{-}\left(k - 2 k_{F}, \omega_{n} - \nu_m\right)\right] \, . \label{eq:F_symmetrization-peierls-loop}
\eea

Let us consider the first term in this expression, where we note that
\begin{align}
k-2 k_{F}-k_{F_{-}} = k-\left(k_+ - k_-\right)-k_-  = k-k_{F_{+}}
\end{align}
Then, we shall evaluate the sum over $k$ as an integral in the continuum limit, and there we shall define the integration variable $\epsilon_{+}$ as follows
\begin{equation}
\epsilon_{+} \equiv \hbar v_{F}\left(k-k_{F_+}\right),
\end{equation}
with $v_{F}=v_{+}$ and $v_{-}=v_{F}+\delta v=v_{F}\left(1+\frac{\delta v}{v_{F}}\right)$. Introducing the parameter
\begin{equation}
    \gamma \equiv 1 + \frac{\delta v}{v_F},
\end{equation}
and the aforementioned change of variable, we have
\bea
    &&\sum_{k} \sum_{\omega_{n}} \mathcal{G}^0_{+}\left(k, \omega_{n}\right) \mathcal{G}^0_{-}\left(k - 2 k_{F}, \omega_{n} - \nu_m\right) \nonumber \\
    &&= \frac{L_z}{2\pi\hbar v_F} \left[ \int_{\Lambda(l)}^{\Lambda} d\epsilon_+ + \int_{-\Lambda}^{-\Lambda(l)} d\epsilon_+ \right]\nn\\ 
    &&\times\frac{1}{\hbar^2} \sum_{\omega_n} \frac{1}{i\omega_n - \epsilon_+/\hbar} \frac{1}{i\omega_n - i\nu_m + \gamma \epsilon_+/ \hbar }.
\eea

Ww first perform the Matsubara frequency sum, to obtain
\begin{widetext}
\begin{align}
    \frac{1}{\hbar^2} \sum_{\omega_n} \frac{1}{i\omega_n - \epsilon_+/\hbar} \frac{1}{i\omega_n - i\nu_m + \gamma \epsilon_+/ \hbar } &= \frac{\beta}{\epsilon_{+}\left(1+\gamma\right)-i\hbar \nu_m}\left[f_{0}\left(\epsilon_{+}/ \hbar\right) - f_{0}\left(i\nu_m-\gamma \epsilon_{+}/\hbar\right)\right] \nonumber \\
    &\hspace{-3.5cm} = -\frac{1}{2}\frac{\beta}{\epsilon_{+}\left(1+\gamma\right)} \frac{1}{1-i\frac{\hbar \nu_m}{(1+\gamma)\epsilon_+}} \left[ \tanh\left( \frac{\beta \epsilon_+}{2} \right) + \tanh\left( \frac{\beta \epsilon_+}{2} \gamma \right) \right] \, ,
    \label{eq_F_sumMats}
\end{align}
\end{widetext}
where we applied the property of the Fermi distribution function $f_0(z + i \nu_m) = f_0(z)$ when $\nu_m = 2\pi m/\beta$ is a Bosonic frequency, since $e^{i\hbar\beta\nu_m}=e^{i2\p m}=1$. Notice that the resulting expression in Eq.~\eqref{eq_F_sumMats} is an even function of $\epsilon_+$. Therefore, for the remaining $\epsilon_+$-integral, consider an even function $F(\epsilon_+)$, and note that
\begin{eqnarray}
    &&\frac{1}{2} \left[ \int_{\Lambda(l)}^{\Lambda} d\epsilon_+ + \int_{-\Lambda}^{-\Lambda(l)} d\epsilon_+ \right] F(\epsilon_+) = \int_{\Lambda(l)}^{\Lambda} d\epsilon_+ \, F(\epsilon_+)\nn\\ &&\approx  F(\Lambda) \int_{\Lambda e^{-dl}}^{\Lambda} d\epsilon_+ = F(\Lambda) \Lambda dl \, , 
\end{eqnarray}
where $\Lambda$ is the energy associated to the momentum cutoff $\Lambda_0$ at the chiral Landau level. Here, we have used the mean value theorem to integrate in the infinitesimal range. Thus,
we obtain the result
\bea
    &&\sum_{k} \sum_{\omega_{n}} \mathcal{G}^0_{+}\left(k, \omega_{n}\right) \mathcal{G}^0_{-}\left(k - 2 k_{F}, \omega_{n}\right) = \frac{-\beta L_z}{2\pi\hbar v_F (1+\gamma)} \nn\\
    &&\times\frac{dl}{1-i\frac{\hbar \nu_m}{(1+\gamma)\Lambda}} \left[ \tanh\left( \frac{\beta \Lambda}{2} \right) + \tanh\left( \frac{\beta \Lambda}{2} \gamma \right) \right] . \label{eq:F_peierls-loop-01}
\eea

On the other hand, for the second term in Eq.~\eqref{eq:F_symmetrization-peierls-loop}, we have
\bea
    &&\sum_{k} \sum_{\omega_{n}} \mathcal{G}^0_{+}\left(k+2 k_{F}, \omega_{n}+\nu_{m}\right) \mathcal{G}^0_{-}\left(k, \omega_{n}\right) \nonumber \\
    &&= \sum_k \sum_{\omega_n} \frac{1}{i\hbar \omega_n + i\hbar \nu_m - \hbar v_+ (k + 2k_F - k_{F_+})}\nn\\
    &&\times\frac{1}{i\hbar \omega_n + \hbar v_- (k - k_{F_-})} \, . 
\eea
In this case, the common momentum variable is $k - k_{F_-}$, so we define the energy variable
\begin{equation}
    \epsilon_- \equiv \hbar v_- (k - k_{F_-}) = \hbar \gamma v_F (k - k_{F_-})
\end{equation}
for integration. Then, the last expression reduces to
\bea
    &&\sum_{k} \sum_{\omega_{n}} \mathcal{G}^0_{+}\left(k+2 k_{F}, \omega_{n}+\nu_{m}\right) \mathcal{G}^0_{-}\left(k, \omega_{n}\right) 
    = \frac{-\beta L_z}{2\pi\hbar v_F (1+\gamma)}\nn\\ 
    &&\times\frac{dl}{1-i\frac{\hbar \nu_m}{(1+\gamma)\Lambda}} \left[ \tanh\left( \frac{\beta \Lambda}{2} \right) + \tanh\left( \frac{\beta \Lambda}{2 \gamma} \right) \right] \, . \label{eq:F_peierls-loop-02}
\eea

Thus, summing \eqref{eq:F_peierls-loop-01} and \eqref{eq:F_peierls-loop-02}, we have
\begin{widetext}
\begin{eqnarray}
    \Pi_P(\nu_m) &=& \frac{1}{2} \sum_{k} \sum_{\omega_{n}}\left[\mathcal{G}^0_{+}\left(k+2 k_{F}, \omega_{n} + \nu_m\right) \mathcal{G}^0_{-}\left(k, \omega_{n}\right) + \mathcal{G}^0_{+}\left(k, \omega_{n}\right) \mathcal{G}^0_{-}\left(k - 2 k_{F}, \omega_{n}-\nu_m\right)\right] \nonumber \\
    &=& -\frac{\beta L_z}{\pi \hbar v_F} \cdot \frac{1}{1+\gamma} \cdot \frac{dl}{1-i\frac{\hbar \nu_m}{(1+\gamma)\Lambda}} \frac{1}{4} \left[ 2\tanh\left( \frac{\beta \Lambda}{2} \right) + \tanh\left( \frac{\beta \Lambda}{2} \gamma \right) + \tanh\left( \frac{\beta \Lambda}{2\gamma}  \right) \right] \nonumber \\
    &=& -\frac{\beta L_z}{\pi \hbar v_F} \cdot \frac{dl}{1+\gamma} \lambda_P(T,\delta v, \nu_m) \nn \\
    &=& -\frac{\beta L_z}{2\pi \hbar v_F} \cdot \frac{dl}{1+\delta v/2v_F} \lambda_P(T,\delta v, \nu_m) \, , \label{eq:F_I_P}
\end{eqnarray}
\end{widetext}
where we have defined
\bea
    &&\lambda_P(T,\delta v, \nu_m,l) \equiv \frac{1}{1-i\frac{\hbar \nu_m}{(1+\gamma)\Lambda(l)}} \frac{1}{4} \left[ 2\tanh\left( \frac{\beta \Lambda(l)}{2} \right)\right.\nn\\ 
    &&\left.+ \tanh\left( \frac{\beta \Lambda(l)}{2} \gamma \right) + \tanh\left( \frac{\beta \Lambda(l)}{2\gamma}  \right) \right] \, . \label{eq:F_lambda_P}
\eea
\section{Polarization insertion in the Cooper channel}
\label{App_Cooper}
In this appendix, we present the details of the calculation of the polarization insertion $\Pi_C(\nu_m)$ in the Cooper channel. For simplicity in the notation,
we start by defining $\nu_m = \omega_{n_1} + \omega_{n_2}$, and then the Matsubara sum and the $k$-integral can be symmetrized following the same procedure explained from \eqref{eq:F_peierls-symmetrization-1} to \eqref{eq:F_symmetrization-peierls-loop}, as follows (using $ k_1 + k_2 \approx k_+ + k_-$)
\bea 
&&\Pi_C(\nu_m) \equiv \sum_{k} \sum_{\omega_{n}} \mathcal{G}^0_{\xi}\left(k_{1}+ k_{2}- k, \nu_m - \omega_{n}\right) \mathcal{G}^0_{\bar{\xi}}\left(k, \omega_{n}\right)  \nonumber \\
&& \approx \sum_{k} \sum_{\omega_{n}} \mathcal{G}^0_{\xi}\left(k_{+}+ k_{-}- k, \nu_m - \omega_{n}\right) \mathcal{G}^0_{\bar{\xi}}\left(k, \omega_{n}\right)  \nonumber \\
&& = \frac{1}{2}\sum_{k} \sum_{\omega_{n}} \left[ \mathcal{G}^0_{+}\left(k, \omega_{n}\right) \mathcal{G}^0_{-}\left(k_{+}+ k_{-}- k, \nu_m - \omega_{n}\right) \right. \nonumber \\
&& \left.+ \mathcal{G}^0_{+}\left(k_{+}+ k_{-}- k, \nu_m - \omega_{n}\right) \mathcal{G}^0_{-}\left(k, \omega_{n}\right) \right] .   \label{eq:G_symmetrization-cooper-loop}
\eea

The first term is
\bea
    &&S_1 \equiv \sum_{k} \sum_{\omega_{n}} \mathcal{G}^0_{+}\left(k, \omega_{n} \right) \mathcal{G}^0_{-}\left(k_+ + k_- - k, \nu_m -\omega_{n}\right) \nonumber \, \\
    &&= \sum_k \sum_{\omega_n} \frac{1}{i\hbar \omega_n - \hbar v_+ (k - k_{F_+})}\nn\\ 
    &&\times\frac{1}{-i\hbar \omega_n + i\hbar\nu_m + \hbar v_- (k_+ + k_- - k - k_{F_-})} \, , 
\eea
we note that
\bea
&&\hbar v_- (k_+ + k_- - k - k_{F_-}) 
= -\hbar v_- \left\{ (k - k_{F_+})\right.\nn\\
&&\left.+ \left[ (k_{F_+} - k_+) + (k_{F_-} - k_-) \right] \right\} \nonumber \\
    &&= -\left( 1 + \frac{\delta v}{v_F} \right) \hbar v_F \left\{ (k - k_{F_+})\right.\nn\\
    &&\left.+ \left[ (k_{F_+} - k_+) + (k_{F_-} - k_-) \right] \right\} \, .
\eea

Therefore, we define the integration variable
\begin{equation}
\epsilon_{+ } = \hbar v_{F}\left(k-k_{F_+}\right), 
\end{equation}
and the parameter
\begin{equation}
    \Delta \equiv \hbar v_F \left[ (k_{F_+} - k_+) + (k_{F_-} - k_-) \right] \, ,
\end{equation}
such that we obtain for $S_1$
\bea
    &&S_1 = \frac{L_z}{2\pi\hbar v_F} \left[ \int_{\Lambda(l)}^{\Lambda} d\epsilon_+ + \int_{-\Lambda}^{-\Lambda(l)} d\epsilon_+ \right]\nn\\
    &&\times\frac{1}{\hbar^2} \sum_{\omega_n} \frac{1}{i\omega_n - \epsilon_+/\hbar} \frac{-1}{i\omega_n - i\nu_m + \gamma (\epsilon_+ + \Delta)/ \hbar } \, .
\eea

We evaluate the Matsubara sum as follows
\bea
    &&\sum_{\omega_n} \frac{1}{i\omega_n - \epsilon_+/\hbar} \frac{-1}{i\omega_n - i\nu_m + \gamma (\epsilon_+ + \Delta)/ \hbar } \nonumber \\
    &&= \frac{-\beta}{\epsilon_{+} + \gamma(\epsilon_{+} + \Delta) - i\hbar\nu_m} \left[f_{0}\left(\epsilon_{+}/ \hbar\right)\right.\nn\\
    &&\left.- f_{0}\left(i\nu_m-\gamma (\epsilon_{+} + \Delta)/\hbar \right)\right] \nonumber \\
    &&= \frac{1}{2}\frac{\beta}{\epsilon_{+}\left(1+\gamma\right) + \gamma \Delta - i\hbar\nu_m} \left[ \tanh\left( \frac{\beta \epsilon_+}{2} \right)\right.\nn\\ 
    &&\left.+ \tanh\left( \frac{\beta (\epsilon_+ + \Delta)}{2} \gamma \right) \right] \, . 
\eea
We perform the integral in $\epsilon_+$ by using the mean value theorem,
\begin{widetext}
\begin{align}
    &\left[ \int_{\Lambda(l)}^{\Lambda} d\epsilon_+ + \int_{-\Lambda}^{-\Lambda(l)} d\epsilon_+ \right] \frac{1}{\epsilon_{+}\left(1+\gamma\right) + \gamma \Delta -i\hbar\nu_m} \left[ \tanh\left( \frac{\beta \epsilon_+}{2} \right) + \tanh\left( \frac{\beta (\epsilon_+ + \Delta)}{2} \gamma \right) \right]  \, \nonumber \\  
    &= \frac{dl}{1+\gamma} \left\{ \frac{1}{1 + \frac{\gamma}{1 + \gamma} \frac{\Delta}{\Lambda} - i\frac{\hbar\nu_m}{(1+\gamma)\Lambda}} \left[ \tanh\left( \frac{\beta \Lambda}{2} \right) + \tanh\left( \frac{\beta (\Lambda + \Delta)}{2} \gamma \right) \right] \right. \nonumber \\
    &\hspace{2cm} \left. + \, \frac{1}{1 - \frac{\gamma}{1 + \gamma} \frac{\Delta}{\Lambda} + i\frac{\hbar\nu_m}{(1+\gamma)\Lambda}} \left[ \tanh\left( \frac{\beta \Lambda}{2} \right) + \tanh\left( \frac{\beta (\Lambda - \Delta)}{2} \gamma \right) \right] \right\}    \, \nonumber \\  
    &= \frac{dl}{1+\gamma} \left\{ \frac{1}{1 + \frac{\gamma}{1 + \gamma} \frac{\Delta}{\Lambda}} \cdot \frac{1}{1-i\frac{1}{1+\frac{\gamma}{1+\gamma} \frac{\Delta}{\Lambda}} \frac{\hbar\nu_m}{(1+\gamma)\Delta}} \left[ \tanh\left( \frac{\beta \Lambda}{2} \right) + \tanh\left( \frac{\beta (\Lambda + \Delta)}{2} \gamma \right) \right] \right. \nonumber \\
    &\hspace{2cm} \left. + \, \frac{1}{1 - \frac{\gamma}{1 + \gamma} \frac{\Delta}{\Lambda}} \cdot \frac{1}{1+i\frac{1}{1-\frac{\gamma}{1+\gamma} \frac{\Delta}{\Lambda}} \frac{\hbar\nu_m}{(1+\gamma)\Delta}} \left[ \tanh\left( \frac{\beta \Lambda}{2} \right) + \tanh\left( \frac{\beta (\Lambda - \Delta)}{2} \gamma \right) \right] \right\}    \, 
\end{align}
\end{widetext}
Then, substituting into $S_1$ we obtain
\begin{widetext}
\begin{align}
    S_1 &= \frac{\beta L_z}{2\pi\hbar v_F} \frac{dl}{1+\gamma} \left\{ \frac{1}{1 + \frac{\gamma}{1 + \gamma} \frac{\Delta}{\Lambda}}\cdot \frac{1}{1- i\frac{1}{1+ \frac{\gamma}{1+\gamma} \frac{\Delta}{\Lambda}} \frac{\hbar\nu_m}{(1+\gamma)\Delta}} \left[ \tanh\left( \frac{\beta \Lambda}{2} \right) + \tanh\left( \frac{\beta (\Lambda + \Delta)}{2} \gamma \right) \right] \right. \nonumber \\
    &\hspace{2cm} \left. + \, \frac{1}{1 - \frac{\gamma}{1 + \gamma} \frac{\Delta}{\Lambda}}\cdot \frac{1}{1+ i\frac{1}{1- \frac{\gamma}{1+\gamma} \frac{\Delta}{\Lambda}} \frac{\hbar\nu_m}{(1+\gamma)\Delta}} \left[ \tanh\left( \frac{\beta \Lambda}{2} \right) + \tanh\left( \frac{\beta (\Lambda - \Delta)}{2} \gamma \right) \right] \right\} \, . \label{eq:G_cooper-loop-01}
\end{align}
\end{widetext}
Similarly, for the second term in Eq.~\eqref{eq:G_symmetrization-cooper-loop} we have
\bea
    &&S_2 =\sum_{k} \sum_{\omega_{n}} \mathcal{G}^0_{+}\left(k_{+}+ k_{-}- k, \nu_m - \omega_{n}\right) \mathcal{G}^0_{-}\left(k, \omega_{n}\right) \nonumber \\
    &&= \sum_k \sum_{\omega_n} \frac{1}{-i\hbar \omega_n - \hbar v_+ (k_+ + k_- - k - k_{F_+})}\nn\\
    &&\times\frac{1}{i\hbar \omega_n + \hbar v_- (k - k_{F_-})} \, .
\eea
In this case, we define the integration variable
\begin{equation}
    \epsilon_- = \hbar v_- (k - k_{F_-}) = \hbar \gamma v_F (k-k_{F_-}) \, .
\end{equation}
Then, following an identical procedure as before, we obtain
\begin{widetext}
\begin{align}
    S_2 &= \frac{\beta L_z}{2\pi\hbar v_F} \frac{dl}{1+\gamma} \left\{ \frac{1}{1 + \frac{\gamma}{1 + \gamma} \frac{\Delta}{\Lambda}} \cdot \frac{1}{1- i\frac{1}{1+ \frac{\gamma}{1+\gamma} \frac{\Delta}{\Lambda}} \frac{\hbar\nu_m}{(1+\gamma)\Delta}} \left[ \tanh\left( \frac{\beta \Lambda}{2} \right) + \tanh\left( \frac{\beta (\Lambda + \gamma \Delta)}{2\gamma} \right) \right] \right. \nonumber \\
    &\hspace{2cm} \left. + \, \frac{1}{1 - \frac{\gamma}{1 + \gamma} \frac{\Delta}{\Lambda}} \frac{1}{1+ i\frac{1}{1- \frac{\gamma}{1+\gamma} \frac{\Delta}{\Lambda}} \frac{\hbar\nu_m}{(1+\gamma)\Delta}} \left[ \tanh\left( \frac{\beta \Lambda}{2} \right) + \tanh\left( \frac{\beta (\Lambda - \gamma \Delta)}{2\gamma} \right) \right] \right\}   \label{eq:G_cooper-loop-02}
\end{align}
\end{widetext}
Finally, adding Eq.~\eqref{eq:G_cooper-loop-01} and Eq.~\eqref{eq:G_cooper-loop-02}, we obtain the polarization insertion for the Cooper channel as follows 
\begin{widetext}
\begin{align}
    \Pi_C(\nu_m) &= \frac{1}{2}(S_1 + S_2)
    = \frac{\beta L_z}{\pi \hbar v_F} \cdot \frac{dl}{1+\gamma} \cdot \frac{1}{4} \left\{ \frac{1}{1 + \frac{\gamma}{1 + \gamma} \frac{\Delta}{\Lambda}}\cdot \frac{1}{1- i\frac{1}{1+ \frac{\gamma}{1+\gamma} \frac{\Delta}{\Lambda}}  \frac{\hbar\nu_m}{(1+\gamma)\Delta}} \tanh\left( \frac{\beta \Lambda}{2} \right) \right. \nonumber \\
    &\hspace{2cm}+\, \frac{1}{1 - \frac{\gamma}{1 + \gamma} \frac{\Delta}{\Lambda}}\cdot \frac{1}{1+ i\frac{1}{1- \frac{\gamma}{1+\gamma} \frac{\Delta}{\Lambda}} \frac{\hbar\nu_m}{(1+\gamma)\Delta}} \tanh\left( \frac{\beta \Lambda}{2} \right)  \nonumber \\ 
    &\hspace{-0.5cm} \left. + \, \frac{1}{1 + \frac{\gamma}{1 + \gamma} \frac{\Delta}{\Lambda}} \cdot \frac{1}{1- i\frac{1}{1+ \frac{\gamma}{1+\gamma} \frac{\Delta}{\Lambda}} \frac{\hbar\nu_m}{(1+\gamma)\Delta}} \cdot \frac{1}{2} \left[ \tanh\left( \frac{\beta (\Lambda + \Delta)}{2} \gamma \right) + \tanh\left( \frac{\beta (\Lambda + \gamma \Delta)}{2\gamma} \right) \right] \right. \nonumber \\
    &\hspace{-0.5cm} \left. + \, \frac{1}{1 - \frac{\gamma}{1 + \gamma} \frac{\Delta}{\Lambda}} \cdot \frac{1}{1+ i\frac{1}{1- \frac{\gamma}{1+\gamma} \frac{\Delta}{\Lambda}} \frac{\hbar\nu_m}{(1+\gamma)\Delta}} \cdot \frac{1}{2} \left[ \tanh\left( \frac{\beta (\Lambda - \Delta)}{2} \gamma \right)  + \tanh\left( \frac{\beta (\Lambda - \gamma \Delta)}{2\gamma} \right) \right] \right\} \nonumber \\
    &= \frac{\beta L_z}{\pi \hbar v_F} \cdot \frac{dl}{1+\gamma} \lambda_C(T,\delta v, \Delta, \nu_m) \nn \\
    &= \frac{\beta L_z}{2\pi \hbar v_F} \cdot \frac{dl}{1+\delta v / 2v_F} \lambda_C(T,\delta v, \Delta, \nu_m) \, , \label{eq:G_I_C}
\end{align}
\end{widetext}
where we have defined
\begin{widetext}
\begin{align}
    \lambda_C(T, \delta v, \Delta, \nu_m,l) &\equiv \frac{1}{4} \left\{ \frac{1}{1 + \frac{\gamma}{1 + \gamma} \frac{\Delta}{\Lambda(l)}}\cdot \frac{1}{1- i\frac{1}{1+ \frac{\gamma}{1+\gamma} \frac{\Delta}{\Lambda(l)}}  \frac{\hbar\nu_m}{(1+\gamma)\Delta}} \tanh\left( \frac{\beta \Lambda(l)}{2} \right) \right. \nonumber \\
    &\hspace{2cm}+\, \frac{1}{1 - \frac{\gamma}{1 + \gamma} \frac{\Delta}{\Lambda(l)}}\cdot \frac{1}{1+ i\frac{1}{1- \frac{\gamma}{1+\gamma} \frac{\Delta}{\Lambda(l)}} \frac{\hbar\nu_m}{(1+\gamma)\Delta}} \tanh\left( \frac{\beta \Lambda(l)}{2} \right)  \nonumber \\ 
    &\hspace{-1cm} \left. + \, \frac{1}{1 + \frac{\gamma}{1 + \gamma} \frac{\Delta}{\Lambda(l)}} \cdot \frac{1}{1- i\frac{1}{1+ \frac{\gamma}{1+\gamma} \frac{\Delta}{\Lambda(l)}} \frac{\hbar\nu_m}{(1+\gamma)\Delta}} \cdot \frac{1}{2} \left[ \tanh\left( \frac{\beta (\Lambda(l) + \Delta)}{2} \gamma \right) + \tanh\left( \frac{\beta (\Lambda(l) + \gamma \Delta)}{2\gamma} \right) \right] \right. \nonumber \\
    &\hspace{-1cm} \left. + \, \frac{1}{1 - \frac{\gamma}{1 + \gamma} \frac{\Delta}{\Lambda(l)}} \cdot \frac{1}{1+ i\frac{1}{1- \frac{\gamma}{1+\gamma} \frac{\Delta}{\Lambda(l)}} \frac{\hbar\nu_m}{(1+\gamma)\Delta}} \cdot \frac{1}{2} \left[ \tanh\left( \frac{\beta (\Lambda(l) - \Delta)}{2} \gamma \right)  + \tanh\left( \frac{\beta (\Lambda(l) - \gamma \Delta)}{2\gamma} \right) \right] \right\} \, .
\end{align}
\end{widetext}
\end{document}